\documentclass[aps, twoside, nofootinbib]{revtex4}
\usepackage{amsfonts}
\usepackage{amssymb}
\usepackage[mathscr]{eucal}
\usepackage[dvips]{graphicx}
\usepackage{amsfonts,amssymb,amsthm}

\def \ni{\noindent}

\let \pp=\partial
\def \be{\begin{equation}}
\def \ee{\end{equation}}
\def \bes{\begin{eqnarray}}
\def \ees{\end{eqnarray}}
\newcommand{\bra}{\langle}
\newcommand{\ket}{\rangle}

\newcommand{\Cb}{{\rm \bf C}}

\def \sl2{SL(2,\Cb)}

\begin{document}
\title{A New Class of Group Field Theories for 1st Order Discrete Quantum Gravity}
\author{{\bf D. Oriti $^a$}\footnote{d.oriti@phys.uu.nl} and {\bf T. Tlas $^b$}\footnote{t.tlas@damtp.cam.ac.uk}}
\affiliation{\small $^a$ Institute for Theoretical Physics and
Spinoza Institute, Utrecht University, Leuvenlaan 4, Utrecht 3584
TD,
The Netherlands\\ $^b$Department of Applied Mathematics and Theoretical Physics, \\ Centre for Mathematical Sciences, University of Cambridge, \\
Wilberforce Road, Cambridge CB3 0WA, UK}
\begin{abstract}
Group Field Theories, a generalization of matrix models for 2d
gravity, represent a 2nd quantization of both loop quantum gravity
and simplicial quantum gravity. In this paper, we construct a new
class of Group Field Theory models, for any choice of spacetime
dimension and signature, whose Feynman amplitudes are given by
path integrals for clearly identified discrete gravity actions, in
1st order variables. In the 3-dimensional case, the corresponding
discrete action is that of 1st order Regge calculus for gravity
(generalized to include higher order corrections), while in higher
dimensions, they correspond to a discrete BF theory (again,
generalized to higher order) with an imposed orientation
restriction on hinge volumes, similar to that characterizing
discrete gravity. The new models shed also light on the large
distance or semi-classical approximation of spin foam models. This
new class of group field theories may represent a concrete
unifying framework for loop quantum gravity and simplicial quantum
gravity approaches.
\end{abstract}

\maketitle
\section{Introduction}
Group field theories (GFTs) \cite{iogft,iogft2,laurentgft, dgftreview} are
quantum field theories on group manifolds, with the group chosen to be the local gauge group of spacetime in D dimensions, i.e. the
Lorentz group, or a suitable extension of it, for models aiming at
a quantization of D-dimensional gravity. They are characterized by
a non-local pairing of field arguments in the interaction term,
designed in such a way as to produce, in perturbative expansion,
Feynman diagrams with a combinatorial structure that are in 1-1
correspondence with D-dimensional simplicial complexes. Because of
these basic properties, GFTs can be understood as a generalization
of matrix models \cite{mm} for 2-dimensional quantum gravity,
obtained in two steps: 1) by passing to generic tensors, instead
of matrices, as fundamental variables, thus obtaining a generating
functional for the sum over 3d simplicial complexes that was the
essence of the dynamical triangulations approach to 3d quantum
gravity\cite{tensor}; 2) adding group structure defining
extra geometric degrees of freedom. The last step is what turns a
generic tensor model into a proper field theory. In fact, the
first example of a GFT was the group-theoretic generalization of
3d tensor models proposed by Boulatov \cite{boulatov}. It was
already at this initial stage, that group field theories allowed a
direct contact between simplicial quantum gravity and what we now
call spin foam models \cite{SF}. In fact the Boulatov model,
sharing the same combinatorics of tensor models and thus
reproducing a 3d sum over triangulations, in perturbative
expansion, produces weights for these triangulations given by the
so-called Ponzano-Regge spin foam model, thanks to the additional
SU(2) group structure. We now know that this is just one example
of a very general result \cite{mikecarlo}: any group field theory
produces Feynman amplitudes that can be re-expressed as spin foam
models, and, conversely, any spin foam model for fixed 2-complex
can be understood as the Feynman amplitude for a given Feynman
diagram of a corresponding GFT. In turn, spin foam models
\cite{SF} have been a very active (and growing) area of quantum
gravity research in the past ten years, for two main (and related)
reasons. First, one obtains a spin foam model when considering,
once more, a discretization of continuum 1st order gravity,
formulated as a constrained BF theory, on a simplicial complex,
and quantizes it via path integral methods. Second, spin foams as
2-complexes with faces labelled by group representations arise
naturally when considering the dynamics of the kinematical quantum
states of geometry as identified by canonical loop quantum gravity
\cite{LQG}; indeed, from the LQG perspective, spin foams
represents the histories of spin networks and are thus the crucial
ingredient of any path integral or covariant formulation of the
quantum gravity dynamics in LQG. From both the simplicial and
canonical perspective, a sum over spin foams, weighted by
appropriate amplitudes, is necessary to define in full the
dynamics of the gravitational field: in simplicial quantum gravity
because only such sum (lacking a suitably defined refinement
procedure) can compensate the truncation of geometric degrees of
freedom that the restriction to a given lattice imposes; in LQG,
because a complete path integral formulation of the dynamics
needs, in general, i.e. except in the purely topological case or
for very special choices of observables, a sum over all the
histories between given spin network states. At present, group
field theories are the only known tool to define uniquely such sum
over spin foams, i.e. with fully specified weights, in a
perturbative expansion of the GFT partition function. In this
property, lies the main reason of interest in GFTs, from the LQG
perspective. And indeed, up to now, group field theories have been
mainly considered and used just as such: as a {\it tool} to define
a sum over spin foams with prescribed weights, i.e. as an
auxiliary formalism to define/construct spin foam models.

\subsection*{Group field theories and quantum gravity}
A different perspective is however possible, even if only
tentatively or as a working hypothesis, at present, given the very
limited understanding and control we have of the whole group field
theory formalism. This perspective has been advocated and
described in more detail in \cite{iogft,iogft2,ioemergence}, to
which we refer.\\

Group field theories can be understood as a 2nd quantized
formulation of both loop quantum gravity and simplicial gravity,
in which wave functions over the space of geometries are turned
into classical fields first and then into operators. In fact, the
quanta of the GFT field correspond to D-valent spin network
vertices \cite{ioetera} and, at the same time, to (D-1)-simplices.
Since both loop quantum gravity and simplicial quantum gravity (in
its quantum Regge calculus as well as in its dynamical
triangulations formulation) are meant to be themselves a
quantization of what is already a classical field theory, i.e.
General Relativity, a 2nd quantization of the same provides a
formalism very similar in spirit to what was (somewhat improperly)
dubbed \lq\lq third quantization\rq\rq of gravity \cite{3rd}. This
was a formal quantum field theory on superspace (the space of all
3-geometries on, say, $S^3$) which would have canonical quantum GR
corresponding to its quantum 1-particle sector, and would produce
a sum over topologies in perturbative Feynman expansion, obtained
as interaction processes of quantum universes (the \lq\lq
particles\rq\rq of this formalism). Clearly, the mathematical
difficulties as well as those regarding the physical
interpretation of the whole formalism are formidable, and consequently they prevented any substantial development.\\

The group field theory formalism \cite{iogft,iogft2,laurentgft, dgftreview}
greatly improves the situation, both at the mathematical and
physical level, by turning to a discrete and local picture of
superspace, in which what is propagated, created and annihilated
are local chunks of a (D-1)-dimensional quantum space (again, spin
network vertices or (D-1)-simplices). One feature of 3rd
quantization is retained, and even generalized, however: on the
one hand these fundamental building blocks of quantum space can be
combined at the kinematical level to represent, in principle, any
spatial geometry and topology, and on the other hand their
interactions still produce, in perturbative Feynman expansion,
quantum spacetimes of any topology.\\

This step from a continuum to a discrete setting has another important consequence. If the quantum gravity Feynman diagrams in the 3rd
quantized formalism were smooth manifolds, and the continuum path
integral for quantum gravity on the given manifold represented, by
construction, the Feynman amplitude for each of them, in the
discrete version of 3rd quantization provided by group field
theories the Feynman diagrams are given by combinatorial and
un-embedded 2-complexes or, dually, by simplicial complexes, and
the Feynman amplitudes have the interpretation of discrete path
integrals (sum over discrete geometries) over each simplicial
complex.\\

This feature allows not only a better mathematical control on the
quantities involved, but it also makes group field theories, at
least in principle, a common framework for several approaches to
quantum gravity, providing the basis for potential
cross-fertilization and mutual enrichment.\\

As mentioned, each group field theory, in fact completely defines a
possible dynamics for loop quantum gravity spin network states,
thus proposing an explicit solution (even if yet to be put to
test) for one outstanding challenge of the LQG programme. At the
same time, it does so by defining a sum over histories that
includes the crucial ingredients of both main simplicial quantum
gravity approaches. As in quantum Regge calculus \cite{williams},
it defines the dynamics of a quantum and classical simplicial
geometry in terms of a sum over discrete geometric data (edge
lengths, areas, etc, according to dimension) for a given base
simplicial complex. In addition to this, it provides a possible
way to encode all the continuum geometric degrees of freedom by a
covariant procedure that is alternative to the infinite refinement
procedure for fixed incidence matrix, that has proven to be
problematic at the quantum level, in quantum Regge calculus: a sum
over inequivalent triangulations. In this it obviously agrees with
the dynamical triangulations (DT) approach \cite{DT}; in fact,
freezing the discrete geometric data attached to each complex
(this can be done in various ways) turns the perturbative
expansion of GFTs into a form very similar to the sum over
triangulations weighted by a purely combinatorial factor that
characterizes the DT approach; such formulation, in its {\it
causal} restriction (corresponding to summing over a certain
restricted class of Lorentzian triangulations) \cite{DT}, has
recently proven to be much more successful in recovering a
continuum spacetime from the quantum theory. As a bonus, with
respect to the dynamical triangulations approach, the GFT approach
permits a better control over the classical and quantum {\it
simplicial} dynamics of geometry, which is here akin to the \lq
few-particle\rq physics in the usual QFTs. This stems from the
possibility of a finer control over simplicial geometric
variables, that puts GFTs and their spin foam amplitudes in closer
relation with LQG, Regge calculus and other formulations of
classical simplicial gravity. This seems also to be confirmed by
recent results on the spin foam/LQG lattice graviton, which match
at least partially the results of simplicial gravity, in the
regime where a connection between the two approaches can be made
precise \cite{graviton}. The links between group field theories
and other approaches to quantum gravity are detailed in
\cite{iogft,ioemergence}, to which we refer\footnote{We would like
to stress at this point that the crucial open issue that {\it all}
these approaches face, at present, is the possibility of
recovering a continuum description of spacetime, in the
appropriate limit, and (a modified version of) Einstein's General
Relativity as the effective dynamics in the same limit.
Understanding the links between various approaches or even
subsuming them within the GFT formalism may be an interesting
exercise in many respects, and it may be very useful in merging
insights and techniques coming from one approach in another, but
it will remain futile if the problem of the emergence of continuum
physics will not be solved or made easier by doing so. Our point
of view on how, in fact, the group field theory formalism, the new
perspective it suggests, as well as the tools it provides, {\it
may} in fact be crucial in order to tackle and solve this issue is
discussed at length in \cite{ioemergence}.}.\\

It should be clear, however, that much more remains to be
understood about these links and that only further work can
confirm or refute the idea that group field theories really
represent in concrete terms a unifying framework for all of them,
as we have been suggesting. As for the relationship with LQG, the
open issues concern both the exact correspondence between GFT
boundary states and their Hilbert space, in specific models, and
that of usual SU(2) LQG spin networks, as well as the relation
between the GFT transition amplitudes and the canonical LQG
Hamiltonian constraint, defining the quantum dynamics. These issues
will not concern us here. The focus of our present work is instead on the relation between GFTs and simplicial quantum gravity, and the
aim is to make the correspondence between the two framework
detailed and clear, with a precise matching between GFT Feynman
amplitudes and simplicial quantum gravity sum over histories. In
order to achieve this, we introduce and analyze in the present
paper a new class of group field theories, characterized by
Feynman amplitudes which have, in any dimension and in any
signature, {\it exactly} the form of a simplicial gravity path
integral. Its amplitudes, that is, are expressed as a (real)
measure part times a phase factor, with phase clearly identified
with a simplicial gravity action. In 3 dimensions, this will mean
obtaining a path integral for 3-dimensional simplicial gravity in
first order formalism, corresponding to a 1st order Regge calculus
action plus higher order ($f(R)$-like) corrections. In higher
dimensions, we will obtain instead a path integral, augmented by a
sum over triangulations of any topology, for what can be
interpreted as topological BF theory with an additional
orientation dependence, and, again, higher order (quantum)
corrections to the action. This work can be understood as the
further development of the line of research on the issue of
causality in spin foam quantum gravity and GFT, and on the
construction of a unified GFT framework for loop quantum gravity,
spin foam models, quantum Regge calculus and dynamical
triangulations, that started from an analysis of the issue of
causality in spin foam models \cite{causal}, continued with the
development of a refined technique for the construction of causal
spin foam models, based on the particle analogy which implicitly
introduced additional variables into the usual spin foam formalism,
in \cite{feynman}, with the explicit construction of
causal spin foam models for pure gravity and gravity coupled to
matter in 3d \cite{causalmatter3d}, and with the construction of a
generalized GFT formalism \cite{generalised} based on the
techniques and variables introduced in \cite{feynman}. The present
GFT construction is indeed, in a sense, a much improved and
further developed version of the one in \cite{generalised}, in a
sense to be detailed in the following.\\

We will detail both the motivation, the basic ideas and the
results of our work in the next section. In section III we present
the general definition of the new class of GFTs, and the general
structure of its Feynman amplitudes. Section IV and V report
instead the detailed form of the amplitudes of these models in 3
and 4 dimensions, in both Riemannian and Lorentzian cases, and a
discussion of their properties. We conclude with a summary of our
results and an outlook on their relevance for further developments
in this area.

\section{Motivation for the new models}
We assume now that the above perspective on group field theory as
a discrete quantum field theory (3rd quantized) of spin networks
or of simplicial geometry, and as a potential common unifying
framework for different approaches to quantum gravity, is agreed
upon, if tentatively only. Let us now focus on the following
questions: what types of amplitudes we then expect or want the GFT
Feynman diagrams to be assigned? That is, what properties and main features our spin foam amplitudes should have? How should they
look like, if they are indeed Feynman amplitudes for a field
theory on a simplicial superspace?

\subsection{Causality, orientation dependence, 3rd quantization and quantum (discrete) gravity path integrals}
From a path integral for a field theory on superspace, continuum
or discrete, we expect to obtain a definition of causal transition
amplitudes between quantum gravity states, i.e. the quantum
gravity analogue of what, in ordinary field theory, is represented
by expectation values in the vacuum state of time-ordered products
of field operators. However, no time coordinate is allowed, in a
fully background independent and diffeomorphism invariant theory
of quantum gravity, to enter the definition of transition
amplitudes, as it does through the Minkowskian time when we define time-ordered 2-point functions in the usual QFT. Therefore, in quantum
gravity, the difference between the various possible \lq\lq
2-point functions\rq\rq can be characterized purely in terms of
symmetry properties and of other formal features of them,
independent of any spacetime coordinates. Let us look at some of
these features. For further discussion on this, see \cite{causal,
causalmatter3d, iogft}, and for a classic treatment of the issue
of causality and different transition amplitudes in quantum
gravity, within the covariant path integral approach, see
\cite{teitelboim,hartlehalliwell}.

\medskip

Consider the simplest case of a 4d spacetime of topology
$\Sigma\times\mathbb{R}$, with compact $\Sigma$. This spacetime
has two boundaries, call them $\Sigma_1$ and $\Sigma_2$, to which
we associate a 3d spatial geometry $h_1$ and $h_2$, respectively.
Assume now that we can uniquely associate (within a canonical
quantum theory) a state $\mid h_1 \ket$ to the geometry $h_1$, and
$\mid h_2\ket$ to the geometry $h_2$. The basic idea underlying
the \lq\lq time-less\rq\rq characterization of the causal quantum
gravity transition amplitudes $\bra h_2 \mid h_1\ket $ is that,
even if these cannot correspond to any time-ordering, they do
implement a \lq\lq time-less ordering\rq\rq or, better, a {\it
causal ordering}. This consists in the requirement that $h_2$ lies
in the causal future of $h_1$, which in turn can be formulated,
when a canonical decomposition of the gravity variables is
possible, as the requirement that the lapse function (which in a
suitable gauge is equivalent to a proper time) between the two
boundaries can only take positive
values \cite{teitelboim,hartlehalliwell}. Note that the formulation
of this criterion does not require any use of coordinates. Notice
also that the above has a direct analogue in the definition of
different Green functions for a relativistic particle
\cite{halliwellortiz}, where it defines indeed the Feynman
propagator. Moreover, this criterion can be generalized to the
situation in which no canonical decomposition is available, for
example, keeping the same boundaries and boundary states, for
spacetimes of non-trivial topology. In such cases it can be
formulated as the requirement that the amplitude is {\it
orientation dependent}, i.e. that it turns into its complex
conjugate if the spacetime orientation is reversed. If the
dynamics is defined by a quantum gravity path integral, in metric
variables, all these requirements are automatically encoded in the
definition of the configuration space as the space of all metrics
up to diffeomorphisms and of the quantum amplitude as the
exponential of $i$ times the Einstein-Hilbert action (or some
higher-derivatives extension), times a diffeo-invariant real
measure: \be\bra h_2\mid h_1\ket =\int_{g \mid
h_1,h_2}\mathcal{D}g\,e^{iS_{EH}(g)} . \label{pathintegral}\ee
Indeed, this corresponds, in a canonical formulation to: \be \bra
h_2\mid h_1\ket = \int\mathcal{D}N\mathcal{D}N^i\left[ e^{i
\int_{\mathcal{M}} d^3x dt \left(\pi^{ij}\dot{h}_{ij} - N
\mathcal{H} - N^i \mathcal{H}_i\right)} \mathcal{D}\pi^{ij}
\mathcal{D}h_{ij}\right], \label{pathintegralcan}\ee with the
integration range $(0,+\infty)$ over the lapse function $N$
\cite{teitelboim,hartlehalliwell}. The above amplitude is {\it
complex}, {\it causal} ($h_2$ is in the causal future of $h_1$)
and {\it orientation dependent} (it turns into its own complex
conjugate under switch of spacetime orientation, as $i
S_{EH}(g)\rightarrow - i S_{EH}(g)$ under this transformation,
equivalent to switching positive to negative lapse).  Moreover, it
defines, at least formally, a {\it Green function} of the
Hamiltonian constraint operator, the dynamical equation of motion
of pure gravity, for trivial spacetime topology, not a solution of
it, i.e. it satisfies: $\mathcal{H}\bra h_2\mid h_1\ket=
\delta(h_1 - h_2)$. Notice that the same definition for the
quantum gravity path integral results in an orientation dependent
transition amplitude also in the case of Riemannian quantum
gravity, i.e. the quantization of Riemannian geometry, even though
then the causality interpretation is not applicable to the same
restriction on the lapse function.\\

A quick comparison with the usual (say, free) quantum field theory
case, or, for that matter, with the relativistic particle case,
shows that these same properties are shared by the time-ordered
product of field operators and by the usual Feynman propagator.
Actually, in these simplified contexts, the above properties {\it
select uniquely} the Feynman propagator or time-ordered n-point
function among the various QFT n-point functions or particle
propagators.

\medskip

In the formal 3rd quantized framework, as in usual QFT, the path
integral of the field theory itself provides, after field
insertions, a definition of the causal transition amplitudes. How
 this is realized in the 3rd quantized formalism is only apparent
in perturbative expansion, and, once more, at a rather formal
level, given the poor understanding of the formalism itself.
However, also in the GFT case, we are at present lacking control
of the theory beyond the perturbative regime. Therefore it is
instructive to recall how the causal transition amplitudes are
characterized in usual QFT perturbative expansion. Consider then
some time ordered product of field operators in the vacuum state;
one inserts the appropriate combination of fields in the path
integral expression for the partition function of the field
theory, and expands in powers of the coupling constant, obtaining
the usual sum over Feynman diagrams weighted by amplitudes
obtained by gluing propagators with interaction vertices. The
choice of field insertion characterizes the boundary states and
the original time ordering is reflected in the presence of the
Feynman propagator in each individual particle line of a Feynman
diagram. This propagator, in position variables, can be expressed
as a sum over histories for the single particle it refers to, i.e.
by a path integral weighted by the usual relativistic particle
action \cite{halliwellortiz}. If one does the same for all the
particles involved in a given Feynman diagram, the whole Feynman
amplitude can be put in the form of a path integral for a discrete
system of particles, weighted by the exponential of the classical
action, and with a constraint on their histories in position
space, representing the particle interactions. The causality
restriction enters, as we said, in the use of the Feynman
propagator, which results from restricting the proper time (or
lapse in canonical variables) along each particle history to be
positive, in turn giving a complex amplitude given by the
exponential of $i$ times the particle action
\cite{halliwellortiz}. The restriction to positive proper times or
lapses is also equivalent to a dependence of the amplitude on the
orientation of the particle trajectories.\\

The same happens in the formal 3rd quantization setting for
gravity \cite{3rd}: each Feynman diagram, a discrete history of
\lq\lq universe interactions\rq\rq, is weighted by an amplitude
given by the path integral (\ref{pathintegral}), the exponential
of the gravity action, i.e. the geometric action for the
particle-universe for each line of propagation, plus appropriate
joining conditions representing the interactions\footnote{Notice
that the same happens as well in the worldsheet formulation of
string theory, critical or not, that indeed can be seen as the 2d
generalization of the expansion of a QFT in 1d Feynman diagrams,
or as the simplified 2d restriction of a possible 3rd quantization
(as QFT on superspace) of 4d gravity (the work of \cite{3rd} was
indeed inspired by string theory considerations and results).}. As
we have discussed above, such path integral for gravity is a Green
function of the dynamical constraint equations (as the Feynman
propagator for the Klein-Gordon equation), and it is causal and
orientation dependent in the sense specified
\cite{teitelboim,hartlehalliwell}. Note also that the end result
of combining the path integrals for gravity on trivial topology
with the joining conditions for interactions is again a path
integral for gravity with the same type of amplitude on a
spacetime with non-trivial topology. The analogy with particle
dynamics and usual QFTs is even clearer if one recalls
\cite{dewitt,greensite} that the Einstein's equations define
indeed the dynamics of a free particle moving in superspace.

\medskip

In a GFT, in light of its proposed physical interpretation, we
would expect a similar structure for the Feynman amplitudes. The
discrete histories of possible interactions for the GFT quanta
are, as discussed, combinatorial 2-complexes dual to simplicial
complexes, because the quanta themselves are identified with spin
network vertices or (D-1)-simplices, as said. Therefore, if the
GFT degrees of freedom and dynamics are to represent a quantum
geometry and its evolution, we would expect the amplitudes
associated to its discrete (virtual) histories, the GFT Feynman
amplitudes, to have the form of a path integral for discrete
gravity, i.e. an exponential of some classical discrete gravity
action. This way, they would have the sought for properties of
causality/orientation dependence and complexity on top of making
the relation with classical and quantum discrete gravity manifest.
Once more, it is the {\it complex nature} of the Feynman
amplitudes and their having the form of an {\it exponential of
some simplicial gravity action}, that would permit the
interpretation of GFTs as 3rd quantized theories of simplicial
geometry and as providing a definition of discrete quantum gravity
transition amplitudes.\\

This expected and, we argue, {\it needed} general feature of GFT
models does not, by itself, select the {\it specific form of the
simplicial action} that has to appear in the GFT Feynman
amplitudes. On the contrary, because GFTs are supposed to describe
quantum gravity {\it at all scales} (unless the future development
of the formalism will turn out to show some unavoidable breakdown
or incompleteness of the same, beyond some energy or distance
scale), one should, a priori, expect a very general formulation of
classical simplicial gravity to describe the classical dynamics of
the GFT quanta. This may correspond, for example, to some
generalized action for Regge calculus, with higher-order terms,
e.g. higher powers of the curvature, appearing as the phase of our
complex GFT Feynman amplitudes. Indeed, we will see how our new
proposed GFTs produce very general types of simplicial actions
inside their {\it complex} Feynman amplitudes, the Regge action
being just one contribution, although the dominant one in
physically interesting limits.

\medskip

This is {\it not} what happens in current GFT models. The Feynman
amplitudes/spin foam models of {\it all} current GFTs are instead:
real, a-causal and orientation independent, so that they do not
reflect the orientation of the underlying simplicial complex nor
allow for the identification of any ordering between the boundary
states. In this sense they define a-causal transition amplitudes.
This structure is due to an underlying $Z_2$ symmetry of the spin
foam amplitudes, first identified and interpreted in
\cite{causal}, which erases the orientation dependence of the same
amplitudes at the level of each dual 2-cell or (D-2)-simplex.
Still in \cite{causal}, and in \cite{feynman}, it was shown how
breaking this symmetry and restoring the orientation dependence
would lead to Feynman amplitudes with a much more direct relation
with 1st order discrete gravity and with the expected exponential
form.\\

Usual spin foam models correspond, then, to a sort of \lq\lq
symmetrized discrete path integrals\rq\rq for gravity in 1st order
form. The continuum (and 2nd order) counterpart of such
symmetrized discrete gravity path integral is given, formally, by
the same formula (\ref{pathintegralcan}), but with the range of
integration over the lapse function extended to the full real
interval $(-\infty,+\infty)$. This formula gives a covariant
definition of the physical inner product between canonical quantum
gravity states and thus of the projector operator onto solutions
of the Hamiltonian constraint operator. In fact the resulting
quantity, morally a definition of \lq\lq $\bra h^2 \mid h^1\ket_H
\simeq \langle h_2 \mid \delta(\mathcal{H})\mid h_1\rangle $\rq\rq
is a {\it solution} of the Hamiltonian constraint equation in both
its arguments \cite{teitelboim, hartlehalliwell}, and it is a real
quantity, as expected from a canonical inner product, as well as
a-causal in that nothing constraints one boundary geometry to lie
in the causal future of the other. This is not so surprising, and
maybe not even un-welcome, given that spin foam models have been
introduced exactly in order to define in a covariant way the
canonical physical inner product \cite{carloprojector}. However,
$\bra h^2 \mid h^1\ket_H$ {\it does not} correspond to the
Lagrangian path integral (\ref{pathintegral}), and thus it is {\it
not} what we would expect to obtain, in perturbative expansion, in
a properly defined field theory on superspace. The difference lies
in the requirement of an additional symmetry on top of the
Lagrangian 4-diffeos \cite{hartlehalliwell}: positive and negative
lapses correspond in fact to the {\it same} class of 4-geometries,
and the difference between the two half-ranges $(0,+\infty)$ and
$(-\infty, 0)$ is only that they correspond to {\it opposite
spacetime orientations}. In other words, the above quantity in
Lagrangian formulation is given by a different {\it symmetric}
choice of quantum amplitude, but for the same space of 4d
geometries: \be \bra h^2 \mid h^1 \ket_H=\int\mathcal{D}g\left(
e^{iS(g)} + e^{-iS(g)}\right). \label{pathintegralsym} \ee Another
insightful way of looking at the difference between the two ranges
of lapse integration, or between (\ref{pathintegral}) and
(\ref{pathintegralsym}), is by recalling the difference between
Lagrangian and Hamiltonian symmetries, the first being the 4d
spacetime diffeomorphisms and the second being the transformations
generated by the canonical operators $\mathcal{H}_i$ and
$\mathcal{H}$ \cite{hartlehalliwell}. The second set of symmetries
is actually {\it larger} than the first, and the two coincide only
after imposition of (some) equations of motion \cite{waldsym}. The
range $N\in (0,+\infty)$ is symmetric under transformations of the
lapse corresponding to 4d spacetime diffeos, while it is not under
canonical symmetries that can connect positive and negative
lapses, thus requiring a symmetric range $N\in (-\infty, +\infty)$
\cite{hartlehalliwell}. Therefore, the Lagrangian path integral
needs a further symmetrization (\ref{pathintegralsym}) to satisfy
the Wheeler-DeWitt equation. In a formulation of quantum gravity
as a quantum field theory on superspace, then, one would expect
the symmetrized amplitudes (\ref{pathintegralsym}) to arise as
Feynman amplitudes only from a {\it restriction} or in a special {\it
subsector} of a more general field theory producing instead
(\ref{pathintegral}) as the typical Feynman amplitude.\\

Once more, all this has a very precise analogue in the
sum-over-histories formulation of the dynamics of a relativistic
particle, with the symmetrized path integral corresponding to the
Hadamard propagator, given in momentum space by $\delta(p^2-m^2)$,
thus imposing the Hamiltonian constraint equation $p^2 -m^2=0$,
instead of the Feynman propagator, given in momentum space by
$\frac{i}{p^2-m^2 + i\epsilon}$, thus relaxing the same
Hamiltonian constraint at the quantum level.\footnote{The
difference between the two types of 2-point functions is not
relevant  at the 'single particle level', where the choice of
boundary states alone encodes the choice of orientation. This
means that in the formal continuum 3rd quantized gravity setting,
one doesn't expect it to be relevant in absence of topology
change, i.e. for the free field theory on superspace, while at the
same time the physical meaning of the symmetrized path integral is
highly questionable in the presence of non-trivial topologies,
where a canonical/Hamiltonian interpretation is more difficult.
The group field theory setting, moreover, introduces a further
ingredient in the picture, that we believe crucial: a
discretization of the virtual spacetimes corresponding to the 3rd
quantized Feynman diagrams; this discretization, in turn, gives a
multi-particle structure and interpretation
\cite{iogft,iogft2,ioemergence} to {\it all} Feynman
diagrams/simplicial complexes, even those of trivial topology, and
thus further supports the idea that an un-symmetrized path
integral expression for the Feynman amplitude is needed.}

\medskip

Some confusion may arise from the fact that usual spin foam model
{\it do} indeed come from a path integral quantization of a
discrete action, at least in 3d. This is, however, BF theory
which, although closely related to gravity, does not coincide with
it. The difference between 3d BF theory and 3d gravity in 1st
order form is that the triad field in a gravity path integral is
summed over those configurations corresponding to positive volume
element only, while the B field in 3d BF theory is summed over
both positive and negative volumes; this has been emphasized in
\cite{laurentkirill3dvolume}. This again is the first order
counterpart of the symmetrization (\ref{pathintegralsym}), and
corresponds to a larger set of symmetries at the quantum level in
BF theory with respect to 1st order gravity. A similar mismatch
between constrained BF-like theories and 1st order gravity arises
in higher dimensions as well \cite{causal,feynman,causalmatter3d}.
Such mismatch also explains
\cite{laurentkirill3dvolume,causal,feynman} why, in the
semiclassical, large representation limit, and thus after
suppression of quantum interference between configurations
associated to opposite orientations at the level of each dual
face, the spin foam vertex amplitude of all known models still
gives the cosine of the discrete gravity action for a single
simplex, as opposed to the exponential of it \cite{asymp}.
Moreover, we note here that the large-$j$ semiclassical limit can
be understood as suppressing, together with interference between
quantum configurations, also all quantum corrections to the Regge
action somehow hidden in usual spin foam models, leaving only the
dominant Regge term, even though still within a sum over opposite
orientations producing the mentioned cosine factor. This second
suppression mechanism, in a slightly different form, will be shown
at work also in our new GFT models, where it will indeed reduce
the GFT Feynman amplitudes to the form of a simple exponential of
the Regge action for simplicial gravity, with its quantum (higher
order) corrections being negligible in the limit.

\medskip

This being the situation, we would then like to identify the true
GFT analogue of the quantum gravity causal transition amplitudes,
or, more precisely, we would like to construct group field
theories for which, as in ordinary QFT, the Feynman expansion of
n-point functions produces Feynman amplitudes given by the
exponential of a discrete gravity action, i.e. with the causality
restrictions implicitly, automatically, but also clearly
implemented. This, for us, would be a clear sign that we are
capturing the causal dynamics of discrete gravity correctly.\\

A subsequent analysis of how the usual spin foam models arise from
a suitable restriction of these generalized amplitudes, or as a
special limit of this new class of GFTs would then shed some
additional light on the exact relation between BF theories and
gravity as well as on the role of the canonical physical inner
product between quantum gravity states within a covariant field
theory on superspace, in the more rigorous setting of both loop
quantum gravity and group field theories.

\subsection{GFTs and simplicial quantum gravity}
An additional motivation for constructing this new class of GFTs
is that they would bring simplicial quantum gravity approaches in
much closer contact with the spin foam formalism for discrete
gravity path integrals, and, via spin foams, with loop quantum
gravity. Actually, there is hope that this new class of GFTs may
represent {\it the} common unified framework in which both
simplicial quantum gravity approaches, quantum Regge calculus and
dynamical triangulations, as well as loop quantum gravity/spin
foam one can be subsumed, for mutual benefit and further
development of each. The general idea of GFTs as a common
framework has been explained in the introduction, and it remains
an interesting and intriguing perspective, in our opinion,
regardless of our present results. However, one crucial step is
needed in order to make such perspective a concrete reality, and
provide a solid basis for understanding the {\it exact} links
between GFTs, loop quantum gravity and simplicial quantum gravity.
This step has the same goal as the motivations we already gave above arising from a purely 3rd quantization
perspective: this goal is to construct GFT models with Feynman amplitudes
given {\it exactly} by the exponential of a simplicial gravity
action, times some appropriate measure. This is the step we take
with the present work.

In fact, both quantum Regge calculus and dynamical triangulations,
although differing in the choice of variables used to encode the
geometry of gravity at a simplicial level (geometric data/edge
lengths in the first case, combinatorics of simplicial complexes
in the second), identify the quantum amplitude to be associated to
each spacetime geometry with the exponential of the Regge action
for simplicial gravity, indeed a beautiful coordinate-free
description of classical simplicial geometry \cite{williams}. A
GFT {\it with the same type of quantum amplitudes} for its Feynman
diagrams and with a sum over both geometric data and
triangulations would represent a unification and a generalization
of both approaches in a very literal, transparent sense.\\

An interesting difference, as for the classical simplicial action
used, between usual simplicial quantum gravity and GFTs, can be
already envisaged. GFTs are based on a 1st order formulation of
gravity and a group theoretic description of geometry. In other
words they refer to a Palatini-like or BF-like formulation of
gravity in terms of a D-bein field and a connection field. We
would then expect that the sought for generalized GFTs would
produce amplitudes given by exponentials of a simplicial gravity
action in 1st order variables as well, i.e. with a double set of
geometric variables: one corresponding to D-beins, and thus
assigning volume information to the simplicial complex, and the
other corresponding to a connection, and thus defining a
group-theoretic notion of curvature, in terms of holonomies. In
pure BF theory, in fact, the B field is just a Lagrange multiplier
and one can achieve a formulation of the quantum amplitudes using
only the connection variables, and basically imposing the flatness
condition everywhere. In gravity, however, one relaxes this
condition and the D-bein is a true dynamical field, that we would
expect to find as a dynamical variable in the path integral as
well as a configuration/momentum variable in the definition of a
field theory on superspace, alongside the connection field.\\

In a simplicial setting, therefore, we would expect to obtain a
formulation of gravity in terms of some 1st order version of the
Regge action, with variables being a discretized D-bein $e_l$,
associated to each link of the simplicial complex, or a
discretized bivector field $B_f$, associated  to each (D-2)-face
of the complex, in a BF-like formulation of gravity, plus a
discretized connection variable, represented for example in terms
of discrete parallel transports (group elements) of the same along
dual links $e^*$ of the simplicial complex, as in all current spin
foam models \cite{SF}. This action would have a general form of
the type: \be S =\sum_{f} V_f(e,B) \, \Theta_f (g_{e*}) + (higher
\;\, order), \label{discreteaction}\ee with $V_f$ being the volume
associated to the (D-2)-face $f$, which is a function of either $e$ variable
or $B$ variables, and $\Theta$ being the corresponding deficit
angle, i.e. the discretized curvature. Similar 1st order
formulations of discrete gravity have been proposed and studied,
e.g., in \cite{magnea}.\\

In dimension $D\geq 4$,  one would have to add to
(\ref{discreteaction}) suitable (non-local) constraints on the
discrete B variables, if a BF-like formulation is the one sought
for, ensuring their geometric interpretation. In absence of such
constraints, in fact, we would just have a discrete version of
classical BF theory, as it clear from the fact that the variation
of the action with respect to $B$ would produce the flatness
condition $\Theta_f =0$ for any face of the complex. In 4
spacetime dimensions, the issue of discrete BF constraints is also
related to the issue of the constraints on area variables in the so-called Area Regge calculus \cite{williams}. We will discuss
these issues in slightly more detail later in this article.\\

With the aim of reproducing the above type of classical action,
plus the hope that this will lead to a more straightforward way of
imposing the above constraint than in the usual spin foam
procedure, and the main motivation of imposing the
causality/orientation dependence condition, which is in fact a
restriction on the integration range over the B field, we are thus
led to introduce additional variables, directly identifiable with
the B field of BF theory, into the usual GFT formalism, which is
based on the connection variables only (group elements).

We will indeed obtain, from our new GFTs, a 3rd quantized version
of discrete BF theory in any dimension and any signature, {\it
with an additional restriction on orientation} automatically
imposed, as well as incorporating what can be interpreted as
quantum corrections to the above classical action (akin to higher
derivative corrections to the Einstein-Hilbert action in effective
approaches to continuum gravity). This means that we will obtain a
3rd quantization of discrete 1st order gravity in 3d, and of an
orientation-restricted BF theory in higher dimension. In 4d the
corresponding GFT model would represent, we believe, the most
suitable framework in which to implement the classical constraints
on the B field that reduce BF theory to gravity, that have been
extensively studied in the construction of spin foam models for 4d
gravity (see e.g. the recent \cite{laurentkirillnew}), because it
may make the identification and implementation of the geometric
constraints at the simplicial level more straightforward, and in a
context in which the needed orientation dependence/causality
restriction is already implemented.

\subsection{New GFT variables and the relation between B and A fields in Lagrangian BF theory}

Before proceeding to the discussion of our results and of the
new GFT formalism, let us conclude by anticipating motivation,
interpretation and advantages of the specific way we have chosen
of introducing the additional variables corresponding to the
continuum B field.\\

In the usual spin foam models, such as the Ponzano-Regge model for
BF theory, as well as in the GFT models that generate them, such
as the Boulatov model, the variable that is interpreted as the
discrete counterpart of the B field of the original continuum BF
theory is the representation label J associated to each
(D-2)-face of the simplicial complex. This is first of all
justified by the way it enters the expression for the spin foam
amplitudes, after Peter-Weyl decomposition of the same. In the Riemannian 3d
case, for example, one indeed gets, for each dual face
\cite{alexmartin}:

\be \int d^3 B_f \exp{ \{i B_f \cdot G_f\} } = \delta(G_f) =
\sum_{j_f} d_{j_f} \chi^{j_f}(G_f) . \label{facebf}\ee

$B_f$ is the original discretized B field, given by an $\mathfrak{su}(2)$ Lie
algebra element, with which one starts from when deriving the spin foam amplitudes from a discrete Lagrangian path integral,
but that does not appear in the corresponding GFT derivation, from
which one just obtains the result of the $B_f$ integral above,
i.e. the delta function over the group, forcing the flatness
condition on the SU(2) holonomy $G_f$. Starting from this delta
function, by harmonic analysis on SU(2) one gets the last
expression in (\ref{facebf}), which indeed resembles the starting
expression but with a discrete replacement for the $B_f$
variables: the representation labels $J_f$. The same happens in
the usual GFTs. Apart from the formal similarities, one physical
rationale for the identification of the $J_f$ with a discretized B
field is the fact that it is conjugate to the connection
variables, i.e. to the group elements $g_{e^*}$, in the sense of
Fourier transforms, just as the B field is canonically conjugate
to the A field in the Hamiltonian formulation of classical BF
theory. This reasoning is of course sensible, and it is indeed
supported by the respective role group representations and group
elements play in loop quantum gravity, again following canonical
analysis in the continuum, but it is also not fully conclusive.
There are a few reasons for being dissatisfied with this
interpretation, even if they are, admittedly, not at all
conclusive either. One is that the $J_f$ corresponds more
precisely to just {\it one} component of the original B field, its
(discretized) absolute value, with the other components still
missing any identification within the formalism. The situation, in
this respect, has been ameliorated somewhat by the recent
development of new spin foam models for BF theory and gravity
\cite{laurentkirillnew,EPR,eterasimone1,eterasimone2,iolaurentkirill}
based on coherent states. Here, the additional parameters
labelling a coherent state basis of vectors in each representation
space $J_f$ are interpreted as the spin foam analogue of the
missing components of the B field with modulus $J_f$. This is
justified by the fact that the {\it expectation value} of a Lie
algebra generator in a coherent state is given by a (bi-)vector
with modulus $J_f$ and components proportional to the coherent
state parameters. However, there are still several questions
unanswered about the relation between a generic B field and the
Lie algebra generators, the exact physical role played by coherent
states, apart from their mathematical convenience, etc. Moreover, since the above is a relation that concerns the expectation values of
quantum operators, one may suspect that it should be understood as
a semi-classical one, holding only in some approximation of the
dynamics of the quantum theory. As we will discuss later on, we
feel that our approach of introducing additional independent
variables playing in a straightforward sense the role of the
discrete B field, and whose relation with Lie algebra generators
for the group considered is governed by the {\it dynamics} of the
theory, may help to clarify, with further work, the role that
coherent states play at the level of spin foam amplitudes.\\

Our main concern with the identification of representation labels,
and, before that, of the generators of the Lie algebra, acting on
connection group elements, with the discrete B field comes from
looking at the issue from a more general Lagrangian, rather than
canonical perspective (which is available only for trivial
topology). Namely, we are looking for a group field theory
discrete realization of the path integral for a {\it gravity}
theory in 1st order form, which, as we have discussed, is likely
to imply a restriction on the configurations of the B field summed
over, that would give a different result for the face amplitudes
with respect to (\ref{facebf}), as it happens, for example, in the
model of \cite{causalmatter3d}. In such a path integral two sets
of variables are present, the geometric B field and the
connection, and the relation between the two is {\it one of the
equations of motion} of the theory (the one imposing metricity of
the connection) and thus should be imposed only by the dynamics of
the theory, and not imposed already at the kinematical level at
the level of each history being summed over in the path integral,
as it appears to be done in current GFT models. We feel that
imposing such condition already at the kinematical level results,
in the usual spin foam models, in freezing a part of the degrees of
freedom of the theory. In particular it may be this restriction is what is responsible for turning what should have been causal transition
amplitudes into rather awkward, from the GFT and 3rd quantization
perspective, a-causal transition amplitudes, which correspond, as
said, to a sort of symmetrized path integral, imposing the
canonical dynamical constraints even in situations, e.g.
non-trivial spacetime topologies, where a canonical interpretation
is problematic and certainly not expected.\\

This feeling, admittedly not much more than this, in absence of
more rigorous arguments, is however confirmed by the way the new
variables we introduce should be restricted in order to reduce our
new GFT models to the usual ones, at the level of their Feynman
amplitudes, i.e. spin foam models.\\

In any case, we would like to have at our disposal a more general
framework, that reproduces the full BF or 1st order gravity path
integral, in which the mutual relations between the discrete
counterparts of the continuum variables mimic more closely the
continuum classical and quantum dynamics. In this way, we may both
confirm that we are indeed reproducing at the discrete level the
features of the continuum dynamics and also, hopefully, shed some light
on the usual spin foam models and procedures.

\section{The New Models}

\subsection{Basic idea behind the construction}
As was explained in the previous section, the new models should be thought of as the causal analogues of the usual GFTs associated with BF theory.\\

BF theory in D dimensions for a group G with a Lie algebra
$\mathfrak{g}$ is a topological field theory defined by the
following action

\be \label{eq:bfaction} S = \int_M Tr (B \wedge F(A)) \ee

where M is a D-dimensional manifold, B can be thought of
locally\footnote{Globally it is a section of the bundle associated
to the principal G-bundle over which A is defined via the adjoint
representation.} as a $\mathfrak{g}$-valued (D-2)-form and F is
the curvature of the G-connection A, so it can
also be thought of locally as a $\mathfrak{g}$-valued 2-form.\\

Let us now describe our main strategy and its rationale,
illustrating it for simplicity in the D=3 case. The extension to
different dimensions is straightforward and follows the same
type of arguments. It will be discussed in detail in the
following.

\begin{itemize}
\item We would like to introduce additional variables,
corresponding to a discrete B field associated to each 1-simplex
in the simplicial complex, in the GFT perturbative expansion. This
means that there should be one such variable for each argument of
the GFT field.

\item We would like the new variables to be
identified with the generators of the Lie algebra of the relevant
group. This implies that they must have the same number of
components. The field should then be a complex function
$\phi(g_1,B_1;g_2,B_2;g_3,B_3) : (G\times \mathbb{R}^3)^{\times 3}
\rightarrow \mathbb{C}$ in this 3d example. The complexity of the
field, together with symmetry under {\it even} permutation of the
arguments, is needed to ensure orientability of the simplicial
complexes arising in perturbative expansion.

\item The identification should follow from some equation of
motion of the theory, so to be part of the dynamics; at the same
time, it should belong to the {\it kinematical} sector of the GFT,
because we would like boundary states to satisfy it, at least
partially, so to have a similar structure to that of loop quantum
gravity spin network states. This condition would then follow from
some sort of {\it asymptotic condition} on boundary states in
computing GFT transition amplitudes (notice however that a GFT
equivalent of the cluster decomposition principle of ordinary QFT
or of the ordinary S-matrix theory has not yet been developed in
full detail).

\item Such equation of motion should then be of the type $B_a^i -
J_a^i =0$, where the indices $a=1,2,3$ label the arguments of the
field, while the indices $i$ are vector indices in $\mathbb{R}^3$,
and $J^i$ are the generators of the $SU(2)$ or $SL(2,\mathbb{R})$
Lie algebra. However, the above equation is not invariant under
group transformations, so we turn it into a covariant form,
obtaining: $B_a^2 - J_a^2=0$, for each argument of the field.

\item The field being a function on the group, the generators of
the corresponding Lie algebra act on the group arguments
of the field as derivative operators, so that the above equation is
actually implemented, in configuration space (with respect to $G$)
as: $B_a^2 - \square_a=0$, where $\square$ is the Laplace-Beltrami
operator acting on the group manifold.

\item After harmonic analysis, the $\square_a$ is turned into
the invariant Casimir of the group $G$, in a given representation
$j_a$, acting as a multiplicative operator on the field now
function on the same representation parameters $j_a$.

\item $B_a^i$ act here as multiplicative operators; however, we can
independently perform Fourier analysis on the $B$ variables as
well, going to conjugate variables $X_a$, also in $\mathbb{R}^3$,
and turn the quantity $B^2_a$ into a differential operator, a new
Laplace-Beltrami operator acting on $\mathbb{R}^3$.

\item By means of this extension of the group field theory
formalism, we want also to reproduce proper simplicial gravity
path integrals in perturbative expansion. Considering that, in the
Regge formalism for discrete gravity, D-simplices are assumed to
be flat and the whole dynamics of geometry comes from the boundary
terms \cite{williams}, we obtain a further motivation for
restricting the modification of the GFT dynamics with respect to
usual models to be confined to only the kinematical term in the
GFT action.

\item The interaction term is only modified by the extension in the
number of variables as well as in a peculiar orientation
dependence, in the variables $X$, Fourier conjugate to $B$, that
is necessary to ensure the proper matching of $B$ variables across
simplices, and encoded in the dependence on the complex structure
of the field $\phi$, as we will see. As for the dependence on the
group $G$, it maintains the same structure of the usual models
describing BF theory.

\end{itemize}

In this way we obtain a new kinetic term given by a differential
operator acting on the field, very similar to the usual
Klein-Gordon operator of scalar field theory, but with a product
structure coming from the independent action of one operator of
the above type acting on each argument of the field: $K = \prod_a
\left( \square_{X_a} - \square_{G_a}\right)$.\\

Notice that there is almost nothing in the above choices that can
select any specific dynamics of the geometric data ($B$ variables
and group elements, say) at the level of the individual Feynman
diagram. The only dynamical ingredient above is the choice of a
certain {\it relation} between them, but nothing seems to dictate,
at the level of the GFT action, the {\it individual dynamics} of
each set of variables. What we put in is then only a) some {\it
complex structure} resulting from the propagator representing the
inverse of the chosen kinetic term, due to its singular nature as
a differential operator, b) the mentioned mutual relation between
$B$'s and $g$'s, and c) the combinatorics of the Feynman diagrams
(dictated by the combinatorics of the variables in the action). It
is only to be expected, then, that the simplicial action
describing their dynamics at the level of each Feynman diagram,
and appearing in the exponent of the phase part of the Feynman
amplitudes (simplicial gravity path integral) will be pretty
generic. The non trivial tests will be to show: 1) that this phase
can be interpreted at all as a simplicial gravity action, because
of the way the GFT variables will enter in it; 2) that this
generalized action will reduce to the usual Regge action (in 1st
order form) in appropriate, clearly identified and physically
meaningful limits. Our proposed formalism passes these tests.

\medskip

The strongest support for the mentioned choices, and for the
resulting form of the GFT action for the new models, is,
therefore, simply the resulting expression for the Feynman
amplitudes of the corresponding GFT, which indeed fulfill {\it
all} the expectations and goals we have stated above. Some
additional nice features of the resulting model can be already
underlined at this point. As we mentioned, the kinetic operator
above is a singular differential operator, which implies that its
inverse has to be defined in the complex domain, just as it
happens in the usual Klein-Gordon case. On the one hand, this
complexification is responsible for the complexity of the
resulting Feynman amplitudes, and ultimately, as we shall see, for
the wanted exponential form of the same amplitudes; on the other
hand, the differential form introduces the dynamical correlations
between simplices that we would expect in a discrete theory of
quantum geometry. Also, the propagator corresponding to the
kinetic operator will introduce quantum corrections, virtual
degrees of freedom, not captured by the on-shell condition $B^i_a
= J^i_a$, thus relaxing it at the quantum level, again matching
our expectations. Finally let us mention that the presence of
derivatives in the GFT kinetic terms allows for the identification
of a non-trivial symplectic structure on the space of fields, and
makes a canonical analysis {\it of the GFT itself} possible. This
nice feature is shared also by the generalized models introduced
in \cite{generalised}, of which the new ones represent a sort of
\lq\lq relativistic extension\rq\rq (fixing some pathologies of
the same arising in perturbative expansion), as we will discuss,
and it is at the basis of the canonical analysis performed in
\cite{iojimmy}.

\subsection{The new models: action and Feynman amplitudes}
We now give the definition of the action for the new class of GFT
models, for general dimension D and general gauge group G.\\

Let G be a semi-simple group (we will deal with the double covers
of the rotation and the Lorentz groups in D dimensions) and let X
be a space isomorphic, as a metric vector space, to the Lie
algebra $\mathfrak{g}$ of G. The basic variable of the theory is a
complex-valued field $\phi$

\[
\phi \, (g_1, g_2, \dots, g_D \, ; \, X_1, X_2, \dots, X_D)  \, :
\, \underbrace{G \times G \times \dots \times G}_{\textrm{D
times}} \times \underbrace{X \times X \times \dots \times
X}_{\textrm{D times}} = G^D \times X^D \rightarrow \mathbb{C},
\]

\ni where D is the dimension of the model, which is the dimension of the generated simplicial complexes (we will concentrate on
the 3 and 4 dimensional cases).\\

\ni The field is interpreted as a (D-1)-simplex, with the group
and Lie algebra variables corresponding to its geometry. The group
elements represent discrete parallel transports of a connection (the discrete analogue of the $A$ of BF theory) from
the centre of the simplex to the boundaries, while the $X$
variables allow us the reconstruction of the volumes of the boundary
(D-2)-simplices, and are thus related to the B field of BF
theory\footnote{As we shall see later on, it is the norms of the
Fourier conjugate variables of the X's, what we call below the
P's, that are to be interpreted as volumes, and are to be
interpreted as the discrete analogue of the B field of BF
theory.}.\\

The field is assumed to be invariant under {\it even} permutation
of the labelling of its (pairs of) arguments $(g_i,X_i$), and to
turn into its own complex conjugate under change of this labelling
by an {\it odd} permutation. In this way, the orientation of the
corresponding (D-1)-simplex is encoded in the complex structure of
the field\cite{DP-P,iogft}.\\

As in usual GFT models, geometric closure of the (D-2)-simplices
which form this (D-1)-simplex translates into invariance of the
field under the global symmetry $\phi \, (g_1 h, g_2 h, \dots, g_D
h \, ; \, X_1, \dots, X_D) = \phi \, (g_1, g_2, \dots, g_D \, ; \,
X_1, \dots, X_D)$ \cite{dgftreview}. We will impose this symmetry
in the usual way by taking the field to be arbitrary and then
projecting it onto the diagonal subspace, i.e. the field $\phi(g_i
; X_i)$ is given by $\phi(g_i ; X_i) = \int_G dh \,
\tilde{\phi}(g_i h ; X_i)$, where $\tilde{\phi}(g_i ; X_i)$ is now
arbitrary. Below, to reduce clutter, we will write the actions in
terms of the $\phi$'s instead o the $\tilde{\phi}$'s.\\

Also, we will denote both the field and its complex conjugate by
$\phi^{\nu}$, with $\nu = \pm 1$ and $\phi^{+1} = \phi$ and
$\phi^{-1} = \phi^*$.\\

\ni The model is defined by the following action

\begin{eqnarray}
S & = & \frac{1}{2} \int_{G^D} \bigg ( \prod_{i=1}^D \, dg_i \bigg ) \,  \int_{X^D} \bigg ( \prod_{i=1}^D \, dX_i  \bigg ) \, \, \,  \phi^*(g_i ; X_i) \underbrace{ \bigg [ \prod_{i=1}^D \Big ( - \Box_{X_i} \, +  \, \Box_{G_i} \,  - \frac{d}{24} \, m^2 \Big ) \bigg ]}_{\textrm{Kinetic Operator}} \phi(g_i ; X_i) + {} \nonumber \\
& & {}  +   \frac{\lambda}{(D+1)!} \sum_{\nu_1, \ldots, \nu_{D+1}} \int_{G^{D(D+1)}} \bigg ( \prod_{i \neq j = 1}^{D+1} \,  dg_{ij}\bigg ) \, \int_{X^{D(D+1)}}\bigg (  \prod_{i \neq j}^{D+1}  dX_{ij}      \bigg ) \, \, \,\underbrace{ \bigg [ \prod_{i < j} \delta(g_{ij} g_{ji}^{-1}) \delta(\nu_i X_{ij} + \nu_j X_{ji}) \bigg ]}_{\textrm{Vertex}} \, \, \, \times {} \nonumber \\
\label{eqnarray:actiongx}
& & {} \times \phi^{\nu_1}(g_{1j} ; X_{1j}) \ldots \phi^{\nu_{D+1}}(g_{D+1 j} ; X_{D+1 j}).
\end{eqnarray}

\ni $\Box_X$ and $\Box_G$ are the Laplace-Beltrami operators on X
and on G respectively, corresponding to the Killing form
\footnote{We would like to draw the reader's attention to the fact
that we are using opposite conventions for the Killing form and
the Cartan-Killing metric. This means that the metric entering the
definition of $\Box_G$, is obtained by extending, using e.g.
left-invariance, the \textit{negative} of the metric used to
define $\Box_X$. The reason for this choice of conventions comes
from the fact that if one uses the same sign for the $\Box_X$ as
the one used in the mathematical literature \cite{camporesi,
marinov} for the $\Box_G$, then one gets a negative-definite
operator in the case G is compact. So, for example, if G is SU(2),
then the corresponding $\Box_X$ would have been given (in the
appropriate coordinates) by $ - ( \frac{\pp^2}{\pp x_1^2} +
\frac{\pp^2}{\pp x_2^2}+ \frac{\pp^2}{\pp x_3^2})$. On the other
hand, with our conventions it is just the usual Laplacian on flat
space.} and the Cartan-Killing metric, and d is the dimension of G
and $X \simeq \mathfrak{g}$.\\

As in usual GFTs, the combinatorics of arguments in the action is
designed in such a way as to mimic the combinatorics of the
(D-2)-faces of a D-simplex in the interaction term, and the gluing
of two D-simplices across their common boundary in the kinetic
term.

The sum over $\nu_i$ in the second term makes the action real.
Interpreting the $\phi$ as representing (D-1)-simplices which are
\lq incoming\rq or \lq in the past boundary\rq, and the $\phi^*$
as representing (D-1)-simplices which are \lq outgoing\rq or \lq
in the future boundary\rq with respect to the D-simplex
corresponding to the GFT interaction vertex, we see that there are
D+2 possible vertices, corresponding to the cases in which (D+1)-n
\lq initial\rq (D-1)-simplices interact to give rise to n \lq
final\rq (D-1)-simplices after the interaction has taken place. In
turn these various terms correspond to the well-known
(D-1)-dimensional Pachner moves. As noticed above, the orientation
of the (D-1)-simplices, inducing a {\it pre-order} \cite{causal,feynman, generalised} also on the set of D-simplices, and
turning the resulting Feynman diagrams into {\it directed graphs}
is encoded in the complex structure of the fields. For simplicity
of presentation, we have chosen the weight the various interaction
terms corresponding to different choices of $\nu_i$'s with the
same coupling constant $\lambda$; it is straightforward to relax
this assumption defining coupling constants $\lambda_{\nu_i}$,
with $\lambda_{\nu_i}= \lambda_{-\nu_i}^*$ in order to ensure
reality of the action, as it was done also in \cite{generalised}.\\

Let us remark that it is possible to choose a different vertex
from the one above. One in which there is no dependence on the
$\nu's$, in the X variables, and this dependence is instead shifted to the P
variables:

\[
\textrm{Vertex} = \bigg [ \prod_{i < j} \delta(g_{ij} g_{ji}^{-1}) \delta( X_{ij} - X_{ji}) \bigg ]
\]

Below, we will call the model given by (\ref{eqnarray:actiongx}) model A, while the one with this new vertex model B. \\

Note also that the the kinetic operator is just a product of D copies of the Klein-Gordon one for a massive scalar field living in $X \times G$, one for
each pair of arguments of the field. \\

Let us write the above action in \lq momentum' space with respect
to the X variables. We will denote the space dual to X as P. Thus
(\ref{eqnarray:actiongx}) is equal to\footnote{Our convention for
the Fourier transform is
\[
f(\vec{p}) = \int d^d \vec{x} e^{i \vec{p}\cdot \vec{x}} f(\vec{x}) \qquad , \qquad f(\vec{x}) = \frac{1}{(2 \pi)^d} \int d^d \vec{p} e^{-i \vec{p} \cdot \vec{x}} f(\vec{p})
\]
Where the vectors denote the coordinates in which the appropriate
Killing form has a canonical form (diagonal matrix with $\pm$1
along the diagonal. }

\begin{eqnarray}
S & = & \frac{1}{2 \, (2 \pi)^{dD}} \int_{G^D} \bigg ( \prod_{i=1}^D \, dg_i \bigg ) \, \int_{P^D} \bigg ( \prod_{i=1}^D  dP_i  \bigg ) \, \, \,  \phi^*(g_i ; P_i) \underbrace{ \bigg [ \prod_{i=1}^D \Big (  \, P_i^2 \, +  \, \Box_{G_i} \,  - \frac{d}{24} \, m^2 \Big ) \bigg ]}_{\textrm{Kinetic Operator}} \phi(g_i ; P_i) + {} \nonumber \\
& & {}  +   \frac{\lambda}{(2\pi)^{dD(D+1)} (D+1)!} \sum_{\nu_1 \ldots \nu_{D+1}} \int_{G^{D(D+1)}} \bigg ( \prod_{i \neq j = 1}^{D+1} \,  dg_{ij} \bigg ) \, \int_{P^{D(D+1)}} \bigg ( \prod_{i \neq j =1}^{D+1} dP_{ij}      \bigg ) \, \, \,\underbrace{ \bigg [ \prod_{i < j} \delta(g_{ij} g_{ji}^{-1}) \delta(P_{ij} - P_{ji}) \bigg ]}_{\textrm{Vertex}} \times{} \nonumber \\
\label{eqnarray:actiongp}
& & {} \times \phi^{\nu_1}(g_{1j} ; P_{1j})  \ldots \phi^{\nu_{D+1}}(g_{D+1 j} ; P_{D+1 j}).
\end{eqnarray}

\ni $P^2$ is the magnitude of P with respect to the Killing form.
The kinetic term can be interpreted as the product of D
Klein-Gordon operators on the group G, and for a particle/field of
(variable) mass square $\frac{d}{24} - P_i^2$.\\

The written action is the one associated with model A (i.e.
equation (\ref{eqnarray:actiongx})). Notice that the orientation
dependence, i.e. the dependence of the vertex term on the
$\nu_i$'s, is apparently lost in going to the $P$ variables (of
course, the vertex has this form in the P variables exactly
because of the $\nu$-dependence in the X variables, thus this
dependence is retained). In model B, instead, the vertex in the
(g,P) variables becomes

\[
\textrm{Vertex} = \bigg [ \prod_{i < j} \delta(g_{ij} g_{ji}^{-1}) \delta(\nu_i P_{ij} + \nu_j P_{ji}) \bigg ]
\]

So the vertex of model B depends explicitly on the $\nu$'s in the
(g,P) variables, and not in the X variables. We will see that the
Feynman amplitudes, when we use the P variables, thus in both the
(g,P) and (J,P) representations, are the same for both models. The
difference between them is apparent only when the X variables are
invoked. Since, as we shall see later on, it is the P variables
that have clear physical significance, and we are going to deal
extensively only with the (g,P) and (J,P) representations, we
shall not draw the distinction between the two versions of the
model in what follows, apart when we briefly report the Feynman
amplitudes in the (J,X) variables at the end of this section.\\

\ni We can also perform the \lq Fourier transform' with respect to
the group variables. Expanding the field harmonically on the group
and using its invariance under the global right shifts
\cite{laurentgft, dgftreview}, we get

\[
\phi(g_i ; P_i) = \sum_{J_i , \Lambda, \alpha_i}  \phi_{\alpha_i}^{J_i \Lambda}(P_i) \, \prod_i \bigg ( D_{\alpha_i \beta_i}^{J_i}(g_i) \, \,  \iota_{\beta_{i_1} \dots \beta_{i_D}}^{J_1 \dots J_D \Lambda} \bigg )
\]

\ni The J's label the representations of the group G. The index J can go over both discrete and continuous values in general as is the case for the Lorentz group. The D's are the representation functions (the components of the representation matrices). $\iota$ is an appropriate normalized intertwiner (between the representations labelled by $J_1, \dots, J_D$), and $\Lambda$ labels the different basis elements of the space of normalized intertwiners.\\

A very important property of the Laplace operator is that it is
multiplicative on the representation functions (see
\cite{camporesi} and references therein). More precisely, $\Box_G
D^J(g) = \mp C_J D^J(g)$ where $C_J$ is the appropriate Casimir
operator and the minus sign is used for compact groups while the
plus sign for the noncompact ones.\footnote{The reason there is
this difference in the sign is because the Casimirs for the
compact group are defined using the negative of the Killing form.
So, for SU(2) the natural Casimir from the point of view of the
Killing form would be $-J_1^2 - J_2^2 - J_3^2$ which is minus the
usual one. The space of Casimirs of the rotation and the Lorentz
groups in 3 dimensions is one dimensional, while in 4 dimensions
it is 2 dimensional. In 4 dimensions, the Casimir that corresponds
to the Laplace-Beltrami operator in the Riemannian case, where the
representations of Spin(4) are labelled by a pair of spins
$(J_1,J_2)$, is proportional to $J_1(J_1+1) + J_2(J_2+1)$, while
for the Lorentzian case, where representations of
$SL(2,\mathbb{C})$ are labelled by an integer $n$ and a real
number $\rho$, it is proportional (in our normalizations) to
$\rho^2 - n^2 +2$.}

Inserting the above into (\ref{eqnarray:actiongp}) we get

\begin{eqnarray}
S   & = & \frac{1}{2 \, (2 \pi)^{dD}} \sum_{J_1, \dots, J_D, \alpha_1, \dots, \alpha_D, \Lambda} \int_{P^D} \bigg ( \prod_{i=1}^D \, dP_i  \bigg ) \, \, \,   \phi_{\alpha_i}^{* J_i \Lambda}(P_i) \underbrace{ \bigg [ \prod_{i=1}^D \Big (   P_i^2  \mp C_{J_i}  - \frac{d}{24} \, m^2\Big)   \bigg ]}_{\textrm{Kinetic Operator}}  \phi_{\alpha_i}^{J_i \Lambda}(P_i) +  {} \nonumber \\
& &  {}  + \frac{\lambda}{(2\pi)^{dD(D+1)} (D+1)!}  \sum_{J_{ij}, \alpha_{ij}, \Lambda_i, \nu_i ; i \neq j = 1, \ldots, D+1}  \int_{P^{D(D+1)}} \bigg ( \prod_{i \neq j = 1}^{D+1} dP_{ij}      \bigg )  \phi_{\alpha_{1j}}^{\nu_1 J_{1j} \Lambda_1}(P_{1j}) \ldots {} \nonumber \\
\label{eqnarray:actionjp}
& & {} \ldots  \phi_{\alpha_{(D+1) j}}^{\nu_{D+1} J_{(D+1) j} \Lambda_{(D+1) j}}(P_{(D+1) j})
 \underbrace{\bigg [  \big \{ J - \textrm{Symbol}\big \}(J_{ij} ; \Lambda_i ) \Big ( \prod_{i < j}^{D+1} \delta(P_{ij} - P_{ji}) \, \delta_{\alpha_{ij} \alpha_{ji}} \Big ) \bigg ]}_{\textrm{Vertex}}.
\end{eqnarray}

\ni The interaction term is essentially the standard one, which is a product of fields whose arguments are contracted in the pattern of a D-simplex multiplied by the appropriate J-symbol, always obtained by the contraction (along pairwise identified tensor indices, following the combinatorics of faces of a D-simplex) of D+1 D-valent intertwiners of the group G. The only difference being that now it is not only the alphas and the J's that are contracted but also the P's as well.\\

For completeness we also list the action in the X,J variables,
which is easily obtained by taking the Fourier transform of
(\ref{eqnarray:actionjp}) with respect to the P variables. Note
that the kinetic term is just the product of D copies of the
Klein-Gordon one on flat $\mathbb{R}^d$, with the metric whose
signature is decided by the appropriate Killing form, and with
(variable) mass square $\frac{d}{24} \pm C_{J_i}$.\\

It is given by:
\begin{eqnarray}
\label{eqnarray:actionjx}
S & = & \frac{1}{2} \sum_{J_1, \dots, J_D, \alpha_1, \dots, \alpha_D, \Lambda} \int_{X^D} \bigg ( \prod_{i=1}^D \, dX_i  \bigg ) \, \, \,   \phi_{\alpha_i}^{* J_i \Lambda}(X_i) \underbrace{ \bigg [ \prod_{i=1}^D \Big (  - \Box_{X_i}  \mp C_{J_i}  - \frac{d}{24} \, m^2\Big)   \bigg ]}_{\textrm{Kinetic Operator}}  \phi_{\alpha_i}^{J_i \Lambda}(X_i) + {} \nonumber \\
& & {}  + \frac{\lambda}{ (D+1)!} \sum_{J_{ij}, \alpha_{ij}, \Lambda_i, \nu_i ; i \neq j = 1, \ldots, D+1}  \int_{X^{D(D+1)}} \bigg ( \prod_{i \neq j = 1}^{D+1} dX_{ij}      \bigg )  \phi_{\alpha_{1j}}^{\nu_1 J_{1j} \Lambda_1}(X_{1j}) \ldots \phi_{\alpha_{(D+1) j}}^{\nu_{D+1} J_{(D+1) j} \Lambda_{(D+1) j}}(X_{(D+1) j})  \times {} \nonumber \\
\nonumber \\
\label{eqnarray:actionjx}
& & {} \times \underbrace{\bigg [  \big \{ J - \textrm{Symbol}\big \}(J_{ij} ; \Lambda_i ) \Big ( \prod_{i < j}^{D+1} \delta(\nu_i X_{ij} + \nu_j X_{ji}) \, \delta_{\alpha_{ij} \alpha_{ji}} \Big ) \bigg ]}_{\textrm{Vertex}}.
\end{eqnarray}

\medskip

It is easy to see that the new models we are presenting are
essentially a sort of \lq\lq relativistic extension\rq\rq of the
generalized group field theories (GFTs) introduced in
\cite{generalised}. In fact, the new models encode the orientation of
the Feynman diagrams/triangulations resulting from the
perturbative expansion of the partition function, that we are
going to discuss in the following, in the action and in the
quantum Feynman amplitudes in almost the same way as the
models in \cite{generalised} (see also the discussion of these
models in \cite{iogft}). The difference from the models outlined
there is the fact that the field is now a function of more
variables, passing, in momentum space, from a variable mass-energy
$M$ valued on the real line to the set of momentum variables
$P_i$. Consequently, the kinetic operator in each argument of the
field turns from a Schroedinger-type one into a Klein-Gordon one.
While this could be considered a somewhat minor modification at
the level of the action alone, the step from a non-relativistic
type of dynamics to a relativistic one has huge consequences at
the level of the Feynman amplitudes and for the whole quantum
dynamics of the corresponding models.

\medskip

We quantize the theory now via the path-integral method. The
partition function is given by

\[
Z = \int \mathcal{D} \phi \, \mathcal{D} \phi^* \, e^{i S[\phi ,
\phi^*]}.
\]

In lack of a better understanding of the quantum theory and of
more powerful tools, we study the quantum dynamics of the theory
in perturbative expansion around the vacuum, expanding the partition function in Feynman diagrams $\Gamma$ in the usual way. We get

\[
Z = \sum_{\Gamma} \frac{\lambda^{V_\Gamma}}{\textrm{sym}(\Gamma)}
Z_{\Gamma},
\]

where $V_\Gamma$ is the number of vertices in the Feynman diagram
$\Gamma$, sym($\Gamma$) is the symmetry factor of the diagram
(order of automorphisms of the diagram/complex), and $Z_{\Gamma}$
is the Feynman amplitude for the graph $\Gamma$ obtained as is
customary by taking the product of vertex functions and Feynman
propagators, obtained by inverting the kinetic operator in the action. \\

We then set out to extract vertex and propagator from our
classical action. \ni Let us begin with the vertex contribution.
It is clear that the interaction term in the (g,p) variables
(equation (\ref{eqnarray:actiongp})) is exactly like the
interaction terms in the usual GFTs for BF theory with the sole
difference being the extra variables (which are contracted in
exactly the same way as the group variables). The vertex amplitude
is then just the usual one, which is nothing but a product of
delta functions connecting the group arguments in the D-simplex
pattern, with the addition of extra delta functions connecting the
P variables paralleling the group ones. In other words, if we
represent the vertex in the standard way \cite{dgftreview} we see
that it consists of (D+1) bundles, of D-strands each, joined
together in a pattern of a D-simplex (in the shaded area of the
picture). Each strand represents a product of a delta function on
the group with a delta function on the Lie algebra. The dark dots
represent the arguments of the delta functions. Since we never
have a situation when several strands meet at a point, it is
obvious that there is no real interaction
enforced by this vertex, at least not in usual local QFT sense,
just a rerouting of the strands. It is in this sense that GFTs are
sometimes referred to
as \lq\lq combinatorially non-local field theories\rq\rq \\

\begin{figure}
\centering
\includegraphics[width=0.7\textwidth]{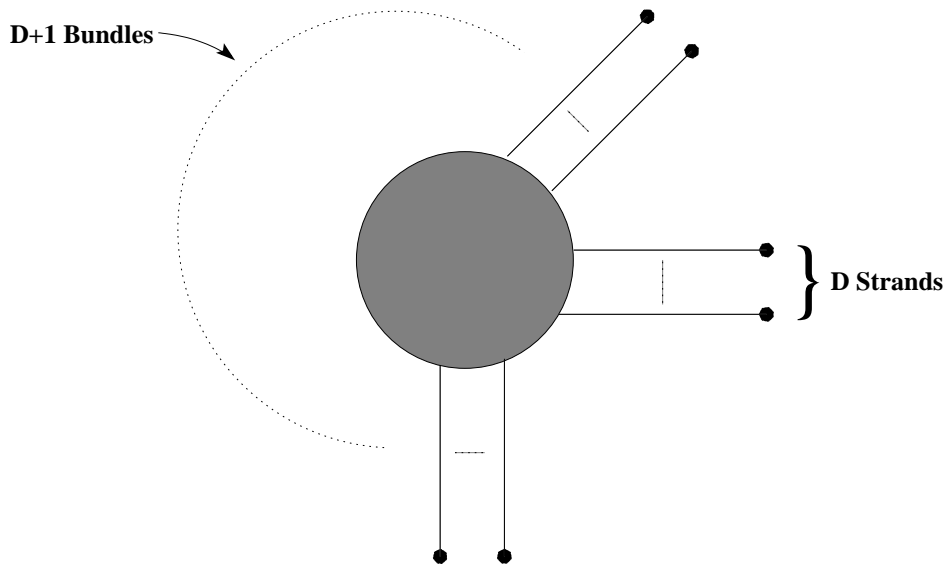}
\caption{\textit{This is the vertex. There are D+1 bundles of D strands each. The strands in the shaded area connect in the pattern of a D-simplex.}}
\end{figure}

\ni We now move on to the propagator. The easiest way to get it is
to use the action written in terms of the (J,P) variables, and the
expansion of a delta function on a group in terms of the
characters. We won't really need the explicit form of the
character functions nor the precise values of the coefficients,
rather just the fact that such an expansion is possible. For all
the groups that we will consider in this work this is indeed the
case.\footnote{The groups we are interested in are the rotation
and the Lorentz groups in D dimensions, SO(D) and SO(1, D-1), and
their double covers. For these groups, the Plancherel expansion
does exist and it has the above general form.} We will use the
following notation for this expansion

\be \label{eq:delta} \delta(g) = \sum_J \Delta_J \, \chi_J(g), \ee

where $\chi_J(g)$ is the character of the representation labelled by J, and as before the index J can go over both discrete and continuous values (the sum standing for the usual sum or for the integral, respectively).\\

From this expansion and from the expression of the kinetic
operator in (\ref{eqnarray:actionjp}) we can immediately read off
the expression of the Feynman propagator\footnote{The integral
over the h follows from the invariance of the field under the
right shifts. This integral will be important when we will compute
the dual face amplitudes as the h's are the only variables that
will be left after all the \lq gluing' integrals are done.}

\be
\label{eq:propagatorjp}
D_F [ g_i , h_i ; P_i , Q_i ]   = \int_G  dh  \prod_{i=1}^D \bigg ( \sum_{J_i} \frac{i \Delta_{J_i}}{ P_i^2 \mp C_{J_i} - \frac{d}{24} \, m^2 + i \epsilon} \, \chi_J(g_i h h_i^{-1}) \, \delta(P_i - Q_i) \bigg ).
\ee

The above expression, indeed, satisfies

\be
\label{eq:propagator equation}
\prod_{i=1}^D \Big (  \, P_i^2 \, +  \, \Box_{G_i} \,  - \frac{d}{24} \, m^2 \Big ) \, D_F [ g_i , h_i ; P_i , Q_i ] = i  \int_G dh \prod_{i=1}^D \Big( \delta(g_i h h_i^{-1}) \, \delta(P_i - Q_i) \Big ).
\ee

The fact that it is the Feynman propagator, as opposed to some other Green's function, follows as usual from the i$\epsilon$ prescription used
in (\ref{eq:propagatorjp}), as it is clear by recalling that the kinetic operator in (\ref{eqnarray:actionjp}), as noticed above, is essentially the
Klein-Gordon operator in momentum variables.\\

Note that by taking the Fourier transform with respect to the P
variables we can obtain the expression for the same propagator in
the (J,X) variables. Instead of computing it this way, which in
fact does not easily lead to an explicit expression, we note that
the kinetic term is perfectly symmetric in the way it treats the
group manifold $G$ and the space $X$. Because of this symmetry, we
can just reproduce the very same steps that will lead us to the
propagator in the (P,g) variables, to obtain instead the same
propagator in the (X,J) variables. Clearly this will be just the
product of Klein-Gordon propagators for a scalar massive (of
mass$^2$ = $\pm C_J + \frac{d}{24} \, m^2$) field in the flat
space X.\footnote{It seems that there is no general simple formula
for this object in a space of arbitrary
dimension and signature.}\\

To sum the series and obtain the propagator in the (g,P) variables
as opposed to the (J,P) above, take advantage of the fact that,
once more, the kinetic term in the (g,P) variables in
(\ref{eqnarray:actiongp}) is just the product of Klein-Gordon ones
on the group G with the mass given by $(P^2 - \frac{d}{24}m^2)$,
and use the Feynman-Schwinger-DeWitt parametrization of the
propagator\cite{feynmanqed,camporesi}. This parametrization
relates the Klein-Gordon propagator of a massive scalar field on a
space to the Schroedinger evolution kernel on that space, in a
fictitious proper time parameter $t$.

\medskip To see this connection between the propagator and the
kernel recall the Schroedinger equation on the group, which is
given by

\be
\label{eq:schrodinger}
i \frac{\pp \psi(g, t)}{\pp t} + \Box_G \, \psi(g,t) = 0.
\ee

The general solution to this equation is given by the
aforementioned Schroedinger evolution kernel $K[g_t, g_0, t]$ which
gives the solution $\psi$ at time $t+t_0$ given the solution at
time $t_0$.

\[
\psi(g_t,t+t_0) = \int_G dg_0 \, \theta(t) \, K [g_t , g_0 ,   t] \, \psi(g_0, t_0),
\]

where $\theta(t)$ is the Heaviside step function. Many properties of $K$ immediately follow from this equation and the fact that (\ref{eq:schrodinger}) is invariant under left and right shifts, notably symmetry, composition and Green function property.\\

By the symmetry property of the kernel, we mean the fact that it
is invariant under shifting both arguments on the group on one
hand and that it is a central function on the other. This latter
fact means that the kernel is expandable in characters, a feature
we will use shortly. In formulae the kernel satisfies

\be
\label{eq:central}
K[g , h, t] = K[g h^{-1} , t]  \quad  \textrm{and} \quad K[g , t]   =   K[h g h^{-1} ,t].
\ee

The composition property of the kernel is the standard one:

\be
\label{eq:composition}
\int_G dg_2  \, K [g_1 , g_2  , t]  \, K [ g_2  , g_3 , s]     =   K [ g_1  , g_3  , t+s].
\ee

This equation will be useful when we will compute the Feynman amplitudes of our model.\\

Finally and most importantly the kernel satisfies the following two equations

\[ i \frac{\pp K[g , h , t]}{\pp t} + \Box_G \, K[ g , h , t] = 0
\; \;  \textrm{and}  \; \; i \frac{\pp \Big ( \theta(t) \, K[g , h
, t] \Big )}{\pp t} + \Box_G \, \Big ( \theta(t) \, K[ g , h , t]
\Big ) = i \, \delta(t) \, \delta(g h^{-1}). \]

These coupled with the boundary condition $\lim_{t \rightarrow 0}
K[g , h , t] = \delta(g h^{-1})$ mean that $\Big ( \theta(t) \, K[
g , h , t] \Big )$ is the retarded propagator for the Schroedinger
equation. It is this last feature that links the Schroedinger
kernel with the Feynman propagator.\\

To see this link take the Fourier transform of the inhomogeneous
equation with respect to t, going to the conjugate variable $\mu$,
which is the mass (square) of the particle in the proper time
parametrization of the Klein-Gordon propagator
\cite{feynmanqed,camporesi}, or the energy of the same in the
usual Schroedinger equation\footnote{We see here, even more
clearly, the strict relation between the new class of GFT models
we are presenting and the generalized GFTs proposed in
\cite{generalised}.} If we denote by $K[g , h , \mu]$ the Fourier
transform of $\Big ( \theta(t) \, K[ g , h , t] \Big )$ then it
follows that

\[
\Big (  \mu  + \Box_G \,  \Big )  \, K [g , h , \mu]   =  i \delta(g h^{-1}).
\]

Comparing this to (\ref{eq:propagator equation}) we see
immediately that K satisfies essentially the same equation as
$D_F$. From this we can easily deduce that

\be
\label{eq:propagatoriskernel}
D_F [ g_i , h_i ; P_i  , Q_i]   =   \int_G dh \prod_{i=1}^D \Bigg( K\bigg [g_i h h_i^{-1} , \Big ( P^2_i - \frac{d}{24} \, m^2 \Big ) \bigg ] \, \delta(P_i - Q_i) \Bigg ).
\ee

Alternatively, we could use the known character expansion of the
$K[g,t]$, which is given by\footnote{It is easy to see that this
is the right expansion by remembering that the Laplacian is
diagonal on the representation functions. The minus in the
exponent is for compact groups, plus for noncompact ones.}
\cite{camporesi}

\[
K [g,t] = \sum_J \Delta_J \, \chi_J(g) \, e^{\mp i C_J t}.
\]

If we Fourier transform this (multiplied by the step function)
with respect to $t$ we get

\be
\label{eq:kernelexpansion}
K [g, \mu] = \sum_J \Delta_J \, \frac{i}{\mu \mp C_J + i \epsilon} \, \chi_J(g).
\ee

Comparing this with the character expansion of the $D_F[g_i , h_i ; P_i , Q_i]$ given in (\ref{eq:propagatorjp}) we re-obtain (\ref{eq:propagatoriskernel}).\\

As was mentioned above, the Feynman propagator for our theory in the (g,P)
variables is just (a product of D copies of) the Klein-Gordon
propagator, here written in terms of the Schroedinger evolution
kernel, for a free particle on the group, with the mass equal to

\[
\mu =  P^2 - \frac{d}{24} \, m^2.
\]

As anticipated, the above procedure can be reproduced in order to
obtain the propagator in the (J,X) variables, which is a product
of propagators (one for each argument of the field) for a scalar
field with mass$^2$ = $\pm C_J + \frac{d}{24} \, m^2$ on the
(flat) space X. Which means that the propagator in these variables is given by\footnote{There are also the $\alpha$'s labelling different field components in (\ref{eqnarray:actionjx}), however they just contribute trivial Kronecker deltas, and since it is customary in the literature to not write them explicitly we do the same here. Their effect on the Feynman amplitude is just to contribute a factor $\Delta_J$ to the weight of every dual face.}

\be
\label{eq:propagatorjx}
D_F [ X_i , Y_i ; J_{1, i} , J_{2,i} ] = \prod_{i=1}^D  \Bigg ( \delta_{J_{1,i} J_{2,i}} \,  K \bigg [  X_i - Y_i , \pm C_J + \frac{d}{24} m^2   \bigg ] \Bigg ).
\ee

The only difference with the (g,P) expression,
that results, as we will see, in a very different form for the
full Feynman amplitudes in the two representations, is the absence of the analogue of the additional integration over h $\in$ G, coming from
the gauge invariance requirement on the GFT field, and which
breaks the symmetry between the X and G spaces in the GFT action.\\

Analogously to the vertex above we represent the propagator by a bundle of D strands as in the picture.\\

\begin{figure}
\centering
\includegraphics[width=0.5\textwidth]{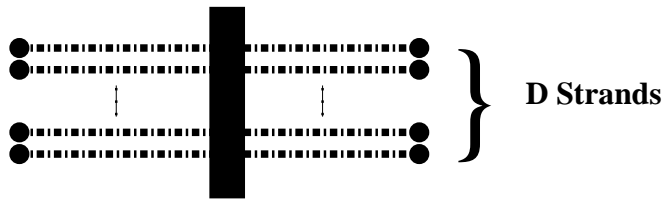}
\caption{\textit{The propagator.}}
\end{figure}

\ni Each strand represents a multiplicand from the right hand side of equation (\ref{eq:propagatoriskernel}), i.e. the i-th strand is

\[
K\bigg [g_i h h_i^{-1} , \Big ( P^2_i - \frac{d}{24} \, m^2 \Big ) \bigg ] \, \delta(P_i - Q_i).
\]

\ni The box across all the strands represents the common integral over the h. The dark dots at the ends of the strands represent the remaining arguments of
the propagator (($g_i , P_i$) on one side and ($h_i , Q_i$) on the other).\\

The reason why we have drawn the strands in the propagator differently from those in the vertex is that in distinction to the situation in the usual models
where the strands represent the same thing (simple delta functions), this is not the case here, as a true propagation of degrees of freedom takes place
between simplices, even though only a re-routing of the same occurs within each simplex.\\

Since we have now both the necessary ingredients, the vertex and
the propagator, we proceed now to construct explicitly the Feynman
amplitudes. A Feynman diagram $\Gamma$ is obtained by gluing
several vertices together using the propagators. If we pick one of
the strands and follow it around the diagram, in absence of
external legs, as in the diagrams merging from the perturbative
expansion of the partition function, the strand closes back on
itself. We can think of this loop as the boundary of a
2-dimensional surface which we assume has the topology of a disk.
The combinatorics of the vertex is such that if we take all these
disks together they form the dual 2-complex $T^*\approx\Gamma$ of
a simplicial D-complex $T$, the original disks being the 2-cells
topologically dual to the (D-2)-dimensional subsimplices in the
simplicial complex (for details consult \cite{iogft, laurentgft, dgftreview}).

\medskip

In the (g,P) variables the Feynman graph amplitude factorizes per
dual face (or, equivalently, per edge of the triangulation), i.e.
the amplitude for a graph $\Gamma$ is a product of dual face
amplitudes:

\[
Z_\Gamma\,=\,Z_{T^*} =  \int_{G^{E^*}} \bigg ( \prod_{i=1}^{E^*}
dh_i \bigg ) \int_{P^{F^*}} \bigg ( \prod_{i=1}^{F^*} dP_i \bigg )
\prod_{f^* \in T^*} A_{f^*} (h_{(e^*\in\partial f^*)} \,  ; \,
P_{f^*}) ,
\]

where, $E^*$ is the number of the dual edges in $T^*$ and $F^*$ is
the number of the dual faces, and  $A_{f^*}$ is the amplitude
assigned to each dual face $f^*$. This amplitude depends on the
group elements $h_{e*}$ that are assigned to the dual edges $e^*$
on the boundary of the dual face $f^*$, and that result from the
gauge symmetry of the field $\phi$ under $G$ (see \cite{iogft}),
and on a single P variable associated to the whole dual face $f^*$
left after doing all the delta functions over intermediate
momenta. More precisely, this amplitude is just a product of
kernels with delta functions, integrated over the common group and
momentum $P$ variables, and for a dual face with N vertices (and
thus N links) it is given by

\be
\label{eq:ampl}
A [h_1, \dots , h_N ; P]  =   \int_{G^{2N}} \bigg ( \prod_{i=1}^N dg_i \, dg_i'  \bigg ) \Bigg ( \prod_{i=1}^N  K\bigg [g_i h_i g_i^{' -1} , \Big ( P^2 - \frac{d}{24} \, m^2 \Big ) \bigg ]  \Bigg )  \Bigg ( \prod_{i=1}^N \delta(g_{i-1}' g_i^{-1})  \Bigg ),
\ee

where $g_0 = g_N$. The first multiplicand is just the the
propagators which are sitting on the dual edges, while the second
multiplicand is the delta functions coming from the
vertices\footnote{We drop the infinite constant $\delta(0)$ which
is a consequence of the translational symmetry in the P variables,
leaving the detailed treatment of this symmetry for future work.}.\\

\begin{figure}
\centering
\includegraphics[width=0.7\textwidth]{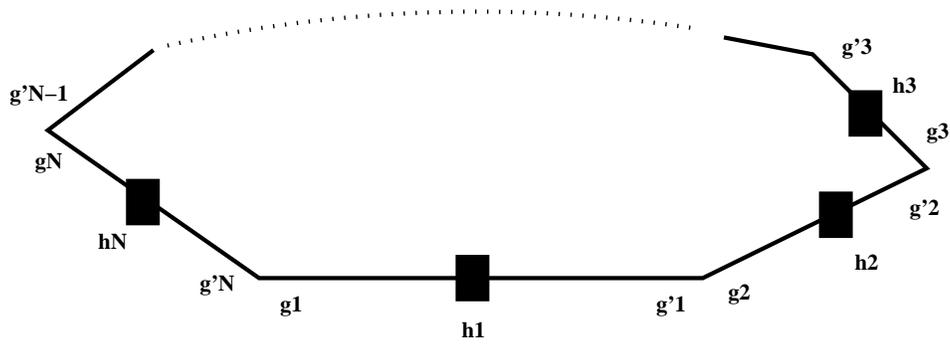}
\caption{\textit{This is a picture of all the variables appearing in (\ref{eq:ampl}) in the amplitude for a dual face. The black boxes represent the integrals over the h's.}}
\end{figure}

We can use the delta functions coming from the vertices to do the
integrals over the $g'$'s obtaining:

\[
A [h_1, \dots , h_N ; P]  =   \int_{G^N}  \bigg ( \prod_{i=1}^N dg_i \bigg ) \Bigg ( \prod_{i=1}^N  K\bigg [g_{i-1} h_i g_i^{-1} , \Big ( P^2 - \frac{d}{24} \, m^2 \Big ) \bigg ]  \Bigg ).
\]

\ni We would like to do the integrals over the remaining g's and
obtain something which depends only on the holonomy around the
dual face, computed through the $h$ variables only, as in usual
spin foam models and GFTs. However, the Schroedinger kernels in the
mass representation in the g-variables do not compose in any
simple way. To bypass this difficulty we use again the
Feynman-Schwinger-DeWitt representation for the kernels in the
previous equation

\[ A [h_1, \dots , h_N ; P]  =   \int_{G^N}  \bigg (
\prod_{i=1}^N dg_i \bigg ) \int_{\mathbb{R}^N}  \bigg (
\prod_{i=1}^N dt_i \bigg ) \Bigg ( \prod_{i=1}^N e^{i t_i ( P^2 -
\frac{d}{24} \, m^2)} \, \theta(t_i) \,  K\bigg [g_i h_i
g_{i+1}^{-1} , t_i \bigg ]  \Bigg ). \]

\ni Now, since the kernels in the proper time representation do
satisfy the composition identity (\ref{eq:composition}) we can
(after interchanging the order of integration) perform the group
integrals obtaining

\[
A [h_1, \dots , h_N ; P]  =   \int_{\mathbb{R}^N}   dt_1 \dots dt_N  \, e^{i ( P^2 - \frac{d}{24} \, m^2) ( t_1 + \dots + t_N)} \, \theta(t_1) \dots \theta(t_N) \, K \bigg [ h_1 \dots h_N \, , \, t_1 + \dots + t_N     \bigg ].
\]

\ni The product of the group elements in the kernel is exactly the holonomy around the dual face which we will denote by H. Thus $A [ h_1 , \dots , h_N ; P] = A_N [ H ; P ]$.\\

\ni To do the integrals over the proper times we change variables

\[
A_N [H ; P]  =   \int_{\mathbb{R}^N}  dT \, dt_2 \dots dt_N  \, e^{i ( P^2 - \frac{d}{24} \, m^2) \, T } \, \theta(T - t_2 - \dots - t_N) \, \theta(t_2) \dots \theta(t_N) \, K \big [ H \, , \, T  \big ].
\]

\ni The integrals over $t_2 , \dots , t_N$ can now be performed as these variables appear only in the step functions giving

\be
\label{eq:amplitudegp}
A_N [H ; P]  =   \frac{1}{(N-1)!} \int_{\mathbb{R}}   dT   \,  \,  e^{i ( P^2 - \frac{d}{24} \, m^2) \, T } \, \, \theta(T) \, \, T^{N-1} \, K \big [ H \, , \, T  \big ].
\ee

What we have shown above is that the dual face amplitude in the
(g,P) variables is the value at $( P^2 - \frac{m}{4})$ of the
Fourier transform of a monomial multiplied by the retarded
Schroedinger kernel in the (proper) time $T$. We will use this
equation repeatedly in what follows. \footnote{As it is well known, multiplication of a function by monomials corresponds to differentiation of its Fourier transform. Thus, by differentiation,
we could obtain the dual face amplitude corresponding to $N>1$ from the case $N=1$. There is a problem with this approach however due to the fact that the
integral above does not converge for $N \neq 1$ \textit{unless} some of the parameters of the kernel are complexified; and while it is possible to find the
complexification needed by a careful analysis using distribution theory, it is much simpler to simply do the integral above explicitly, since then the
required complexification is then easy to see.} The explicit form of this object depends on the details of the group under consideration \cite{marinov}.
We will give the explicit formulae for the rotation and Lorentz groups in three and four dimensions in the next sections.\\

The above discussion gives the Feynman amplitude $Z_{T^*}$ in
terms of the (g,P) variables. To make connection with the usual
spinfoam we want to write this amplitude in terms of the (J,P)
variables as well. This is done by returning to the general
expression of the face amplitude (\ref{eq:ampl}), inserting the
character expansion of the propagator (\ref{eq:propagatorjp}) and
using the fact that the characters satisfy\footnote{As usual, the
indices can go over discrete and continuous values. $\delta_{JK}$
is the Kronecker delta in the discrete case and the Dirac delta in
the continuous. The easiest way to see that this equation is true
comes from seeing that it follows from the fact that the delta
functions on the groups compose, i.e. that $\int_G dg_2  \,
\delta(g_1 g_2^{-1}) \, \delta(g_2 g_3^{-1}) = \delta(g_1
g_3^{-1})$.}

\[
\int_G dg_2 \, \Delta_J \, \chi_J(g_1 g_2^{-1})  \, \,  \Delta_K
\, \chi_K(g_2 g_3^{-1}) = \delta_{JK} \, \Delta_J \, \chi_J(g_1
g_3^{-1}).
\]

It is then easy to see that the dual face amplitude is given
by\footnote{Note that this reconfirms equation
(\ref{eq:amplitudegp}) as if we take the given character expansion
of the kernel, plug it into the right hand side of
(\ref{eq:amplitudegp}) and evaluate the integral over T, we obtain
exactly the answer given in (\ref{eq:amplitudejp}).}:

\be \label{eq:amplitudejp} A_N[ H ; P] =  \sum_{J} \Bigg [
\frac{i^N \Delta_{J}}{\Big ( P^2 - \frac{d}{24} \, m^2  \mp C_{J}
+ i \epsilon \Big ) ^N}  \Bigg ]  \, \chi_{J}(H), \ee where N is
again the number of dual edges (vertices) in the dual face $f^*$.
Going through the standard computations
\cite{SF} of group integrals, we can obtain
from this formula the spin foam picture of our model.  The
amplitude of the dual 2-complex (the Feynman amplitude) obtained
from our model is given by

\be \label{eq:spinfoamjp} Z_{T^*} = \Bigg ( \prod_{f^*\in T^*
}\sum_{J_{f^*}} \,\int_P dP_{f^*} \Bigg ) \Bigg ( \prod_{f^* \in
T^*} \Bigg [ \frac{i^{N_{f^*}} \Delta_{J_{f^*}}}{\Big ( P_{f^*}^2
- \frac{d}{24} \, m^2  \mp C_{J_{f^*}} + i \epsilon \Big )
^{N_{f^*}}} \Bigg ] \prod_{v^* \in T^*} \{ \textrm{J-Symbol} \}
\Bigg ). \ee

The sum goes over all labellings of the dual 2-complex by representations of G, and J-Symbol stands for the appropriate symbol coming from the
representation theory of G (it is the 6-J symbol in 3 dimensions and 15-J symbol in 4 dimensions). Note that the Feynman amplitude in these variables is
now factorized differently, as it is no longer just a product of amplitudes assigned to dual faces, but, as a result of the group integrations, there are
contributions coming from the dual vertices.\\

It is easy to see that, in the spin foam representation, i.e. in
momentum space, from the GFT perspective, the difference between
the new models and the usual ones lies in the amplitudes assigned
to the dual faces. These amplitudes are just just products of the
coefficients of the character expansion of the propagators above.
However, albeit limited, this difference is crucial and has many
consequences: 1) it makes the Feynman amplitudes complex; 2) it
produces truly dynamical propagating quantum degrees of freedom,
as the usual Feynman propagator of QFT does; 3) it selects as
dominant contributions to the amplitudes the solutions of the
kinematical QFT equations of motion, i.e. those for which
$P_{f^*}^2 - \frac{d}{24} \, m^2  \mp C_{J_{f^*}}=0$, which, up to
a constant $\frac{d}{24} \, m^2$, implies the identification of
the $P_i$'s with the Lie algebra generators for the group $G$,
which, as explained in the previous section, is what we want to
mimic the structure of a BF path integral, given the
identification (that we will confirm in detail in presenting the
3d and 4d models) of the P variables with the discrete analogue of
the B field of BF theory .

\medskip
We report here for completeness also the expression for the
Feynman amplitudes in the (J,X) representation:

The propagator was already given above (\ref{eq:propagatorjx}). The vertex is completely analogous to the one we gave in the (g,P) variables (under the substitutions g $\rightarrow$ X and P $\rightarrow$ J), with the sole difference that the whole expression is now multiplied by the appropriate $\{ J - Symbol \}$. Now, however, there is a difference between models A and B mentioned at the beginning of this section, as the delta functions on the X variables are different. \\

\[
D_F [ X_i , Y_i ; J_{1, i} , J_{2,i} ] = \prod_{i=1}^D  \Bigg ( \delta_{J_{1,i} J_{2,i}} \,  K \bigg [  X_i - Y_i , \pm C_J + \frac{d}{24} m^2   \bigg ] \Bigg ),
\]

where K is the Schroedinger kernel in the \lq mass' representation. Note that this expression is basically the same as the propagator in the (g,P) variables (\ref{eq:propagatoriskernel}), showing that the formulations of the theory in the (g,P) variables and in the (J,X) ones are dual to each other. There is an important difference however, which spoils this duality which is the extra symmetry the field satisfies in the g variables (shift invariance) which has no analogue in the X ones. This is the reason there is no integral in the above expression over the space X analogous to the integral over h in (\ref{eq:propagatoriskernel}).\\

The vertex is completely analogous to the one we gave in the (g,P) variables (under the substitutions g $\rightarrow$ X and P $\rightarrow$ J), with the sole difference that the whole expression is now multiplied by the appropriate $\{ J - Symbol \}$.\\

If we use now these ingredients to calculate the Feynman amplitudes $Z_{\Gamma} = Z_{T^*}$, we get the following

\[
Z_{T^*}^{(A), (B)} = \prod_{f^* \in T^*} (\sum_{J_{f^*}} \int_X dX_{f^*} )  \prod_{f^* \in T^*} A^{(A), (B)}_{N_{f^*}}(J_{f^*} , X_{f^*} ) \prod_{v^* \in T^*} \{J - Symbol\},
\]

with the dual face amplitude $A^{(A), (B)}_{N_{f^*}}(J_{f^*} ,  X_{f^*})$ being given by either\footnote{We are dropping the infinite factor which is equal to $\sum_{J} \delta_{J J}$ arising from contracting all the representation Kronecker deltas around a dual face.}

\[
A^{(A)}_{N_{f^*}} (J_{f^*} , X_{f^*}) = \frac{\Delta_{J_{f^*}}}{(N-1)!} \int_0^{\infty} dT \, T^{N-1} e^{i (\pm C_J + \frac{d}{24} m^2) T} \, K[  (1 + (-1)^{|\nu|}) X_{f^*} \, , \, T],
\]

or

\[
A^{(B)}_{N_{f^*}} (J_{f^*}) = \frac{\Delta_{J_{f^*}}}{(N-1)!} \int_0^{\infty} dT \, T^{N-1} e^{i (\pm C_J + \frac{d}{24} m^2) T} \, K[ 0  \, , \, T],
\]

where, as mentioned above, K is just the Schroedinger kernel on X.
$|\nu|$ is a function, with value either $0$ or $1$, of the
combinatorics of the triangulation and the assignment of the
orientation data (the $\nu$'s) to the triangulation whose exact
form is not important here, as we are not going to use or discuss this particular representation of the Feynman amplitudes. \\

Notice the marked difference with the same quantity in the (g,P)
representation (which is especially clear in model B where the
above amplitudes are completely independent of the X variables).
This is due to the absence of the analogue of the \lq shifting'
gauge integrals. 

\medskip

Let us summarize what we have discussed so far. We have defined a
new class of generalized GFT models in (\ref{eqnarray:actiongx},
\ref{eqnarray:actiongp}, \ref{eqnarray:actionjp},
\ref{eqnarray:actionjx}). We then analyzed the Feynman rules of
the theory. The vertex is easily seen to be almost the standard
one. The propagator for the theory (which in a sense encodes most
of the new features of the model) is obtained using the
Schwinger-DeWitt parametrization. We have then constructed the
Feynman amplitudes of the model in both the (g,P), (J,P) and (J,X)
variables. Some general features of the new models are already
apparent at this stage, such as the complexity of the amplitudes,
the presence of propagating degrees of freedom at the quantum
level, the relaxation at the quantum level of the relation between
(the discrete analogue of) the B field and the generators of the
Lie algebra of the group G. We will now move on, and present in
detail the model one obtains from this general definition in the
3d and 4d cases, in both Riemannian and Lorentzian settings. In
doing so, the above features will become even clearer, as in
particular it will become clearer the geometric interpretation of
both the P and the g variables. Moreover, we will see that the
Feynman amplitudes of the new models, in the (g,P) variables, have
indeed the form of path integrals for simplicial quantum gravity of the form of a BF theory restricted to positive orientation. This extra condition is what makes the Feynman amplitudes we get not triangulation independent.

\subsection{New vs. Conventional Models}

Here we discuss the relation between the new and the usual models.
There are two ways in which one re-obtains the more traditional
GFTs and spin foam models for BF theory, as an appropriate
restriction, from these generalized ones (the same was true for
the models proposed in \cite{generalised}).

\begin{itemize}
\item The conventional GFTs are obtained when we take the static-ultra-local limit \cite{rivers} of the action
(\ref{eqnarray:actiongx}), and for a specific choice of the mass
parameter $m^2=1$ (which however does not play any role in the
resulting amplitudes). In this limit one gets rid of the
propagation in the theory by replacing the derivatives in the
kinetic term with delta functions:

\[ \bigg [ \prod_{i=1}^D \Big ( - \Box_{X_i} \, +  \, \Box_{G_i}
\,  - \frac{d}{24} \, m^2 \Big )\delta(g_i,g_i')\delta(\nu_i X_i +
\nu_i' X_i') \bigg ]\;\;\rightarrow\;\; \bigg [
\prod_{i=1}^D\,\delta(g_i,g_i')\delta(\nu_i X_i + \nu_i' X_i')
\bigg ]. \]

 If we do this in
(\ref{eqnarray:actiongx}) we will obtain essentially the usual GFT
model but with the sole difference of having extra arguments which
the field depends on.

How are the Feynman amplitudes affected by these extra variables?
Since there is no coupling between the group and the X (or P)
variables, they are just propagated in parallel around the Feynman
graph.
The upshot of this is that the extra variables X (or P) contribute just
an overall (infinite) constant and thus do not affect the amplitudes, that reduce then to the usual spin foam models.\\

\item Another way of looking at the relation between the new model
and the conventional, which clarifies the fact the new model is
the causal analogue of the usual ones, comes from considering the
theory in the (J,P) variables.

Take a single propagator and look at its character expansion
(\ref{eq:propagatorjp}). As is clear from this equation that the
coefficients of the characters are just the usual Klein-Gordon
propagators on a flat space X, whose dimensions is equal to the
dimension of the group G and which has a metric which is the
Killing form. Also, it is clear that it is from here that the
complexity (thus the causal nature, as we discussed) of the
amplitudes comes. Using Sohozki's formula $\frac{i}{x+i \epsilon}
= \pi \delta(x) + i P(\frac{1}{x})$ and the reality of
characters\footnote{Technically this is not true for noncompact
groups. Fortunately the characters enter into the delta function
expansion (\ref{eq:delta}) (from which the propagator is derived)
in a symmetric way, such that the coefficients of characters which
are complex conjugates of each other are real and equal.}, it
follows that the real part of the propagator is given by

\be \label{eq:onshell} D_F [ g_i , h_i ; P_i  , Q_i]   =
\prod_{i=1}^D \Bigg ( \, \delta(P_i - Q_i) \, \int_G dh \, \Big[
\sum_{J_i} \delta \big ( P^2_i - \frac{d}{24} \, m^2 \mp C_{J_i}
\big ) \, \chi_{J_i}(g_i h h_i^{-1}) \Big ]  \Bigg). \ee Notice
that taking the real part of the propagator is the same as going
on-shell with respect to the corresponding equation of motion,
which is the classical relation between the P variables and the
Lie algebra generators of the group G, or, as we will confirm in
the next sections and we have discussed in the previous, between
the B and the A field of BF theory (metricity of the connection).
If we do now the integrals over the P variables it is immediate
that we just get the propagator and thus the whole spin foam
amplitudes of the usual GFTs, as the delta functions integrate to
one.

\end{itemize}

Let us stress once more that the \lq\lq causal\rq\rq nature of the
new models, i.e. the fact that their amplitudes are complex
functions of the geometric data, interpretable, as we will confirm
in the next sections, as discrete gravity path integrals, results
exactly from the lifting of a classical equation of motion (a
discrete analogue, we argue, of the relation between B and A in BF
theory) to allow for off-shell propagation. One could go further
and argue that it is this quantum lifting of a classical condition
that allows us to go beyond usual BF theory, where it is only the
classically allowed flat configurations that have non-zero
amplitude in the path integral.\\

Notice also that the situation here is entirely analogous to what
happens in the case of a free relativistic quantum particle. The
real part of the propagator $\frac{i}{p^2 - m^2 + i \epsilon}$ is
simply the on-shell condition $\delta(p^2 - m^2)$. It is crucial
to allow the momentum to go over the classically
disallowed value in order to have genuine quantum behaviour of the system. This is also what characterizes time-ordered products of field
operators with respect to other 2-point functions in scalar field theory\\

However, even though the above interpretation is intriguing and in
line with what our initial motivations were and our results of the
next sections will show, we feel that there is much more left to
understand about the physics behind the above outlined relation
between the new models and the traditional BF ones, as well as
about the relation between the two ways, discussed here, in which
one can re-obtain the traditional models from these new ones. We
leave this for future work.
\section{New 3d GFT models}

\subsection{Riemannian 3d gravity}

We now specialize the class of models considered above to the case
D=3 and G=SU(2). The usual models (with the trivial kinetic term),
for this
choice of dimension and group, give Euclidean 3-d BF theory, augmented by a sum over topologies, in perturbative expansion.\\

The action (\ref{eqnarray:actiongp}) becomes

\begin{eqnarray*}
S & = & \frac{1}{2 \, (2 \pi)^{9}} \int_{G^3} \bigg ( \prod_{i=1}^3 \, dg_i \bigg ) \, \int_{P^3} \bigg ( \prod_{i=1}^3  dP_i  \bigg ) \, \, \,  \phi^*(g_i ; P_i)  \bigg [ \prod_{i=1}^D \Big (  \, P_i^2 \, +  \, \Box_{G_i} \,  - \frac{m^2}{8} \Big ) \bigg ] \phi(g_i ; P_i) + {}  \\
& & {}  +   \frac{\lambda}{(2\pi)^{36} 4!} \sum_{\nu_1 \ldots \nu_4} \int_{G^{12}} \bigg ( \prod_{i \neq j = 1}^{4} \,  dg_{ij} \bigg ) \, \int_{P^{12}} \bigg ( \prod_{i \neq j =1}^{4} dP_{ij}      \bigg ) \, \, \, \bigg [ \prod_{i < j} \delta(g_{ij} g_{ji}^{-1}) \delta(P_{ij} - P_{ji}) \bigg ] \times{} \\
& & {} \times \phi^{\nu_1}(g_{1j} ; P_{1j})  \ldots \phi^{\nu_4}(g_{4 j} ; P_{4 j}).
\end{eqnarray*}

SU(2) is a compact group of rank one, hence the kernel depends on
a single periodic parameter. It is convenient to choose this
parameter to be the \lq angle of rotation' in the usual
representation of SU(2). More precisely, if H $\in$ G then $H =
e^{i \frac{\theta(H)}{2} \vec{n} \cdot \vec{\sigma}}$ where
$\theta(H)$ is the angle of rotation, $\vec{n} \in S^2$ is the
axis of rotation and $\vec{\sigma}$ are the Pauli matrices. The
angle $\theta(H)$ is a \textit{multivalued} function of the group
element. This should be clear as $\theta(H)$ and $\theta(H) + 4
\pi n$ for $n \in \mathbb{Z}$ correspond to the same group
element. In other words, any choice of $n$ in the expression
$\theta(H) + 4 \pi n$ provides a possible definition of the angle
characterizing the holonomy $H$. What this means is that from a
geometrical point of view, the angle of rotation, is intrinsically
an equivalence class of real numbers modulo addition of $4 \pi$.
We will denote this equivalence class by $[\, \theta(H) \, ] =
\theta(H) \, \textrm{mod} \, 4 \pi$, and identify $[\theta(H)]$,
i.e. $\theta(H) + 4 \pi n$ for any choice of $n$, with the
holonomy angle. However, since the equivalence class is not a
number, to write any formula involving the angle of rotation, one
should pick a representative of the equivalence class (i.e. choose
a specific $n$, for example $n=0$ thus restricting oneself to the
$[0, 4 \pi]$ range). This random choice, does not matter if the
function is automatically periodic when $\theta \rightarrow \theta
+ 4 \pi$ (e.g. the character function). However, when the
functional expression one is dealing with is not periodic (the
evolution kernel below), one needs to sum over all the equivalence
classes (all possible $n$ to obtain a
function with the correct boundary conditions, i.e. a function on the group. \\

The explicit form of the evolution kernel on SU(2) in (proper)
time $T$ is given \cite{camporesi} by the following
formula

\be
\label{eq:su2kernel}
K [ H \,  , \, T ]   =   \frac{1}{(4 \pi i T )^{\frac{3}{2}}} \sum_{n=- \infty}^{\infty} \Bigg( \frac{\theta(H) + 4 \pi n}{2 Sin\Big(\frac{\theta(H)}{2}\Big)}\, \, \, \, Exp \bigg [  \frac{i}{2T} \Big( \theta(H) + 4 \pi n \Big )^2  + \frac{i T}{8}      \bigg ] \Bigg ).
\ee

Note the sum enforcing periodicity in $\theta \rightarrow \theta +
4 \pi$.  To avoid writing the sums which enforce periodicity in
what follows, we adopt the following notation: whenever we have a
sum which enforces periodicity of a certain function, i.e.
whenever have an expression of the form $\sum_{n= -
\infty}^{\infty} f(\theta + 4 \pi n)$, we will just write $f([\,
\theta \, ])$. The sum, which is required to convert an expression
involving $[\, \theta \, ]$ to a legitimate one involving just
real numbers, will be kept implicit. This is perfectly reasonable
from the geometric point of view as well, as it is exactly the
entire equivalence class that has the meaning of an angle of
rotation. This sum has also the meaning of a sum over all geodesics over the group (i.e. $S^3$) connecting the same two points \cite{camporesi}. \\

Once more, we define the partition function of the model as a
perturbative expansion in Feynman diagrams:
\[
Z = \sum_{\Gamma} \frac{\lambda^{V_\Gamma}}{\textrm{sym}(\Gamma)}
Z_{\Gamma},
\]
and, again, the Feynman amplitudes factorize per dual face:
\[
Z_\Gamma\,=\,Z_{T^*} =  \int_{G^{E^*}} \bigg ( \prod_{i=1}^{E^*}
dh_i \bigg ) \int_{P^{F^*}} \bigg ( \prod_{i=1}^{F^*} dP_i \bigg )
\prod_{f^* \in T^*} A_{f^*} (h_{(e^*\in\partial f^*} \,  ; \,
P_{(f^*)}) ,
\]

According to (\ref{eq:amplitudegp}), to get the amplitude
$A_{f^*}$ for a dual face with N vertices in the (g,P) variables,
we should multiply the expression for the evolution kernel by
$\frac{\theta(T) \, T^{N-1}}{(N-1)!}$ and then take the Fourier
transform of the result, with respect to $T$, at the value $P^2 -
\frac{m^2}{8}$\footnote{We are using the normalizations of
\cite{marinov}. The Killing form in our conventions is given by $2
\mathbb{I}$, where $\mathbb{I}$ is the 3x3 identity matrix. As a
consequence, since the metric on the dual to X is given in terms
of the inverse of this Killing form, $P^2 =  \frac{1}{2}
\vec{P}^2$.}. \\

Thus

\be
\label{eq:su2amplitude}
A_N \big [H \, , \, P \big ] = \frac{1}{(4 \pi i )^{\frac{3}{2}}(N-1)!} \sum_{n=- \infty}^{\infty} \Bigg( \frac{[ \, \theta(H) \, ] }{2 Sin\Big(\frac{[ \, \theta(H) \, ]}{2}\Big)}  \int_0^{\infty} dT \, \,  T^{N - \frac{5}{2}}\, \, Exp \bigg [  \frac{i}{2T} [ \, \theta(H) \, ]^2  + i T \Big ( \frac{\vec{P}^2}{2} - \frac{m^2-1}{8}    \Big)      \bigg ] \Bigg )
\ee

The integral can be evaluated explicitly \cite{grad} using the formula

\be \label{eq:hankel} \int_0^{\infty} dT \, T^{\nu-1} \, Exp \Big
[ \, \frac{i p}{2} \big ( \, T \, + \, \frac{q^2}{T}\,  \big  )
\Big ] =  i^{\nu +1} \pi \, q^{\nu} \,  H_{\nu}^{(1)}(q p), \ee

where $H_{\nu}^{(1)}(z)$ is a Hankel function of the first kind of
order $\nu$. The two coefficients p and q are complex numbers in
general, but what is very important is that they should satisfy
(Im$(p) > 0$ and Im$(pq^2) > 0$). It should be obvious that this
should be the case as the integrals will simply not converge
otherwise. Note that while the left hand side has $q^2$ in it, the
right hand side has $q$. The fact that we have to take a square
root will be very important in the Lorentzian
case.\\
For us \be \label{eq:pq} \nu = N - \frac{3}{2}\qquad  \, , \,
\qquad \,    p =  \vec{P}^2 - \frac{m^2-1}{4}  \, \qquad \,
\textrm{and} \, \qquad \, q^2 =  \frac{[ \, \theta(H) \, ]^2}{\vec{P}^2 - \frac{m^2-1}{4}}. \ee

It is clear that $(\textrm{Im}(p) > 0$ and $\textrm{Im}(pq^2) >
0$)  imply that both the $\vec{P}^2$ and the $\theta(H)$ should be
complexified and given small positive imaginary parts. This
complexification is nothing but the usual Feynman $i \epsilon$
prescription. The root of $q^2$ is defined in the usual way, by
taking a cut along the negative real axis, letting $\sqrt{1} = 1$
and extending by continuity. As both the numerator and denominator
have small phases (both are positive), their ratio also has a
small phase. Thus the square root of $q^2$ is very close to the
real axis and is very nearly equal to $\frac{| [ \, \theta(H) \, ] |}{|\vec{P}^2 - \frac{m^2-1}{4}|}$.\\

Plugging (\ref{eq:hankel}) into (\ref{eq:su2amplitude}) we get

\[
A_N \big [H \, , \, P \big ] = \frac{i^{N-2}}{16 \sqrt{\pi} (N-1)!}  \Bigg [ \, \,  \frac{[ \, \theta(H) \, ]}{\textrm{Sin}\Big(\frac{[ \, \theta(H) \, ]}{2}\Big)} \Bigg ( \frac{| \, [ \, \theta(H) \, ] \, |}{\sqrt{\vec{P}^2 - \frac{m^2 - 1}{4}}} \Bigg )^{N- \frac{3}{2}} \Bigg ] \, H^{(1)}_{N-\frac{3}{2}} \Big ( \sqrt{\vec{P}^2 - \frac{m^2 - 1}{4}} \, | \, \, \, [ \, \theta(H) \, ] \, |  \Big ).
\]

The Hankel function of half-integer order can be given explicitly
in terms of elementary functions via

\be
\label{eq:polynomial}
H_{N - \frac{3}{2}}^{(1)}(z) = \sqrt{ \frac{2}{\pi z}} i^{- (N-1)} \Bigg \{ \, \sum_{K=0}^{N-2} (-1)^K \frac{(N + K -2)!}{K! (N-K-2)!} \, \frac{1}{(2 i z)^K} \Bigg \}\, \, e^{iz}.
\ee

Using this expression we get that the dual face amplitude has the form

\be
\label{eq:ramplitudemeasure}
A_N \big [ H \, , \, P \big ] = \mu \Big ( [ \, \theta(H) \, ]  \, , \, |\vec{P}| \, , \, N   \Big ) \, e^{i \sqrt{\vec{P}^2 - \frac{m^2 - 1}{4}} \, \, \, | \, [ \, \theta(H) \, ] \, |},
\ee

with $\mu$ being given by

\begin{eqnarray}
\mu \Big ( [ \, \theta(H) \, ]  \, , \,
|\vec{P}| \, , \, N   \Big ) & = & \frac{- i \sqrt{2}}{16 \pi
(N-1)!} \Bigg [ \frac{1}{\textrm{Sin}([ \, \theta(H) \, ])} \Bigg
( \frac{|[ \, \theta(H) \, ]|}{\sqrt{\vec{P}^2 - \frac{m^2 - 1}{4}}}\Bigg )^{N-1} \Bigg ] \times{} \nonumber \\
\label{eqnarray:measure}
& &{} \times \Bigg \{ \sum_{K=0}^{N-2} \Big ( -1 \Big )^K  \frac{(N + K
-2)!}{K! (N-K-2)!} \frac{1}{\Big ( 2 i \sqrt{\vec{P} - \frac{m^2 - 1}{4}} \, \, | [ \,
\theta(H) \, ] | \Big )^K} \Bigg \}.
\end{eqnarray}

Above, we have given the amplitude for just one dual face, or
recalling that in 3d a dual face is dual to an edge of the
triangulation, it is the amplitude for a single edge. However, as
was mentioned in the previous section, the amplitude of the dual
complex in the (g,P) variables is just the product of the
dual-face amplitudes, or in the 3d context the product of edge
amplitudes. Thus, we can easily write the amplitude for the whole
triangulation (Feynman graph) $Z_T$. It is

\be
\label{eq:partition}
Z_T = \int_{G^{E^*}} ( \prod_{e^* \in T^*} dg_{e^*})  \int_{P^{E}} ( \prod_{e \in T} d\vec{P}_e \,    ) \,  \, \mu( g_{e^*} , \vec{P}^2_e, N_e) \, \, \, e^{i \sum_e \sqrt{\vec{P}^2_e - \frac{m^2 - 1}{4}} \, \, \, | [ \, \theta_e \, ] |  },
\ee

where the products go over all edges in the triangulation $e \in
T$ and all dual edges in the dual 2-complex $e^* \in T^*$, and the
factor $\mu(g_{e^*}, \vec{P}^2_e, T)$ is a product of all the
$\mu$'s coming from each dual face i.e. $\mu = \prod_{e \in T}
\mu_e$ with $\mu_e$ given by (\ref{eqnarray:measure}).\\

Now, consider the exponent in the above expression. We see
immediately that it is just the Regge action for Euclidean 3d
gravity

\[
S_{Regge} =  \sum_{edges} L_e  \, \theta_e ,
\]

in 1st order form,  after identification of $\sqrt{\vec{P}^2_e  -
\frac{m^2 - 1}{4}}$ with $L_e$. Here the sum goes over all edges
of the triangulation. $L_e$ stands for the length of the edge e
and $\theta_e$ for the deficit angle, i.e. the discretized
curvature, around the dual edge, which coincides with the angle of
rotation $ [ \, \theta_e (H_e) \, ]$ that characterizes our holonomies $H_e$ (again, equivalent to $\theta(H) + 4 \pi n$ for any choice of $n$).\\

This reconfirms and makes precise the interpretation for the new
variables, the $P$'s, which was proposed in the introduction, as that they give the length of the edges to which they are associated, and thus as
representing the discretized triad (B field) associated with these
edges, while the group elements are confirmed as a discretization
of the Lorentz connection field $A$. Indeed, we obtain an
expression for the simplicial gravity action of the same type as
the ones in \cite{magnea}, and, as there, with the edge lengths
(hinge
volumes) restricted to have a positive orientation. Note that this identification of the length with the variable P becomes especially nice if we set $m^2 =1$, as then it is the length of $\vec{P}_e$ directly, $|\vec{P}_e|$, which coincides with the length of the edge $L_e$. For this reason, as well as to simplify the formulae, we will adopt this choice for $m^2$ in the following discussion of the amplitude in the (g,P) variables.\\

It is clear that the variation of the above action with respect to
the edge lengths, or the variables $\vec{P}_e$, gives the
classical equation $\theta_e=0$, i.e. imposes flatness of the
discrete geometry as the only {\it classically} allowed
configuration, as we expect from 3d gravity. The variation with
respect to the connection variables is more involved, and we would expect it to provide a discrete analogue of the continuum conditions enforcing metricity of the  connection. We leave
its analysis for future work \cite{renorm}.\\

The amplitude for the triangulation $Z_T$ is then just the
partition function for discrete 3d Euclidean gravity, in 1st order
form, with a measure factor $\mu(g_{e^*}, \vec{P}^2_e, N_e)$, as
desired.

\medskip

\ni Let us now consider the measure factor $\mu(g_{e^*},
\vec{P}^2_e, N_e)$, in more detail. This factor is a complex
number in general as should be evident from (\ref{eqnarray:measure}).
Thus if we write

\[
\mu(g_{e^*}, \vec{P}^2_e, N_e) = |\mu(g_{e^*}, \vec{P}^2_e, N_e)|
\, \, e^{i S_{c}(g_{e^*}, \vec{P}^2_e, N_e)} ,
\]

with

\be S_c (g_{e^*}, \vec{P}^2_e, N_e) = \sum_e \Bigg [ -
\frac{\pi}{2} + \textrm{arctan} \Bigg  ( \frac{
\sum_{K=0}^{\lfloor \frac{N_e-3}{2}   \rfloor}
\frac{(-1)^K}{2^{2K+1}}   \frac{(N_e+2K - 1 )!}{(2K+1)!
(N_e-2K-3)!}  \frac{1}{ ( |\vec{P}_e| \, | [ \, \theta_e \, ] |
)^{2K+1}}    }{   \sum_{K=0}^{\lfloor \frac{N_e-2}{2}   \rfloor}
\frac{(-1)^K}{2^{2K}}   \frac{(N_e+2K-2)!}{(2K)! (N_e-2K-2)!}
\frac{1}{( |\vec{P}_e| \, | [ \, \theta_e \, ] | )^{2K}} } \Bigg )
\Bigg] \label{Sc}, \ee

we see that the full Feynman amplitude for the whole triangulation
has the form

\[ Z_T = \int_{G^{E^*}} ( \prod_{e^* \in T^*} dg_{e^*} )  \int_{P^E} (  \prod_{e \in T}
d\vec{P}_e     ) \,  \, |\mu( g_{e^*} , \vec{P}^2_e, N_e)| \, \,
\, e^{i \big [ S_{Regge}( g_{e^*} , \vec{P}^2_e) + S_c( g_{e^*} ,
\vec{P}^2_e, N_e)  \big ] }. \]

The modulus of the quantum measure $\mu(g_{e^*}, \vec{P}^2_e,
N_e)$, i.e. $|\mu( g_{e^*} , \vec{P}^2_e, N_e)|$ is then what
should be considered as a proper quantum measure factor in our
path integral, while the phase $e^{i S_{c}(g_{e^*}, \vec{P}^2_e,
N_e)}$ gives what can be interpreted as quantum corrections to the
Regge action (hence the subscript). We thus see that the
amplitudes of our model, more precisely, have the form of a path
integral (with an explicitly defined measure) of an extended 1st
order Regge calculus, in which the Regge action is extended by
(also explicitly computable) quantum corrections.\\

Let us then study in more detail these quantum corrections, in
order to confirm their geometric meaning and thus their proposed
interpretation. We then study the explicit formula (\ref{Sc}) for
$S_c$ as well as the expression (\ref{eqnarray:measure}). Also, we focus
on the dependence on the geometric data $P$ and $\theta$,
neglecting constant factors, which give a constant contribution
to the phase at every edge (equal to $- \frac{\pi}{2}$).\\

One of the most important properties of this part of expression
(\ref{eqnarray:measure}) is that it depends on $[ \, \theta_e \,]$
and $\vec{P}_e$ solely through the combination ($| \vec{P}_e | \,
| [ \, \theta_e \, ] |$). This also implies that it can be
expanded in (in general, positive and negative) powers of the same
combination ($| \vec{P}_e | \, | [ \, \theta_e \, ] |$), weighted
by factors that will necessarily be purely combinatorial, i.e.
dependent on $N_e$ only.\\

Under the interpretation discussed above for the P variables and
for the $\theta$, a first possible interpretation of the powers of
the expression ($L_e\theta_e$) is that they represent the discrete
analogues of higher order corrections to the Einstein-Hilbert
action, given by powers of the Ricci scalar \cite{ruthhamber}. One
could then expect the correspondence \[\sum_e C_e \Big (|
\vec{P}_e| \, | [ \, \theta_e \, ] | \Big )^K \sim \, \,\int
R^K(g) \, \, \textrm{Vol}   , \] where $C_e$ is the mentioned
combinatorial factor, Vol is the volume form and the
aforementioned correspondence holds in the continuum approximation
(in the sense of measures) \cite{ruthhamber}.\\

However, the simplicial geometry of such higher powers of the
Regge term is subtle (see again \cite{ruthhamber} for an extensive
and detailed analysis). In particular, for the square power of the
above expression, another plausible interpretation is provided by
the square of the Riemann tensor, giving:

\[\sum_e C_e \Big (| \vec{P}_e| \, | [ \, \theta_e \, ] | \Big )^2 \sim \, \,\int  R_{\mu\nu\rho\sigma}(g) R^{\mu\nu\rho\sigma}(g) \, \, \textrm{Vol}   . \]

In general, in fact, higher order curvature terms, as
traditionally defined in simplicial gravity, involve an additional
geometric ingredient, a normalization of the hinge volumes, that
gives them the correct dimensionless character. This is taken to
be the contribution of the D-simplex volume associated to the
specific hinge considered, $V_h$, giving a complete quadratic term
of the form $\Big ( \frac{ | \vec{P}_h| \, | [ \, \theta_h \, ]
|}{V_h} \Big )^2$. Its exact form could be argued, by universality
arguments, to be most likely irrelevant for the continuum
correspondence, but of course this is not at all obvious. With
this choice of normalizing factor, one can indeed show that (the
discrete analogue of) both $R^2$ and $(R_{\mu\nu\rho\sigma})^2$
agree when restricted to a single hinge. Therefore the difference
between the two types of higher order terms depends exclusively on
how different hinges are coupled, each being weighted individually
by the quadratic expression above. The simplest choice of coupling
$\sum_h V_h \Big ( \frac{ | \vec{P}_h| \, | [ \, \theta_h \, ]
|}{V_h} \Big )^2$  gives then a contribution to the action
corresponding to the square of the Riemann tensor. Other
constructions are however possible for both the Riemann tensor
itself and the quadratic terms that can be constructed from it
\cite{ruthhamber}. Also, we are not aware of similar detailed
analyzes for higher powers, thus for curvature invariants beyond
the quadratic order.\\

In our model, the normalizing volume factor can be interpreted as
being given by the Planck length to the appropriate power and
multiplied by our purely combinatorial factor $C_e$, a function of
$N_e$. Therefore a more complete interpretation scheme for the
higher order corrections to the Regge action provided by our GFTs
does involve a careful analysis of these combinatorial factors and
in particular of the way they couple different hinges in the same
D-simplex and beyond. This analysis will be performed and reported
elsewhere \cite{renorm}.\\

From a more general perspective, however, these corrections to the
Regge action, predicted by our model(s) share two main features:
1) they involve, as mentioned, both positive and negative powers
of the curvature invariants, and 2) they depend on two independent
sets of geometric variables, the (discrete analogues of) the
D-bein and the connection fields. This implies, therefore, that
the corrections to the bare Regge action produced by the model are
of the general f(R) type in the metric affine formalism \cite{fR}.\\

We would like to emphasize once more that these corrections are
not arbitrary, rather their form, including relative coefficients
weighting them, and their behaviour in the various regimes of the
theory are fully determined by the our choice of the original GFT
action. This also means of course that one can modify the exact
dependence on them of the simplicial action appearing in our
Feynman amplitudes, by modifying the same GFT action, thus
constructing different specific models within the general class of
GFTs we have defined.\\

Let us analyze further the physics behind the corrections $S_c$.
We are most interested in two approximations, both of which can be
given a clear physical interpretation.\\

The first regime is when the lengths becomes large, i.e. when
$|\vec{P}| \gg 1$ (remember that we are working in Planck units).
Equivalently, this is the regime of large actions, in units of the
Planck's constant, due to the way in which the edge lengths enter
the discrete Regge action. This approximation can thus be
considered as a \lq semiclassical approximation' as it corresponds
to the case where the relative size of the quantum fluctuations of
the action (and of the edge lengths) is small. This is the
analogue, for our models, of the asymptotics usually
considered in the standard spin foams (the large J asymptotic).\\

The second regime is approached when the edge lengths and discrete
curvatures become small, and the triangulation becomes finer and
finer, i.e. when $( |\vec{P}| | [ \, \theta \, ] | ) \rightarrow
0$ and $N \rightarrow \infty$. This can be thought of as the \lq
continuum approximation\rq. \\

Let us first look at the behaviour of the measure and thus of the
quantum corrections $S_c$ at the heuristic level. Consider then
the explicit expression for the (complex) measure in
(\ref{eqnarray:measure}), and in particular to the part of it within
curly brackets.\\

In the first case (large lengths $|\vec{P}_e|$) it is the first
term in the sum in (\ref{eqnarray:measure}) that dominates, and since
this term is real, it means the Regge action remains the dominant
contribution to the phase of the path integral amplitude. We
expect then the phase, including corrections, to be of the general
form $S_{Regge} + O(\sum_e \frac{1}{\mid P_e\mid \theta_e})$, thus
with inverse powers of the curvature to play the role of quantum
corrections to the Regge action, and the full Feynman amplitude
(discrete gravity path integral) to be approximated by

\begin{eqnarray*} Z_T
& \sim & \int_{G^{E^*}} ( \prod_{e^* \in T^*} dg_{e^*} ) \int_{P^{E}} (  \prod_{e \in T}
d\vec{P}_e    ) \,  \, \Bigg \{  \prod_e \Bigg [ \frac{- i \sqrt{2}}{16 \pi
(N_e-1)!} \Bigg ] \Bigg [ \frac{[\theta_e]^{2(N_e-1)}}{\textrm{Sin}([ \,
\theta_e \, ])} \Bigg ( \frac{1}{|\vec{P}_e| [
\theta_e]}\Bigg )^{N_e-1} \Bigg ] \times{} \\
& & {} \times  \Bigg [ 1 + O \bigg ( \frac{1}{|\vec{P}_e| [ \theta_e] } \bigg ) \Bigg ] \Bigg \} \, \, e^{i \big [ S_{Regge}(
g_{e^*} , |\vec{P}_e|) + O( 1 / |\vec{P}_e| [ \theta_e])  \big ]
}.
\end{eqnarray*}

In the second case (small edge lengths and very fine
triangulation, i.e. high $N_e$) it is the last term in the sum in
(\ref{eqnarray:measure}) that dominates. This term also contributes just
a constant to $S_c$ (equal to $(N-2) \frac{\pi}{2}$). We expect
then the phase, including corrections, to be of the general form
$S_{Regge} + O(\sum_e ( | \vec{P}_e| \theta_e)^2)$, thus with
positive powers of the curvature to play the role of quantum
corrections to the Regge action, and the full Feynman amplitude
(discrete gravity path integral) to be dominated by a term like:

\[
Z_T \sim \int_{G^{E^*}} ( \prod_{e^* \in T^*} dg_{e^*} )  \int_{P^E} ( \prod_{e \in
T} d\vec{P}_e   ) \,  \,\Bigg \{ \prod_e \Bigg [ \frac{(C_{N_e}
[\theta_e])^{2(N_e-1)}}{\textrm{Sin}([ \, \theta_e \, ])} \Bigg
( \frac{1}{|\vec{P}_e| [ \theta_e]}\Bigg )^{2N_e-3} \Bigg ] \Bigg \} \,
\, e^{i \big [ S_{Regge}( g_{e^*} , \vec{P}^2_e) \,  + \, O(\sum_e ( | \vec{P}_e| \theta_e)^2)  \big ] }.
\]

\medskip

We would like now to go beyond the naive heuristic considerations
and analyze the form of the quantum measure, and of $S_c$ in
particular, in
more detail.\\

This can be done with full confidence for the semiclassical
approximation. The reason for this is that (\ref{eqnarray:measure}), and
thus the full Feynman amplitude, is regular at the limiting point
$\mid P\mid \rightarrow \infty$ (it goes simply to zero), for a generic triangulation. Also, the proper analysis involves the
asymptotic expansion of the Hankel function for large values of
the argument, but, for half-integer order, this {\it coincides}
with the expression (\ref{eq:polynomial}) that we have used. This
allows us to obtain full understanding of the way the phase
behaves in the large length limit. We can then use directly the
expression (\ref{Sc}) and, expanding the arctangent in powers of
$\frac{1}{\theta_e |\vec{P}_e|}$, we get that

\[
S_c (g_{e^*}, \vec{P}^2_e, N_e) = \sum_e  \Bigg [ {N_e \choose 2} \frac{1}{\theta_e |\vec{P}_e|} + o \bigg (\frac{1}{( \theta_e |\vec{P}_e|)^2} \bigg) \Bigg ].
\]

Of course, all the coefficients in the expansion can, in principle
be computed within our model. As said, we can think of $\sum_e C_e
\frac{1}{\theta_e |\vec{P}_e|}$ as the inverse of the scalar
curvature. Thus $S_c \sim \int [ \frac{1}{R} + o(\frac{1}{R^2}) ]
\textrm{Vol}$. Since the corrections are inverse in the curvature,
they are of the infrared type, as it is intuitively to be expected
as we are doing a large scale approximation to out model. Thus we
see that the new model predicts long-distance effects, at the
simplicial level, of the same type as those predicted by effective
$f(R)$-extended gravity models, and that have been found relevant
in cosmological applications (most notably for modelling dark
energy effects)
\cite{fR}.\\

The other case of interest (the \lq continuum' limit) is much more
involved to analyze, and the purely heuristic argument can be
trusted as a limited indication of the relevant physics (it is
intuitively obvious that in the small distance regime one gets
quantum corrections of the ultraviolet type $O(R^2)$), but one
that cannot be easily confirmed by a detailed analysis, at this
point.\\

The reason for this is that, as is not difficult to see from
(\ref{eqnarray:measure}), the Feynman amplitude has a badly singular
point in ($| \vec{P}_e |=0$): 1) it diverges in the limit like
$\frac{1}{|\vec{P}|^{2N-3}}$; 2) the Hankel function has a branch
point at 0, which poses extra problems one needs to deal with due
to the Stokes' phenomenon\footnote{The phenomenon that the asymptotic expansion around a point of a nonanalyticity of a function depends on the sector chosen for the approach to the given point.}, whose main consequence is, in this
context, that the expression for the amplitude around this point
depends heavily on how exactly the limit is taken, i.e. which path
one takes in the complex domain to approach the singular point.
Finally,
the limit $( |\vec{P}| | [ \, \theta \, ] | ) \rightarrow 0$ by
itself is not very physically meaningful. It acquires its
importance when combined with the limit $N \rightarrow \infty$.
However, it is not difficult to see that the way the amplitude
behaves is sensitive to the way these two limits are combined. Due
to the above reasons we defer the detailed treatment of this
regime of the model, as well as of the corresponding formulation
of simplicial geometry for future work \cite{renorm}.\\

Finally, let us note that the fact that the amplitude diverges as
$\frac{1}{|\vec{P}|^{2N-3}}$ is very appealing intuitively, as it
implies that for larger triangulations it is the small values of
$|\vec{P}|$ that are the most relevant ones and that they become
more and more dominant as we take larger and larger triangulations
(this is because the higher powers are more divergent).\\ Since we
have interpreted the P variables as giving the lengths of the
edges of the triangulation, this looks exactly like the behaviour
one would want in order to recover a good continuum limit: for a
triangulation consisting of a large number of tetrahedra, the
dominant histories are those for which the basic simplices are
small, corresponding moreover to a singularity in the quantum amplitudes. \\

\medskip

The new model is a causal one in the sense of \cite{causal,
feynman, generalised} and it shares many features of the 3d model
presented in \cite{causalmatter3d}. Let us briefly recall the
model proposed there. The action used in \cite{causalmatter3d} is
a discretized version of (\ref{eq:bfaction}). The B field is
replaced with a Lie algebra element $P = \vec{P} \cdot \vec{J}$
associated to every edge of the triangulation, and the connection
A is substituted by its holonomy around the dual face $H =
\textrm{Exp}(\theta \, \vec{n} \cdot \vec{J})$. The discrete
action is then given by

\[
S' = \sum_{e \in T} \vec{p} \cdot \vec{n}\, \textrm{Sin}(\theta_e).
\]

The model is quantized via the path integral method in the usual
way the only crucial difference being that the product $\vec{P}_e
\cdot \vec{n}_e$ is restricted to be nonnegative. This is because,
as was argued in \cite{causalmatter3d}, this corresponds to
restricting the discretzied \lq volume' to be positive, and thus
it represents the wanted implementation of the \lq causality\rq
restriction in quantum gravity transition amplitudes. Thus the
partition function is given by\footnote{The reader should not confuse the $\theta_e$ which is the deficit angle around the edge with the Heaviside step function $\theta(\vec{P}_e \cdot \vec{n}_e)$.}

\be
\label{eq:causalpartition}
Z = \int_{G^{E^*}} \bigg ( \prod_{e^* \in T^*}    dg_{e^*} \bigg ) \int_{P^E} \bigg ( \prod_{e \in T}  d \vec{P}_e  \bigg ) \theta(\vec{P}_e \cdot \vec{n}_e) \, e^{i \, \vec{P}_e \cdot \vec{n}_e \, \textrm{Sin}(\theta_e)}.
\ee

The new model, which generates amplitudes given by
(\ref{eq:partition}), is causal in the same sense as (\ref{eq:causalpartition}) due to the simple fact that the
$|\vec{P}_e|$ is always positive. Thus, keeping the interpretation
of the P's in mind, in our GFT model the integral over the
discretized field is also restricted to be such that the hinge
volumes are positive. This restriction results
\cite{causalmatter3d} in the causal analogues of usual spin foams
in both the free and matter coupled
cases.\\

There are several differences, however, between the model proposed
here and the one proposed in \cite{causalmatter3d}, i.e. between
(\ref{eq:partition}) and (\ref{eq:causalpartition}).

\begin{itemize}
\item First, the discretizations used in the two cases are somewhat different. Although, both use the holonomies to represent the curvature and both average
the B field over an edge (and get a vector), the way these two
objects enter into the discrete action is slightly different.
Notably, the two variables are totally independent in the new
model and interact simply through multiplicative coupling, at
least before one uses the equations of motion resulting from the
variation of the simplicial action. In the old model however, the
variables mix more substantially: a) there is extra coupling
introduced by the dot product $\vec{P}_e \cdot \vec{n}_e$ ($\vec{n}_e$ is completely absent from the action in (\ref{eq:partition})); and
b) the domains of integration of the two variables are
interdependent, due to the step function. With regards to both
these points the new model is \textit{simpler} than the old one.
It is well possible, however, that one can get a 3d model, in the
same new class of GFTs we are proposing, that is closer to the one
in \cite{causalmatter3d} by imposing additional (symmetry)
conditions on the variables appearing in the GFT action.
\item The measure factor $\mu( g_{e^*} , \vec{P}^2_e, N_e)$ present in (\ref{eq:partition}) is absent from (\ref{eq:causalpartition}).  These are,
as discussed above, corrections to the bare Regge action (and thus
to the 3d BF action) that have been here deduced from first
principles and not added in an ad hoc way (which of course could
be done with (\ref{eq:causalpartition})). Thus the new model is
significantly richer than the old one. Also with respect to this
point, we notice that there is still freedom left in choosing
specific GFT actions within the general class of GFTs we
introduced, and thus obtaining models with modified (and possibly
simpler) path integral measures in the perturbative expansion.
\item Due to the fact that the factor $\mu( g_{e^*} , \vec{P}^2_e, N_e)$ depends on $N_e$, it should be clear that if we perform the integrals
over the P's in (\ref{eq:partition}) we will get dual face
amplitudes which depend on the number of vertices in each dual
face, i.e. each dual face amplitude is a function of $N_e$. This
however is not the case in the old model where the dual face
amplitudes, which were computed explicitly in
\cite{causalmatter3d} were independent of this factor. The reason
for this can be traced to the following fact. At the spin foam
level, and in the construction of \cite{causalmatter3d}, the basic
building block of the model was considered to be the dual face. At
the GFT level, it is necessarily the wedge (i.e. the portion of
the dual face contained within a D-simplex) from which everything
else is constructed \cite{mikecarlo, kirill}. The causal
restriction advocated for in \cite{causalmatter3d} was a dual-face
one, and this is the reason for the independence of the resulting
amplitudes from the number of wedges (vertices) making the dual
face. One would expect that if the construction in
\cite{causalmatter3d} is repeated but with the causality restriction being imposed at the level of each wedge, one would obtain a model which is closer to the one reproduced here.
\item Finally, in \cite{causalmatter3d} the causal restriction, although shown plausible, was implemented essentially by hand simply by inserting the step
function into the partition function (\ref{eq:causalpartition});
therefore one could be left wondering about the possibility of
different ways of implementing the same type of causal
restriction. In the new model(s) we are proposing such freedom is
absent, at least for given choice of GFT action: the amplitudes
are built, in a {\it unique} manner, from the same building block,
the propagator, and it is exactly the propagator that has the
information about causality, orientation dependence and the
propagation of quantum degrees of freedom.
\end{itemize}

\medskip

Consider now the model in the (J,P) variables (i.e. equation
(\ref{eq:amplitudejp}) specified for $D=3$ and G= SU(2))

\be
\label{eq:rspinfoamjp}
Z_{\Gamma}\,=\,\bigg (\prod_{f*\in
T^*}\sum_{J_{f^*}} \int_P d^3 \vec{P}_{f^*}\,\bigg ) \prod_{f^* \in
T^*} \Bigg [ \frac{i^{N_{f^*}} (2J_{f^*}+1)}{\Big (
\frac{\vec{P}_{f^*}^2}{2} - \frac{m^2}{8}  -
\frac{J_{f^*}(J_{f^*}+1)}{2} + i \epsilon \Big ) ^{N_{f^*}}} \Bigg
] \prod_{v^* \in T^*} \left\{ 6-J \right\},
\ee

where of course now the representations are labelled by
half-integers $J_{f^*}$. Since it is the P's that give the
lengths, the interpretation of the J variables is not as
straightforward as it is in the usual models. Looking at the
expression above we can see that the J's label the different poles
of the dual face amplitude. Since the expression for the dual face
amplitude is essentially a product of Feynman propagators we can
think of the J variables as labelling the different
semi-classical, on-shell values of $|\vec{P}_e|$. The poles are

\[
\vec{P}^2  =   J (J+1) + \frac{m^2}{4}.
\]

If we make the same choice for $m^2$ as before, i.e. if we set $m^2 = 1$, then we see that $\vec{P}^2 = \frac{(2J + 1)^2}{4} =
\frac{\Delta_J^2}{4}$.\\

Notice that if we plug this back\footnote{In other
words, if we restrict the P variables to these discrete values by
inserting $\prod_e \sum_{J_e} \delta(|\vec{P}_e| -
\frac{\Delta_{J_e}}{2})$ into the path integral. We can heuristically interpret this restriction as imposing the connection metricity equation of motion (i.e. the equation obtained by varying the connection) into the path integral.} into
(\ref{eq:partition}) we see that it becomes

\[ Z_T = \sum_{J_1, \ldots, J_{F^*}}\int  ( \prod_{e^* \in
T^*}dg_{e^*} ) \,  \, \mu( g^* , \Delta_{J_e}, N_e) \, \, \, e^{i
\sum_e  \frac{\Delta_{J_e}}{2} \theta_e   }, \]

from which we see that the exponent is just the Regge action with
the edge length restricted to be $\frac{\Delta_J}{2}$. This
matches nicely with the expression obtained in \cite{alekseev}
(see also \cite{laurenteteracarlo}) for the eigenvalues of the
length operator in 3d canonical quantum gravity.\footnote{Apart
from the factor of half, which is a consequence of the
normalization we chose for the P field.} \\

Of course, this does not
really mean that the \lq lengths' are quantized in our model. This
is because the \lq length' information is given by $|\vec{P}_e|$'s
and these are unconstrained, in general. Just like in the Feynman
propagator
for a scalar particle the momentum is not constrained, in the quantum theory, to the mass shell.\\

\medskip

We can now obtain a pure spin foam expression for the Feynman
amplitudes of our model, i.e. one involving only the
representation variables. It is not difficult to perform the
integrations over $\vec{P}_f$'s in (\ref{eq:rspinfoamjp}). The easiest way
to do this is by using Cauchy's formula\footnote{This is done by
changing to polar coordinates, extending the radial integral to go
from $-\infty$ to $\infty$ and then closing the contour in the
complex plane. By Jordan's lemma, since $N_{f^*} \geq 2$, the
integral of the expression we have along a semicircle centered at
the origin of radius R, goes to zero as R $\rightarrow \infty$.
This allows us to add this bit to the integral closing the
contour.}. The result of these integrations is given by

\[  Z_T = \sum_{J_1, \ldots, J_{F^*}}
\,\Bigg ( \prod_{f^* \in T^*} A_{f^*} (J_{f^*}, N_{f^*})
\prod_{v^* \in T^*} \left\{6-j \right\} \Bigg ), \]

where the dual face amplitude is given by

\be \label{eq:rspinfoamj} A_{f^*} ( J_{f^*} , N_{f^*} ) =  4 \pi^2
\Bigg [  \frac{-2i}{\Delta_{J_{f^2}}^2} \Bigg ]^{N_{f^*}-1} \,
\Bigg [ \frac{ (2N_{f^*}-2)!}{\big ( (N_{f^*}-1)! \big)^2} \Bigg ]
\, \, \, \Bigg [ \frac{N_{f^*}-2}{2N_{f^*}-3} - \frac{1}{2} \Bigg
]. \ee

Let us now try to extract some physical information on the model,
and in particular how it depends on the combinatorics of the
underlying triangulation, starting from this expression for the
amplitudes.\\

Consider the regime of large $N_{f^*}\gg 1$, i.e. consider the
triangulations which are composed of many tetrahedra, which we
have argued is one ingredient for approximating continuum physics
in this setting. This should be combined with a small $|\vec{P}|$
approximation; however, having integrated out the P's, we can only
expect to read out from the amplitudes what are the dominant
configurations in the $J$ variables. Using Stirling's
formula\footnote{e is the Euler number.} $n! \sim \sqrt{2 \pi n}\,
(\frac{n}{e})^n$  we can easily see that the second multiplicand
in (\ref{eq:rspinfoamj}) is asymptotic to
$\frac{4^{N_{f^*}-1}}{\sqrt{\pi (N_{f^*} -1)}}$. Thus for large
$N_{f^*}$'s

\be \label{eq:asymptotic} A_{f^*} ( J_{f^*} , N_{f^*} )  \sim
\frac{1}{F(N_{f^*})}
\,  \,  \Bigg [  \frac{8}{\Delta_{J_{f^*}}^2}    \Bigg
]^{N_{f^*}-1},   \ee

where F is a function of polynomial growth.\\

We conclude that the amplitude consisting of a large number of
tetrahedra is dominated (as this is when
$\frac{8}{\Delta_{J_{f^*}}^2}
> 1$) by the two lowest values of J's,
$J_{f^*}=0$, which can be thought of as the vacuum configuration,
and $J_{f^*}=\frac{1}{2}$, which is some sort of lowest excited
state. So, if we interpret the values of $J$ as edge lengths, as
in usual spin foam models, it is the shortest values that are the
dominant ones for fine triangulations, as we would expect. In the
limit of finer and finer triangulations (which, again, we would
expect to lead to a continuum approximation of the discrete path
integral, then, the partition function can be reasonably well
approximated by a purely combinatorial sum, with amplitudes given
by the above quantities evaluated at $J=0$, i.e. for purely
equilateral triangulation with edge lengths $L_e= l_P
(2J_e+1)\mid_{J_e=0}=l_P$. In other word, in this regime, the
model would effectively, and {\it dynamically}, reduce to a pure
dynamical triangulations model \cite{DT}. \\

Consider the regime of large $J_{f^*}$'s. Again, having integrated
out the \lq true\rq edge length variables $P$, we can
heuristically interpret this regime as a large distance
approximation. Looking again at the same expression
(\ref{eq:rspinfoamj}), it is clear that it is the lowest values of
$N_{f^*}$'s that are most relevant in the limit. What this means
is that if we look at the large length limit the most important
Feynman diagrams are represented by the simplest triangulations,
more precisely those with least number of vertices for each dual
face. In other words, {\it if one is interested only in large
distance and semi-classical physics}, then considering simple
triangulations would suffice, as the GFT partition function, in
perturbative expansion, is anyway dominated by such
configurations. This leads further support to the nice results
obtained in the calculation of the lattice graviton propagator in
\cite{graviton}, working indeed in the context of the large-J
limit of spin foam models, and using semi-classical boundary
states based on simple boundary triangulations, as well as very
simple bulk triangulations (low order in the GFT coupling constant
$\lambda$).\\

These considerations should however be taken with care, since the
$\Delta_{J_f}$ are not, strictly speaking edge lengths, as we have
stressed above, this role being in fact played by the $P$
variables. We hope that the above results underlie the fertility
and potential usefulness of the proposed model in understanding
quantum geometry.

\subsection{Lorentzian 3d gravity}

We now move on to the case when D=3 and G=SL(2,$\mathbb{R}$) $
\simeq $ SU(1,1), i.e. G is the double cover of the Lorentz group
in three dimensions. Thus, this model corresponds to the
Lorentzian gravity in 3d.\\

SU(1,1) has two nonconjugate Cartan subgroups.\footnote{For a short summary of the group theoretic facts used in this work, see the Appendix.} This is easy to see
as $\mathfrak{su}(1,1)$ can be obtained by complexifying two of
the generators of $\mathfrak{su}(2)$. Thus we would obtain a
generator of rotation and two generators of boosts. The Cartan
subgroups are thus the two subgroups generated by these different
elements. One Cartan subgroup is just U(1) generated by the
uncomplexified element, we will denote its conjugacy class by R
(for rotation). The other Cartan subgroup is generated by one of
the complexified elements and it is a noncompact group (isomorphic
to $\mathbb{R}$) whose conjugacy class we will denote by B (for
boost).\\

The fact that they are Cartan subgroups means that any element of
SU(1,1) is conjugate to either an element of R or of B apart from
a set of elements of measure zero in the Haar measure.\footnote{At
a technically level, let g $\in$ SU(1,1). If $|Tr(g)| > 2$ then g
$\in$ B. If $|Tr(g)| < 2$ then g $\in$ R. The set of elements
which satisfy $|Tr(g)| = 2$ is a set of measure zero.} The
conjugacy classes of the elements of R will be parametrized by a
periodic parameter $\theta$ (angle) for which we choose a
normalization such that its period is $4\pi$. While the conjugacy
classes of the elements of B will be parametrized by a real number
$\psi$ (the boost parameter, rapidity). \\

The explicit formula for the evolution kernel in proper time is
given by the following formula \cite{marinov}

\begin{eqnarray}
K [ H \,  , \, T ]  & = &   \frac{1}{(4 \pi i T )^{\frac{3}{2}}}  \frac{[ \, \theta(H) \, ] }{2 Sin\Big(\frac{[ \, \theta(H) \, ]}{2}\Big)}\, \, \, \, Exp \bigg [  \frac{i}{2T} [ \, \theta(H)  \,]^2  + \frac{i T}{8}      \bigg ]  \qquad \qquad \qquad \textrm{when} H \in R  \nonumber \\
\label{eqnarray:su11kernel}
& & \\
&   =   &  \frac{1}{(4 \pi i T )^{\frac{3}{2}}}  \frac{\psi(H) }{2 Sinh\Big(\frac{\psi(H)}{2} \Big)}\, \, \, \, Exp \bigg [ - \frac{i}{2T} \psi^2  + \frac{i T}{8}      \bigg ] \qquad \qquad \qquad \quad \, \textrm{when} H \in B \nonumber ,
\end{eqnarray}

where we have used the same notation for the periodic parameter $\theta$ as in the previous subsection.\\

Note that when the holonomy group element is a rotation then the
SU(1,1) evolution kernel has exactly the same form as the SU(2)
one (\ref{eq:su2kernel}). The crucial difference between the
rotation and the boost cases is the different sign sitting in
front of the parameters in the two cases; plus in the rotation
case and minus in the boost case.\\

Once more we are interested in the partition function of the theory, expanded perturbatively in Feynman diagrams

\[
Z = \sum_{\Gamma} \frac{\lambda^{V_\Gamma}}{\textrm{sym}(\Gamma)}
Z_{\Gamma},
\]

and, again, the Feynman amplitudes factorize per dual face

\[
Z_\Gamma\,=\,Z_{T^*} =  \int_{G^{E^*}} \bigg ( \prod_{i=1}^{E^*}
dh_i \bigg ) \int_{P^{F^*}} \bigg ( \prod_{i=1}^{F^*} dP_i \bigg )
\prod_{f^* \in T^*} A_{f^*} (h_{(e^*\in\partial f^*} \,  ; \,
P_{(f^*)}) .
\]

According to the general formula (\ref{eq:amplitudegp}) we get the
dual face amplitude $A_{f^*}$ by multiplying the above expression
for the kernel (\ref{eqnarray:su11kernel}) by $\frac{\theta(T)
T^{N-1}}{(N-1)!}$ and evaluating its Fourier transform at $P^2 -
\frac{m^2}{8}$. Note that the Killing form (which enters into the
definition of $P^2$) has now signature $(+--)$. Thus there is one
\lq timelike' direction (the generator of the compact subgroup)
and two \lq spacelike' ones. Using the same normalizations as in
the case of SU(2), we get $P^2 = \frac{1}{2} ( P^1_1 - P^2_2 -
P^2_3) = \frac{1}{2} \vec{P}^2$. So, the amplitude for a dual face
with N vertices is given by

\[
A_N  [H \, , \, P ] =\frac{1}{(N-1)!} \int_0^{\infty} dT \, T^{N-1} \, e^{i T \,( P^2 - \frac{m^2}{8}) } \, K[H \, , \, T].
\]

Now, consider the case when H is a rotation (H $\in$ R). Then
since the formula for the kernel (\ref{eqnarray:su11kernel}) is
exactly the same as the one we used in the SU(2) calculation
(\ref{eq:su2kernel}) we can just write down the answer. Thus

\begin{eqnarray}
A_N \big [ H \, , \, P \big ] & = & \mu_R \Big ( [ \, \theta(H) \, ]  \, , \, \sqrt{\vec{P}^2} \, , \, N   \Big ) \, e^{i \sqrt{\vec{P}^2 - \frac{m^2 - 1}{4}} \, | \, [ \, \theta(H) \, ] \, |}\\
\mu_R \Big ( [ \, \theta(H) \, ]  \, , \, \sqrt{\vec{P}^2} \, , \, N   \Big ) & = & \frac{- i \sqrt{2}}{16 \pi (N-1)!} \Bigg [ \frac{1}{\textrm{Sin}([ \, \theta(H) \, ])} \Bigg ( \frac{|[ \, \theta(H) \, ]|}{\sqrt{\vec{P}^2 - \frac{m^2 - 1}{4}}}\Bigg )^{N-1} \Bigg ]  \times{} \nonumber \\
\label{eqnarray:rsu11amplitudegp} & & {} \times   \Bigg \{
\sum_{K=0}^{N-2} \Big (-1 \Big )^K  \frac{(N + K -2)!}{K!
(N-K-2)!} \frac{1}{\Big ( 2 i \sqrt{\vec{P}^2 - \frac{m^2 - 1}{4}}
\, | [ \, \theta(H) \, ] | \Big )^K} \Bigg \}.
\end{eqnarray}

These are exactly the same formulae as before
(\ref{eq:ramplitudemeasure}), (\ref{eqnarray:measure}), with the
difference being that $\vec{P}^2$ is calculated with the Minkowski
metric and not the Euclidean one, and that this formula is not
valid for arbitrary SU(1,1) element H, rather only when H is a \lq
rotation' (H $\in$ R). Note the exponential factor in the
amplitude. Restricting for the moment to the case when $\vec{P}^2
- \frac{m^2 - 1}{4} > 0$, it should be clear that if we interpret,
analogously to the Riemannian case, $\vec{P}^2 - \frac{m^2 -
1}{4}$ to be the square of the Minkowski length of the edge dual
to the dual face under consideration, then the above exponent
gives exactly the expected contribution to the Regge action coming
from the edge under consideration. In order to simplify the
following formulae and discussion we will set $m^2 =1$. This of
course also has the effect of making the length of $\vec{P}^2$ to
be directly the square of the edge length.\\

Now, the crucial difference between the Riemannian and the Lorentzian cases lies in the fact that $\vec{P}^2$ can now go over both positive and negative values. As mentioned above the case when $\vec{P}^2 > 0$ one just gets the exponent in the amplitude above becomes $e^{i |\vec{P}| | [ \, \theta(H) \, ] | }$. A simple oscillating phase. \\

When $\vec{P}^2$ goes negative, clearly $\sqrt{\vec{P}^2} = \pm i
|\vec{P}|$, and we have to choose a sign. As was mentioned in the
previous section we know that $\vec{P}^2$ should have a small
positive imaginary part. This means that when $\vec{P}^2$ goes
from positive to negative values it does so above the origin in
the complex. This means that $\vec{P}^2$ has values above the cut
we used to define the square root in the previous section. Thus we
have to choose the \lq positive' square root, i.e.
$\sqrt{\vec{P}^2} = + i |\vec{P}|$ (see the picture). Plugging
this into our exponent we see that it is equal to $e^{-
|\vec{P} | | [ \, \theta(H) \, ] }$.\\

\begin{figure}
\centering
\includegraphics[width=0.7\textwidth]{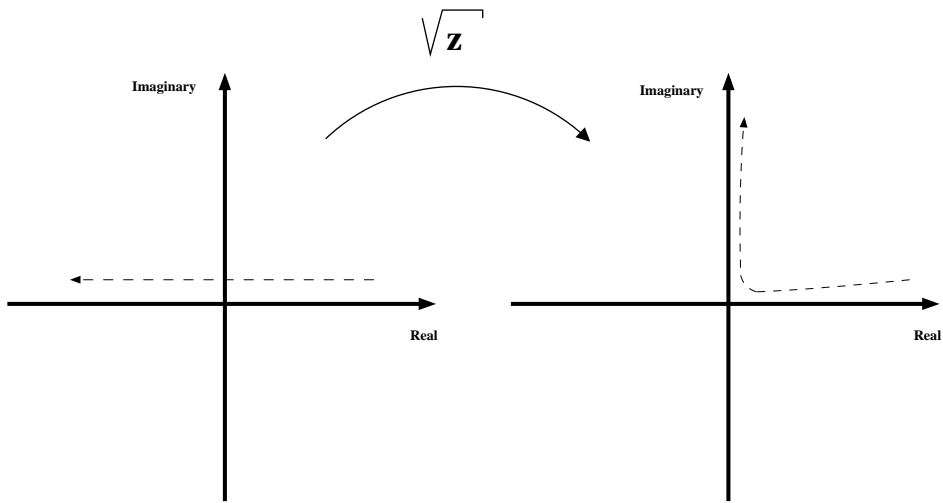}
\caption{\textit{This figure clarifies the way the square root behaves in the amplitude. The dotted path on the right is the image of the one on the left under the square root map.}}
\end{figure}

Keeping in mind the interpretation of the P variables as that
$\vec{P}^2$ is the length of the corresponding edge of the
triangulation, we see a very interesting phenomenon happening. The
exponent as we said earlier coming from an edge contributes a
summand towards the Regge action of the triangulation, with $[ \,
\theta \, ]$ being the deficit angle and $|\vec{P}|$ being the
length of the relevant edge. Now, as long as the \lq length' is
positive, i.e. $\vec{P}_e$ is timelike we get an oscillating phase
in the partition function. On the other hand when the $\vec{P}_e$
goes spacelike making the \lq length' negative, we get an
exponential suppression of the amplitude.\\

Classically, in the Regge action when the edge of the
triangulation is timelike, the curvature defect around it has to
be a rotation (think of a massive point particle). We see that
quantum mechanically this is not true. The edge corresponding to a
rotational defect can be both timelike and spacelike, however the
spacelike case is suppressed exponentially in the path integral.
This is similar to the behaviour exhibited by the Feynman
propagator of the relativistic point particle (which is not
surprising as we have essentially the same mathematics here). The
probability for the particle to propagate inside the lightcone is
given by an oscillating phase. The particle can also leak outside
the light cone (despite being relativistic). But, the probability
of doing so is exponentially suppressed.\\

This of course is an intuitively satisfying feature of the model.
However, the discussion above was limited to the case when the
holonomy H around the dual face is a rotation (lies in R).\\

When H is a boost we can of course repeat the same calculation as
before\footnote{The formula analogous to (\ref{eq:hankel}) gives a
Hankel function of the second kind.}. However, there is no real
need to do this. Look at the two expressions in
(\ref{eqnarray:su11kernel}). Note that apart from the factor in
front and a phase factor ($e^{i \frac{T}{8}}$), the case when g
$\in$ B is just the complex conjugate of the case when g $\in$ R,
due to the difference in the sign in front of the $\theta$ and
$\psi$. As, from the mathematical point of view, in order to get
the dual face amplitude we are taking a Fourier transform, we can
apply the general theorem that relates the Fourier transform of a
function to the Fourier transform of its conjugate. Namely, if we
denote the Fourier transform of a function f by $\mathcal{F}(f)
[k]$, then $\mathcal{F}(f^*) [k] = \big ( \mathcal{F}(f)[-k] \big
)^*$. The Fourier transform of a complex conjugate of a function
is the complex conjugate of the Fourier transform evaluated at the
negative of the argument. Using this we can immediately write down
the dual face amplitude in the case when g $\in$ B. It is given by

\begin{eqnarray*}
A_N \big [ H \, , \, P \big ] & = & \mu_B \Big (  \psi(H)   \, , \, \sqrt{ \vec{P}^2} \, , \, N   \Big ) \, e^{- i \sqrt{ - \vec{P}^2} \, | \,  \psi(H) } \\
\mu_B \Big (  \psi(H) \, , \, \sqrt{\vec{P}^2} \, , \, N   \Big ) & = &   \frac{ \sqrt{2}}{16 \pi (N-1)!} \Bigg [ \frac{1}{\textrm{Sinh}( \psi(H) )} \Bigg ( \frac{ \psi(H) }{\sqrt{- \vec{P}^2}}\Bigg )^{N-1} \Bigg ]  \times{}  \\
& & {} \times   \Bigg \{ \sum_{K=0}^{N-2} \Big (-1 \Big )^K
\frac{(N + K -2)!}{K! (N-K-2)!} \frac{1}{\Big ( - 2 i \sqrt{-
\vec{P}^2} \, | \psi(H)  | \Big )^K} \Bigg \}.
\end{eqnarray*}

The formula for $\mu_B$ is obtained from $\mu_R$ by letting
$\vec{P}^2 \rightarrow -\vec{P}^2$, replacing $[ \, \theta(H) \,
]$ with $\psi(H)$, switching the trigonometric sine for the
hyperbolic one and finally taking the complex conjugate. Of
course, by doing the whole calculation from scratch along the same
lines as in the Riemannian case, one gets the same result.\\

Now we can easily see that the behaviour of the amplitude when H
is a boost with respect to the different two possibilities of the
sign of the $\vec{P}^2$ is opposite of that when H is a rotation,
due to the minus sign in front of $\vec{P}^2$ in the formula
above. In other words, when $\vec{P}^2$ is positive, i.e.
$\vec{P}$ is a spacelike vector, then we just have an oscillating
phase. On the other hand when $\vec{P}^2$ goes negative, or
equivalently, when $\vec{P}$ becomes timelike, amplitude becomes a
decaying exponent.\footnote{Let us remark that we could have
arrived at the same conclusions by being careful with the square
root in the formula above. Since $-P^2$ has to have a small
positive imaginary part, $\vec{P}^2$ has a small negative
imaginary part. Thus when we go from the positive values to the
negative ones, we are doing so under the cut, thus choosing the
\lq negative' square root $\sqrt{\vec{P}^2} = - i |\vec{P}|$.}
Again, this is in full agreement with expectations as classically
the curvature defect around a spacelike edge is a boost.\\

Summarizing, if we put together all the dual face amplitudes and form the amplitude for the whole triangulation then what we get is

\be \label{eq:lamplitude} Z_T = \int_{G^{E^*}} ( \prod_{e^* \in
T^*} dg_{e^*} ) \int_{P^E} (  \prod_{e \in T} d\vec{P}_e \,    )
\,  \, \mu( g^*_{e^*} , \vec{P}_e^2, N_e) \, \, \, e^{i  S_{Regge}
} \ee

Where as before $\mu$ is the quantum measure factor, being a
product of $\mu_R$'s and $\mu_B$'s as appropriate, and $S_{Regge}$
is given by

\[
S_{Regge} = \sum_e  \alpha_e L_e  \, \Theta_e
\]

Here $L_e$ stands for the absolute value of the length of the edge e ($|\vec{P}_e|$) and $\Theta_e$ stands for the deficit parameter sitting at the edge e (an angle or a boost). Note that they are varied independently of each other showing that we have is 1st order theory. $\alpha_e$ is a function of both $L_e$ and $\Theta_e$ and is given by the following table

\be
\label{eq:table}
\begin{array}{c|c|c}
& Rotation &  Boost \\
\hline
Timelike  & +1   & +i  \\
\hline
Spacelike & +i   & - 1
\end{array}
\ee

Thus as we've said above when the variables are such that one is
off-diagonal in this table (Rotation-Spacelike or Boost-Timelike)
one gets exponential suppression of the amplitude. While when one
is on the diagonal then one gets an oscillating phase. This means
that the configurations that do not allow for a simultaneous
classical geometric interpretation for both the discrete B field
and the discrete connection, i.e. those configurations that would
be classically disallowed, are not forbidden but still
exponentially suppressed. We would like to stress the fact that
this causal behaviour is not put into the model by hand, but
rather emerges naturally from its very
definition as there were no arbitrary choices made anywhere in the construction (once the GFT action has been chosen). \\

Since the formulas in the Lorentzian case are so close to those in
the Riemannian one we can easily carry over all the results from
there. So, it is not difficult to see that
(\ref{eqnarray:measure}) carries over without much change. In
fact, there is no change when H is a rotation apart from the
definition of P. When H is a boost the angle becomes a boost
parameter, the trig sine goes to a hyperbolic one as well as a few
sporadic minus signs. The conclusions deduced from the measure
factor carry through without any change in the case in which we
have an oscillatory contribution to the partition function (i.e. a
complex exponential). The only difference being that when H is a
boost, all the phases go to their conjugates, which of course does
not affect the qualitative behaviour.\\

When we are off-diagonal in the table, and we have then an
exponential suppression, the integrand is, apart from an overall
factor (a power of i), real. This is easiest to see from the fact
that, as is evident from (\ref{eq:polynomial}), the Hankel
function for purely imaginary arguments is a (multiple of) real
function. Thus strictly speaking one just has the measure factor
in the path integral and no complex exponential (whose phase is to
be interpreted to be the action). However, we find it far more
clear, intuitively, and more insightful from the physical
perspective to split again the integrand into a \lq measure
factor' and an exponent as we did above. Applying this philosophy
to $\mu( g_{e^*}, P_e, N_e)$, we get corrections to the action
$e^{-S_{Regge}}$ of the form $e^{-S_c}$, exactly in
accordance with expectations, and in complete similarity with the results obtained in the other cases.\\

As before, in the large Minkowski length limit the quantum
corrections $S_c$ coming from the phase of the factor $\mu$ are of
the inverse scalar curvature type ($\sim \frac{1}{R}$), indicating
the infrared corrections to the bare Regge action in the
semiclassical limit.\\

Moreover, in all cases, we still get an amplitude that diverges
like $\frac{1}{|\vec{P}|^{2N-3}}$ as $|\vec{P}| \rightarrow 0$,
which means that when the number of tetrahedra in the
triangulation increases, the shorter \lq lengths' become more and
more dominant ones, which is what one would expect if the model is
to have a good continuum limit. The point $\vec{P}^2 = 0$ is a
branch point of the amplitude which diverges there, thus requiring
a much more detailed  treatment deferred for future work
\cite{renorm}.

\medskip

Let us now move on to the Lorentzian analogue of the (J,P)
representation (\ref{eq:rspinfoamjp}) for the quantum amplitudes.
To do this note that SU(1,1) has two types of representations
\cite{lfreidel}:

\begin{itemize}
\item Discrete ones labelled by a positive half integer J. The Casimir $C_J$ for these representations is negative and is equal to\footnote{Note that due
to a difference in normalizations our Casimir is half the one in
\cite{lfreidel}. Also, we are not differentiating here, as it is
not necessary nor very useful, between positive and negative
discrete series of representations as they both enter in exactly the same way into the partition function (\ref{eqnarray:lspinfoamjp}). This duplicity of representations is responsible for the factor of 2 in front of the sum over the discrete representations.} $C_J = - \frac{1}{2} J(J+1)$.
The constant $\Delta_J$ appearing in the character expansion of
the delta function is $\Delta_J = 2J+1$.
\item Continuous ones labelled by a positive real number $\rho$. The Casimir for these representations is positive and is equal to $C_{\rho} = \frac{\rho^2}{2} + \frac{1}{8}$. The constant $\Delta_{\rho}$ is just $\Delta_{\rho} = 2 \rho$
\end{itemize}

If we plug these expressions into (\ref{eq:spinfoamjp}) (note that we have to pick the positive sign in front of the Casimir as SU(1,1) is noncompact) we get

\be
\label{eq:lspinfoamjp}
Z_{\Gamma}\,=\,\bigg (\prod_{f*\in
T^*} \Big [  2 \sum_{J_{f^*}}  + \int_0^{\infty} d \rho_{f^*} \Big ] \int_P d^3 \vec{P}_{f^*}\,\bigg ) \prod_{f^* \in
T^*} A_{f^*}(J_{f^*}/ \rho_{f^*} \, , \, P_{f^*} \, , \, N_{f^*})   \prod_{v^* \in T^*} \left\{ 6-J \right\},
\ee

where now we get two types of the dual face amplitude in the (J,P) variables

\[
A_{f^*} [J_{f^*} \, , \, P_{f^*} \, , \, N_{f^*}   ]  =  \frac{i^{N_{f^*}} (2J_{f^*}+1)}{\Big ( \frac{\vec{P}_{f^*}^2}{2} - \frac{m^2}{8}  - \frac{J_{f^*}(J_{f^*}+1)}{2} + i \epsilon \Big ) ^{N_{f^*}}},
\]

when the representation is of discrete type, and

\[
A_{f^*}  [ \rho_{f^*} \, , \, P_{f^*} \, , \, N_{f^*} ] = \frac{i^{N_{f^*}} ( 2 \rho_{f^*} )}{\Big ( \frac{\vec{P}_{f^*}^2}{2} - \frac{m^2}{8}  + \frac{\rho_{f^*}^2}{2} + \frac{1}{8} + i \epsilon \Big ) ^{N_{f^*}}},
\]

when it is of the continuous type.\\

It is obvious from these two expressions that we get poles of two
types, timelike and spacelike. What we mean by this is that there
are two sets of poles, one when $\vec{P}^2$ is positive, i.e. when
$\vec{P}$ is timelike; and one when it is negative, i.e. $\vec{P}$
is spacelike. The first type of poles occurs when the
representation labelling the dual face is of discrete type, as
then we have in the denominator of (\ref{eq:lspinfoamjp}) the
following expression $\Big ( \frac{\vec{P}^2}{2} - \frac{m^2}{8} -
\frac{1}{2} J(J+1) \Big )$, which vanishes when

\[
\vec{P}^2 = J(J+1) + \frac{m^2}{4},
\]

which, if we set $m^2 =1$, gives $\vec{P}^2 = \frac{\Delta_J^2}{4}$ as in the Riemannian case. \\

The other type of poles occurs when the relevant representation is
of a continuous type as then we have in the denominator of the
same equation the expression $\Big (\frac{\vec{P}^2}{2} -
\frac{1}{8} +  \frac{\rho^2}{2} + \frac{1}{8} \Big )$. Which
vanishes when

\[
\vec{P}^2 = \rho^2 + \frac{m^2 - 1}{4},
\]

which, on setting $m^2=1$, gives $\vec{P}^2 = - \rho^2 = - \frac{\Delta_{\rho}^2}{4}$.\\

If we interpret these formulae as giving the semiclassical values
of the \lq length', we arrive at the intriguing fact that there
are no preferred spacelike lengths as $\rho$ is continuous and
thus the \lq poles' at the $\Delta_{\rho}$'s fill the line. In
contrast, there are preferred timelike \lq lengths', which are the
discretely spaced $\Delta_J$'s.\\

Finally, by doing a \lq Wick rotation\rq $(P_1, P_2, P_3 )
\rightarrow (P_1, i P_2, i P_3)$ \footnote{We can do it because we
are essentially sitting in 3d Minkowski with the usual metric.},
we can perform the integrals over the $\vec{P}_f$'s along the
lines this was done in the Riemannian case. The asymptotic formula
(\ref{eq:asymptotic}) goes through essentially unchanged, and we
get

\[
 A_{f^*} ( J_{f^*} , N_{f^*} )  \sim
\frac{1}{F(N_{f^*})} \,
\,  \,  \Bigg [  \frac{8}{\Delta_{J_{f^*}}}    \Bigg
]^{N_{f^*}-1}  \quad \textrm{or} \quad  A_{f^*} ( J_{f^*} , N_{f^*} )  \sim
\frac{1}{F(N_{f^*})} \,
\,  \,  \Bigg [  \frac{8}{ \Delta_{\rho_{f^*}}^2}    \Bigg
]^{N_{f^*}-1}
\]

the only relevant difference being that the factor second factor
(which dictates the behaviour of the asymptotic) is now either $[
\frac{8}{\Delta_{J}^2}   ]^{N-1}$ (as before) or alternatively
equal to $[  \frac{8}{\Delta_{\rho}^2}   ]^{N-1}$. The conclusion
is the same as before : for large $N_f$'s it is only the lowest
$J$'s and $\rho$'s that contribute ($J_f < 1$ and $\rho_f <
\sqrt{8}$).\footnote{Strictly speaking the function F here, when the representation is of continuous type, is different from the F in (\ref{eq:asymptotic}). However, the exact form of F is of no importance for us here as long as it is still a function of polynomial growth. We will continue to use this notation for this prefactor in 4d as well.}\\

Note that we could have performed the mentioned \lq Wick
rotation\rq anywhere in the above discussion. Most importantly, we
could have done it in the triangulation amplitude
(\ref{eq:lamplitude}). Since, this amplitude is just a partition
function for gravity, we thus see that there is a straightforward
way of performing the \lq Wick rotation\rq in the gravity
partition function coming from the new model which does not rely
on the existence of any particular time slicing. We would like to
point out however, that this \lq Wick rotation\rq (although very
similar to the rotation in the squares of the edge lengths
performed in causal dynamical triangulations \cite{DT}) is not the
complete story, in the sense that it does not turn the action for
Lorentzian gravity into one for Riemannian gravity, nor it turns
complex exponentials into real ones (thus quantum mechanical
amplitudes into statistical weights). This is due to the fact that
we have a first order theory with the B and A fields being totally
independent. Thus, while we Wick rotate the B field to a Euclidean
one, we do not touch the connection. In this sense, the label \lq
Wick rotation' is a slight abuse of language, as it really
corresponds to some sort of partial or \lq half-performed Wick
rotation\rq, from a geometric perspective, hence the quotation
marks. However, we find it very intriguing that even this partial
transformation can be performed in such a natural way, and believe
it can be a good starting point for a similarly natural, but this
time complete definition of a geometric Wick rotation in
simplicial quantum gravity. \\

Summarizing, we see that the Lorentzian case is not particularly
different from the Riemannian one. There is essentially only one
major, qualitative difference, which stems from the fact that the
Lorentzian geometry is richer than the Euclidean one. Due to the
first order nature of the theory, in the Lorentzian setting one
gets additional, classically forbidden, histories, which have \lq
mismatching' B and A fields. These histories are, as is customary
in quantum mechanics, exponentially suppressed. As for the rest
the same simplicial gravity path integral interpretation for the
Feynman amplitudes of our GFT applies, and similar types of
quantum corrections to the 1st order Regge action are identified.

\section{New 4d GFT models}
\subsection{Riemannian BF theory}

We now come to deal with the four dimensional case (D=4). Our
discussion in this subsection and the next will be rather brief
as, if we stick to (causally restricted) BF theory (as opposed to
gravity, in higher dimensions), there is little difference between
the 3d and 4d cases. Our main aim in the present section is indeed
to show explicitly that there are no qualitative new features
added to the model by going to the fourth dimension, in neither
the Riemannian nor the Lorentzian signatures, which shows how our
proposed new class of GFTs behaves similarly in any dimension. As
we shall see below, the four dimensional models are essentially carbon copies of the three dimensional ones.\\

The only \textit{absolutely crucial} difference between 3 and 4
dimensions appears, of course, when one tries to convert BF theory
into a gravitational one. At the continuum level, this is done by
imposing the so-called simplicity constraints on the B field, in a
Plebanski-like formulation of gravity. Since we're interpreting
the P variables as a discretized B field, the difference between
having BF and gravity lies in these variables, and indeed we
expect the discrete analogue of the Plebanski constraints to be
imposed on them, when passing to gravity \cite{iosimplicity}, as
they indeed have the needed component structure (see the
discussion in the third subsection below). In this work, however,
we will treat the P's as being just elements of a metric vector
space isomorphic to the Lie algebra of G, neglecting any further
constraint. Thus our discussion will be rather and will amount to
little more than a presentation of the results. \\

The action becomes

\begin{eqnarray*}
S & = & \frac{1}{2 \, (2 \pi)^{24}} \int_{G^4} \bigg ( \prod_{i=1}^4 \, dg_i \bigg ) \, \int_{P^4} \bigg ( \prod_{i=1}^4  dP_i  \bigg ) \, \, \,  \phi^*(g_i ; P_i)  \bigg [ \prod_{i=1}^4 \Big (  \, P_i^2 \, +  \, \Box_{G_i} \,  - \frac{m^2}{4} \Big ) \bigg ] \phi(g_i ; P_i) + {} \\
& & {}  +   \frac{\lambda}{(2\pi)^{120} 5!} \sum_{\nu_1 \ldots \nu_5} \int_{G^{20}} \bigg ( \prod_{i \neq j = 1}^5 \,  dg_{ij} \bigg ) \, \int_{P^{20}} \bigg ( \prod_{i \neq j =1}^{5} dP_{ij}      \bigg ) \, \,  \bigg [ \prod_{i < j} \delta(g_{ij} g_{ji}^{-1}) \delta(P_{ij} - P_{ji}) \bigg ] \times{}  \\
& & {} \times \phi^{\nu_1}(g_{1j} ; P_{1j})  \ldots \phi^{\nu_5}(g_{5 j} ; P_{5 j}).
\end{eqnarray*}

The group that we are using for the Riemannian version of the 4d
theory is the double cover of the rotation group in 4 dimensions
SO(4), which is just Spin(4) $\simeq$ SU(2)xSU(2). The fact that
the group is a direct product of two copies of the group we used
for the 3d Riemannian case allows us to carry over easily
essentially all the results we discussed in that case to the 4d
setting. The reason for this is the fact that the Schroedinger
kernel on $G_1 \times G_2$ is just the product of the kernels on
$G_1$ and $G_2$. This in turn follows from the fact that the
Laplacian on the direct product of two groups is just the sum of
the two Laplacians $\Box_{G_1 \times G_2} = \Box_{G_1} +
\Box_{G_2}$. This allows us to write down the kernel on
SU(2)xSU(2) right away, essentially by squaring the expression
given in (\ref{eq:su2kernel})

\be
\label{eq:so4kernel}
K [ H \,  , \, T ]   =   \frac{1}{(4 \pi i T )^3}  \frac{ [ \,  \theta_1(H) \, ] [ \, \theta_2(H) \, ]}{4 \textrm{Sin}\Big(\frac{[ \, \theta_1(H) \, ]}{2}\Big) \textrm{Sin}\Big(\frac{[ \, \theta_2(H) \, ]}{2}\Big)}\, \, \, \, \textrm{Exp} \bigg [  \frac{i}{2T}\bigg ( [ \, \theta_1(H) \, ]^2 + [ \, \theta_2(H) \, ]^2 \bigg )  + \frac{i T}{4}      \bigg ].
\ee

We are of course using the same notation as before with respect to
the periodic parameters $\theta_1$ and $\theta_2$. As before, we
want to calculate the Feynman graph/triangulation amplitude
$Z_{\Gamma} = Z_{T^*}$. Since this amplitude when written in terms
of the (g,P) variables factorizes per dual face, we concentrate on
the amplitude for a single dual face. To get the dual face
amplitude for a face with N vertices $A_N[ H \, , \, P ]$, we
multiply the expression of the kernel by $\frac{\theta(T)
T^{N-1}}{(N-1)!}$ and take the Fourier transform evaluated at $P^2
- \frac{m^2}{4}$, i.e.

\[
A_N [ H \, , \, P ] = \frac{1}{(N-1)!} \int_0^{\infty} dT \, e^{i (P^2  - \frac{m^2}{4}) T} \, T^{N-1} \, K [ H \, , \, T].
\]

Since the group is compact, its Killing form, in our conventions,
is positive definite. Also, since the space P is isometric to the
(dual of) $\mathfrak{spin}(4) \simeq \mathfrak{su}(2) \oplus
\mathfrak{su}(2)$ we have $P = P_1 \oplus P_2$, with $P_1 \simeq
P_2 \simeq \mathfrak{su}^*(2)$. Thus, (with our normalizations)
$P^2 = \frac{1}{2} ( \vec{P}_1^2 + \vec{P}_2^2   ) = \frac{1}{2}
\vec{P}^2$. Also, below we will denote the combination $\sqrt{[ \,
\theta_1(H) \, ]^2 + [ \, \theta_2(H) \, ]^2}$ as $[ \Theta(H)]$.
As in the 3d case, this (equivalence class of) parameter(s) has
the geometric interpretation as the square distance between the
origin and the point on the group manifold corresponding to the
holonomy $H$, measured along a geodesic.\\

Using the formula (\ref{eq:hankel}), we get

\[
A_N \big [H \, , \, P \big ] =    \frac{i^{N-1}}{(16 \pi)^2
(N-1)!} \, \, \frac{ [ \,  \theta_1(H) \, ] [ \, \theta_2(H) \,
]}{ \textrm{Sin}\Big(\frac{[ \, \theta_1(H) \, ]}{2}\Big)
\textrm{Sin}\Big(\frac{[ \, \theta_2(H) \, ]}{2}\Big)} \, \, \Bigg
[  \frac{[\Theta(H)]}{\sqrt{\vec{P}^2 - \frac{m^2}{4} }}  \Bigg
]^{N-3}  \, \, H^{(1)}_{N-3} \Big ( \sqrt{\vec{P}^2 - \frac{m^2 -1
}{2}}  \, [ \Theta(H) ]\Big ),
\]

with the same analytic continuation in the variables as in the 3d
case. The Hankel function of integer order does not have an
expression in terms of elementary functions analogous to
(\ref{eq:polynomial}). Instead it is given in terms of the
following non-elementary integral\footnote{Strictly speaking this
formula is valid only when the order of the Hankel function is
greater than 0, i..e when $N \geq 3$. However, there is a very
simple relation between a Hankel function of a negative order with
the one of a positive one, which is $H_{- \nu}^{(1)}(z) = e^{i \nu
\pi} H_{ \nu}^{(1)}(z)$. This means that all we need to do when
N=2 (this is the only allowed value for N which is less than 3,
since any dual face has at least two vertices) is multiply the
given formula by a sign.}

\[
H_{N-3}^{(1)} (z) = - \frac{2^{N-2} \, i \, z^{N-3}}{\Gamma(N - \frac{5}{2}) \sqrt{\pi}}\Bigg \{ \int_0^{\frac{\pi}{2}} ds \, \frac{\textrm{Cos}^{N - \frac{7}{2}}(s) \, e^{- i ( N - \frac{7}{2}) s}    }{ \textrm{Sin}^{2N - 5}(s)  } \, \textrm{Exp} \Big (- 2 z \textrm{Cot(s)} \Big ) \Bigg \} \, \, e^{i \, z}.
\]

As we see, the formula above still furnishes a natural split of the amplitude into an exponential piece and a \lq measure' piece. Thus the dual face amplitude is equal to

\[
A_N \big [H \, , \, P \big ] =  \mu ([\Theta(H)] , \vec{P}, N ) \, e^{i \sqrt{\vec{P}^2 - \frac{m^2 -1}{2}} | [ \Theta(H) ] |},
\]

with $\mu$ given by

\begin{eqnarray}
\mu ([\Theta(H)] , \vec{P}, N ) & = & - \frac{i^N 2^{N-10}}{\pi^{\frac{5}{2}} (N-1)! \Gamma(N - \frac{5}{2})} \, \frac{ [ \,  \theta_1(H) \, ] [ \, \theta_2(H) \, ]}{ \textrm{Sin}\Big(\frac{[ \, \theta_1(H) \, ]}{2}\Big) \textrm{Sin}\Big(\frac{[ \, \theta_2(H) \, ]}{2}\Big)} \,  \Big [ \Theta(H) \Big ]^{2(N-3)} \, \times {} \nonumber \\
\label{eqnarray:rmeasure}
& & {} \times \Bigg \{ \int_0^{\frac{\pi}{2}}ds \frac{\textrm{Cos}^{N-\frac{5}{2}}(s) \, \, e^{- i (N - \frac{7}{2}) \, s} }{\textrm{Sin}^{2N - 5}(s)} \, \textrm{Exp} \Big [ -2 \, \textrm{cot}(s) \, \big ( \Theta(H) \sqrt{\vec{P}^2 - \frac{m^2 -1 }{2}}   \Big ] \Bigg \}.
\end{eqnarray}

As before, we multiply together all the dual face amplitudes and obtain the amplitude for the Feynman diagram/triangulation

\[
Z_T = \int_{G^{E^*}} ( \prod_{e^* \in T^*} dg_{e^*}  )
\int_{P^{T}} ( \prod_{t \in T} d\vec{P}_t    ) \,  \, \mu( g_{e^*}
, \vec{P}_t^2, N_t) \, \, \, e^{i  S_{CBF}   },
\]

the products go over all the dual edges $e^*$ of the dual complex
$T^*$ and over all the triangles $t$ in the triangulation $T$, the
$\mu$ is the product of all the $\mu$'s coming from all the edges.
The expression $S_{CBF}$ in the exponent is

\[
S_{CBF}=\sum_t \sqrt{\vec{P}^2_t- \frac{m^2}{4}} | [ \Theta_t ] |
\]

Thus the Feynman amplitudes of the model are partition functions
for an action of discretized BF theory type. Classically, the
theory given by the action $S_{CBF}$ coincides with the one given
by the usual BF action, as the equations of motion that they
produce are the same (zero curvature)\footnote{Notice also the
similarity with the action appearing in the asymptotic (large-J)
approximation of the Barrett-Crane spin foam vertex amplitude.}.
However, quantum mechanically, there is a significant difference
between the two theories. The difference being that for the usual
BF theory the integral over the B field is unrestricted, with the
integration producing the usual a-causal, real partition function.
For the model given by $S_{CBF}$ the fact that the variable P,
which represents the discretized B field, enters only through its
length (which is always positive), means that what we have is the
\lq causal' analogue of the usual BF theory (hence the subscript).
\\

It is tempting to call $\sqrt{\vec{P}^2_t - \frac{m^2 -1}{2}}$ the
area of the triangle t of the triangulation. However, this is
untenable as the variable P, being generic and non-simple (i.e.
not itself a wedge product of 4-vectors), does not have an
interpretation of defining the geometry of the triangle to which
is associated, as one needs a simple bivector to do this. As
before, the identification is cleanest if we set $m^2 =1$ which we
do in what follows to simplify the discussion and formulae. It is
clear, however, that we are setting the stage for obtaining a
proper causal spin foam model for 4d gravity, to be defined from
the above by imposition of suitable simplicity constraints on the
P variables.

\medskip

As before we write the $\mu$ in terms of magnitude and phase

\[
\mu( g_{e^*} , \vec{P}_t^2, N_t)  = | \, \mu( g_{e^*} , \vec{P}_t^2, N_t) \, | e^{i \, S_c(g_{e^*}, \vec{P}_t^2, N_t)}.
\]

Again, we interpret the $| \, \mu( g_{e^*} , \vec{P}_t^2, N_t) \, |$ as a quantum measure factor, while the phase $e^{i \, S_c}$ gives quantum corrections to the pure BF action $S_{CBF}$.\\

It is straightforward to extract the explicit expression for the phase from (\ref{eqnarray:rmeasure}). It is given by,

\[
S_c(g_{e^*}, \vec{P}_t^2, N_t) = \sum_t \Bigg [  \big ( (N+1) \textrm{mod} 4 \big ) \frac{\pi}{2}  + \textrm{arctan} \Bigg \{  \frac{\int_0^{\frac{\pi}{2}}d s \,  \frac{Cos^{N-\frac{9}{2}}(s) \, Sin \big (\frac{s}{2} - (N-4) s \big ) e^{  -2 \, cot(s) \, |\vec{P}_t| \, \Theta_t    }  }{Sin^{2N - 7}(s)} }{\int_0^{\frac{\pi}{2}}d s \, \frac{Cos^{N-\frac{9}{2}}(s) \, Cos \big (\frac{s}{2} - (N-4) s \big ) e^{  -2 \, cot(s) \, |\vec{P}_t| \, \Theta(H)  } }{Sin^{2N - 7}(s)   } }           \Bigg \}    \Bigg ].
\]

Although this expression looks totally different from the one we
had in 3d (\ref{eqnarray:measure}) many of the features of the
three dimensional model carry through without any change. Most
importantly, it still depends on the $\Theta_t$ and $|\vec{P}_t|$
solely through the combination $(| \vec{P}_t| \, \Theta_t)$,
which, at least when $\vec{P}_t$ is simple, can be interpreted to
be the discrete analogue of the Ricci scalar R. Which means that
the quantum corrections arising from the \lq measure' $\mu$ are of
the general form f(R), just like in 3d.\\

It is possible to analyze asymptotically the expression for the
phase above, along the lines done in 3d, and compute the {\it
exact} coefficients and combinatorial factors weighting the
corrections to the Regge action.\footnote{Technically speaking, it
is easier to use the asymptotic expansion of the Hankel function
\cite{grad}. This expansion looks very much like
(\ref{eq:polynomial}), which, in the half-integer-order case,
terminates and provides an explicit expression.} In the large \lq
area' asymptotic ($|\vec{P}| \rightarrow \infty$). The result is
the same as before. One gets inverse scalar curvature corrections
($ \int [ \frac{1}{R} + O(\frac{1}{R^2}) ] \, \textrm{Vol}$) to
the bare BF action, i.e. one gets infrared terms arising from the
factor $\mu$ in the large distance and semiclassical regime. \\

Also, just as is the case in three dimensions, it is possible to
see that the dual face amplitude goes like $\frac{1}{|\vec{P}|^{2N
- 8}}$ when $|\vec{P}| \rightarrow 0$.\footnote{This simply
follows from the fact that $H_{\nu}^{(1)}(z) \sim
\frac{1}{z^{\nu}}$ when z is close to zero.} As before, we would
like to draw the reader's attention to the fact that this type of
behaviour is at least consistent with, if not suggestive of, the
existence of the
continuum limit.\\

Let us move on to the J variables. Since our group is a product of
two copies of SU(2), its representation theory follows from that
of the SU(2). More precisely, each irrep of SU(2) $\times$ SU(2),
is characterized by a pair of half-integers ($J_1 , J_2$). The
dimension of such an irrep is $\Delta_{J_1} \Delta_{J_2} = (2J_1 +
1) (2J_2 + 1)$, finally the Casimir that concerns us\footnote{This
is a somewhat technical point. The space of Casimirs for this
group is two dimensional. It is spanned by, for example, the sum
and the difference of the two Casimirs coming from each factor of
SU(2). However, among all these Casimirs there is one special
which comes from the Killing form (often called, the scalar
Casimir, in the literature). It is exactly this one that
corresponds to the Laplace-Beltrami operator that we have and
which is equal to this operator's eigenvalues.} is just the sum of
the two Casimirs coming from the two SU(2) factors $\frac{1}{2}
\big ( J_1(J_1+1) + J_2(J_2+1) \big )$.\\

The 4d case corresponding to equation (\ref{eq:spinfoamjp}) is:

\[
Z_{T^*}\,=\,\sum_{J_1, \ldots, J_{F^*}} \,\Bigg ( \prod_{f^* \in
T^*} \Bigg [ \int d^6 \vec{P}_{f^*} \frac{i^{N_{f^*}}
(2J_{f^*,1}+1)(2J_{f^*,2}+1)}{\Big ( \frac{\vec{P}_{f^*}^2}{2} -
\frac{m^2}{4}  - \frac{J_{f^*,1}(J_{f^*,1}+1) +
J_{f^*,2}(J_{f^*,2}+1)}{2} + i \epsilon \Big ) ^{N_{f^*}}} \Bigg ]
\prod_{v^* \in T^*} \left\{ 15-j \right\} \Bigg ),
\]

from which we immediately the semiclassical values of $|\vec{P}|$. They are

\[
|\vec{P}| = \frac{\sqrt{(2J_1+1)^2 + (2J_2 + 1)^2 + m^2 - 1}}{2},
\]

which as before have the nicest form when $m^2 =1 $.\\

Finally, it is not difficult to perform the integrals over the P
variables, and obtain the analogue of equation
(\ref{eq:asymptotic}). The result is

\[ A_{f^*} ( J_{f^*} , N_{f^*} )  \sim
\frac{1}{F(N_{f^*})} \,
 \,
 \,  \Bigg [  \frac{8}{\Delta_{J_1}^2 + \Delta_{J_2}^2}    \Bigg
]^{N_{f^*}-1}.
\]

As we see it is entirely analogous to the one before, with the
crucial factor $[  \frac{8}{\Delta_{J}^2}   ]^{N-1}$, which
determines the asymptotic being replaced by $[
\frac{8}{\Delta_{J_1}^2 + \Delta_{J_2}^2}   ]^{N-1}$ . The
dominant contributions come from the representations for which
$\Delta_{J_1}^2 + \Delta_{J_2}^2 \leq 8$. This is satisfied only
when, neither J exceeds $\frac{1}{2}$. In other words, the two
allowed values are those corresponding to the \lq\lq vacuum\rq\rq
and to the \lq\lq lowest excited state\rq\rq. Also, note that if
we impose, by hand, the simplicity constraint at this level, in
the way it is imposed in usual spin foam models, i.e. if we set
$J_1 = J_2$ then there are exactly two dominant contributions: the
vacuum $J_1 = J_2 = 0$, once more, and the configuration with $J_1
= J_2 = \frac{1}{2}$. Once again, one can think of this as an
indication of a dynamical reduction of the model to a purely
combinatorial one of the dynamical triangulations-type.

\subsection{Lorentzian BF theory}

Finally, let us consider the case when D=4 and
G=SL(2,$\mathbb{C}$). The technical difference between the (double
cover of the) Lorentz group in 4 dimension and the one in 3 is
that in 4d the group SL(2,$\mathbb{C}$) has just one Cartan
subalgebra. Thus, apart from a set of measure zero\footnote{The
set of elements whose trace is equal to 2.}, all elements in the
group are conjugate to the elements of the Cartan subgroup, which
is the image of the Cartan subalgebra under the exponential map.
As $\mathfrak{s l}(2,\mathbb{C})$ is spanned by three rotations
and three boosts one can take the Cartan subalgebra spanned by a
rotation and a boost along the same direction, i.e. one \lq
compact' and one \lq noncompact' element. Thus the Schroedinger
kernel will be parametrized by one periodic parameter $\phi$ (with
period $4\pi$) and one non-periodic one $\psi$. One can think of
them as giving the angle of rotation and the boost parameter of
the given group element. Intuitively, since $\mathfrak{s l}(2,
\mathbb{C})$ can be thought of to be a complexification of
$\mathfrak{su}(2) \oplus \mathfrak{su}(2)$ \cite{hermann}, we will
see that what happens in the Lorentzian domain can be guessed by
complexifying the results obtained in the Riemannian one. For
example, the formula for the kernel on SL(2,$\mathbb{C}$) is
effectively a complexification of that on SU(2) $\times$ SU(2)
given in (\ref{eq:so4kernel})

\[
K [ H \,  , \, T ]   =   \frac{1}{(4 \pi i T )^3}  \frac{[ \, \theta(H) \, ] \psi(H)}{4 \textrm{Sin}\Big(\frac{[ \, \theta(H) \, ]}{2}\Big) \textrm{Sinh}\Big(\frac{\psi(H)}{2}\Big)}\, \, \, \, \textrm{Exp} \bigg [  \frac{i}{2T}\bigg ( [ \, \theta(H) \, ]^2 -\psi^2(H) \bigg )  + \frac{i T}{4}      \bigg ].
\]

As should be easy to see, the above expression is obtained by picking one of the $\theta$'s in (\ref{eq:so4kernel}) and analytically continuing it to $i \psi$.\\

Again, we want to compute the Feynman graph amplitude $Z_{\Gamma}
= Z_{T^*}$ in the (g,P) variables. Since in these variables the
total amplitude is just a product of dual face amplitudes it is
sufficient to calculate a generic amplitude $A_N[ H \, , \, P]$ of
a dual face with N vertices. According to what should be the
standard procedure by now, to get $A_N[ H \, , \, P]$ we multiply
the kernel by $\frac{\theta(T) T^{N-1}}{(N-1)!}$ and take the
Fourier transform at $(P^2 - \frac{1}{4})$, where we have set $m^2
= 1$ as this simplifies the formulae on one hand, and gives the
cleanest interpretation of the variable P on the other.

\[
A_N [ H \, , \, P ] = \frac{1}{(N-1)!} \int_0^{\infty} dT \, e^{i (P^2  - \frac{1}{4}) T} \, T^{N-1} \, K [ H \, , \, T].
\]

The Killing form on $\mathfrak{s l}(2, \mathbb{C})$ has signature $(+++---)$, thus in our normalization

\[
P^2 = \frac{1}{2} \vec{P}^2 = \frac{1}{2} \Big ( P_1^2 + P_2^2 + P_3^2 - P_4^2 - P_5^2 - P_6^2    \Big ).
\]

Now, there are two ways to do the needed integral. Either we use
the Hankel function (\ref{eq:hankel}) formula and plow ahead with
the algebra, paying attention to how we approach the cut when we
take the square root. Or we use similar arguments to what we used
when we discussed the Lorentzian case in 3 dimensions, using the
fact that mathematically we are just performing a 1-d Fourier
transform, which allows us to rely on the relation between the
Fourier transforms of the function and its complex conjugate.
Either way, the dual face amplitude is given by

\[
A_N \big [H \, , \, P \big ] =  \mu ([\Theta(H)] , \vec{P}, N ) \, e^{i \, \alpha (  [\Theta(H)],  \vec{P}^2)  \, \,  |\vec{P}^2| \, \, | [ \Theta(H) ] |},
\]

where $[\Theta(H)]^2$ is equal to $ [\phi(H)]^2 - \psi^2(H)$ which
is just the (square of the) length of the Cartan subalgebra
element parametrizing the conjugacy classes, or equivalently it is
the length of a geodesic on the group manifold joining the point
given by the element H to the identity. The $\alpha$ is given by
the following table, which is a carbon copy of the one in 3d
(\ref{eq:table})

\[
\begin{array}{c|c|c}
& Rotation &  Boost \\
\hline
Timelike  & +1   & +i  \\
\hline
Spacelike & +i   & - 1
\end{array}
\]

\ni The columns are labelled by the two possible cases of the $\Theta^2(H)$. The \lq Rotation' is when $\Theta^2(H)$ is positive, as it is easy to see that the H is then conjugate to a rotation; while the \lq Boost' is when $\Theta^2(H)$ is negative as this is when H is conjugate to a boost.\\
The rows, on the other hand, are labelled by the two possible cases of $\vec{P}^2$. \lq Timelike' is when this vector has positive length and \lq Spacelike' when this vector has negative length.\\

Finally, the $\mu$ is, apart from sporadic signs and factors of i, just the the analytic continuation ($\theta \rightarrow i \psi$) of the measure in the Riemannian case (\ref{eqnarray:rmeasure}). For completenss we give the exact formula here\\

\begin{eqnarray*}
\mu ([\Theta(H)] , \vec{P}, N ) & = &  \frac{ \mp i^{\pm N} 2^{N-10} \alpha^{N-3} (  [\Theta(H)],  \vec{P}^2)}{\pi^{\frac{5}{2}} (N-1)! \Gamma(N - \frac{5}{2})} \, \frac{ [ \,  \theta_1(H) \, ] [ \, \theta_2(H) \, ]}{ \textrm{Sin}\Big(\frac{[ \, \theta_1(H) \, ]}{2}\Big) \textrm{Sinh}\Big(\frac{[ \, \theta_2(H) \, ]}{2}\Big)} \,  \Big [ \Theta(H) \Big ]^{2(N-3)} \, \times {} \nonumber \\
& & {} \times \Bigg \{ \int_0^{\frac{\pi}{2}}ds \frac{\textrm{Cos}^{N-\frac{5}{2}}(s) \, \, e^{ \mp i (N - \frac{7}{2}) \, s} }{\textrm{Sin}^{2N - 5}(s)} \, \textrm{Exp} \Big [ -2 \, \textrm{cot}(s) \, \alpha (  [\Theta(H)],  \vec{P}^2) \, | \vec{P} | \Theta(H)   \Big ] \Bigg \}.
\end{eqnarray*}

It is straightforward to compute the Feynman graph amplitude
$A_{\Gamma} = A_{T^*}$ now. It is

\[
Z_{T^*} = \int_{G^{E^*}} ( \prod_{e^* \in T^*} dg_{e^*} )
\int_{P^T} ( \prod_{t \in T} d\vec{P}_t \,    ) \,  \, \mu( g^* ,
\vec{P}^2, N_t) \, \, \, e^{i  S_{CBF}   },
\]

where as before $\mu$ is a product of all the $\mu$'s coming from
each dual face and $S_{CBF}$ is given by

\[
S_{CBF} = \sum_t  \alpha_t  \, |\vec{P}_t|  \, |\Theta_t|,
\]

where $\alpha$ is given in the table above. Once again, we get a
\lq causal' BF action in our partition function, i.e. we get a
theory whose classical equations of motion are just like those of
the standard BF, while there are profound differences at the
quantum level.\\

As was the case in 3d we get exponential suppression of the \lq
wrong' type of correlation between the variables. More precisely,
had it not been for the fact that the variables P are in general
not simple, we could have said that in the situation when the
triangle corresponding to a holonomy given by a rotation is
timelike or alternatively when it is spacelike when the holonomy
is a boost, then this triangle contributes a phase to the
partition function. On the other hand, if there is a mismatch
between the $\Theta$ and $\vec{P}$ (rotation-spacelike or
boost-timelike), this triangle contributes an exponential
suppression factor to the partition function. The behaviour of the
model in the Lorentzian case is unaffected by dimension. \\

Also, since the \lq measure' factor is effectively the same as in
the Riemannian case, its phase depends on the deficit parameter
$\Theta$ and on the \lq area' $|\vec{P}|$ in the same way as the
bare BF action does, i.e. the phase of $\mu$ is a function of
$\alpha_t  \, |\vec{P}_t|  \, |\Theta_t|$ (as well as the $N_t$'s
characterizing the triangulation). This fact is interpreted as
before to mean that there are quantum corrections arising from the
factor $\mu$ of the general f(R) type.\\

The semiclassical analysis is exactly the same as in the
Riemannian case and one sees that in the limit of large \lq
areas', we get inverse scalar curvature corrections to the bare BF
action as before. Finally, the amplitude is as divergent as before
in the neighborhood of the $\vec{P}^2 = 0$. As in the previous
section, we consider this fact to be a necessary condition for the
existence of the continuum limit.\\

It is not difficult to write down the full Feynman amplitude in
the (J,P) variables. The relevant representations of $SL(2,
\mathbb{C})$ are labelled by two parameters. A half integer J and
a real positive parameter $\rho$. The relevant Casimir and
normalizations is equal to $C_{J , \rho} = \frac{1}{4} - \frac{J^2
- \rho^2}{8}$. The analogue of (\ref{eq:spinfoamjp}) is now

\be
\label{eqnarray:lspinfoamjp}
Z_{\Gamma}\,=\,\bigg (\prod_{f*\in
T^*} \Big [  \sum_{J_{f^*}} \int_0^{\infty} d \rho_{f^*} \Big ] \int_P d^3 \vec{P}_{f^*}\,\bigg ) \prod_{f^* \in
T^*}  \frac{i^N (J_{f^*}^2 + \rho_{f^*}^2)}{\Big ( \frac{\vec{P_{f^*}}^2}{2} - \frac{m^2}{4} + \frac{1}{4} - \frac{J_{f^*}^2 - \rho_{f^*}^2}{8}    \Big )^{N_{f^*}} }    \prod_{v^* \in T^*} \left\{ 15-J \right\},
\ee

The poles in the expression (\ref{eq:spinfoamjp}) are obviously located at

\[
\vec{P}^2 = \frac{J^2 - \rho^2}{4} + \frac{m^2 - 1}{2}
\]

so these particular values are the preferred semiclassical \lq areas'. \\

Finally, as in 3d there exists a simple way of performing a
(partial, as explained) Wick rotation in this model, by
analytically continuing (some components of) the P variables. This
shows that also the existence of a good Wick rotation in our model
is independent of the dimension. We can use this Wick rotation to
perform the integrals over the P's in the amplitude in the (J,P)
variables (equation (\ref{eq:spinfoamjp})) and obtain the 4d
Lorentzian analogue of (\ref{eq:asymptotic}), which is given by

\[ A_{f^*} ( J_{f^*}, \rho_{f^*} , N_{f^*} )  \sim
\frac{1}{F(N_{f^*})} \,
\,  \,  \Bigg [  \frac{8}{J_{f^*}^2 - \rho^2_{f^*}}    \Bigg
]^{N_{f^*}-1}.
\]

The relevant difference from the Riemannian case is that the
factor $[  \frac{8}{\Delta_{J_1}^2 + \Delta_{J_2}^2}   ]^{N-1}$,
which controls the way the asymptotic behaves, gets replaced with
$[ \frac{8}{J^2 - \rho^2}   ]^{N-1}$. This means that the most
dominant contributions are those which satisfy $|J^2 - \rho^2| <
8$. This does not of course force the J and the $\rho$ to each be
small (as was the case in 3D). However, it does force the
Minkowski \lq length' (or more appropriately area) to be small.
Note however, that if we restrict -by hand- to representations
which are simple (i.e. those for which either J or $\rho$ is
zero), the expression above \textit{does} force each of the
parameters to be small, hinting again at a dynamical reduction to
a dynamical triangulations-like sector.

\subsection{Discussion: a new route from BF to gravity?}

We have presented above a new GFT model for a BF-type formulation
of quantum simplicial gravity, in 4 dimensions, in the spin foam
formalism. The spin foam amplitudes (GFT Feynman amplitudes) have
the form, modulo a quantum measure, of the exponential of a
classical 1st order action based on two types of variables: a set
of bivectors associated to 2-simplices of the simplicial complex
and a set of Lorentz group elements representing parallel
transports of a Lorentz connection. The action has a Regge
calculus expression, augmented by higher order terms that can be
interpreted as quantum corrections, that become negligible in the
semi-classical limit. In this generalized simplicial gravity
action, the areas\footnote{This terminology represents a slight
abuse of language, as the geometric interpretation of $\mid
B_f\mid$ as area of the corresponding triangle is, strictly
speaking, not applicable until the (discrete) bivector is
constrained to be a quadratic function of a (discrete) tetrad
vector, as noted earlier.} of the triangles are functions of the
bivectors and the deficit angle associated, again, to each triangle
obtained from the holonomy of the same Lorentz connection, and
thus a function of the corresponding group elements. The equations
of motion following variation of the dominant contribution to the
action, in the semiclassical (large distance) limit, restrict the
holonomies to be flat, just as in ordinary BF theory, but the
integration over the bivectors in the path integral does not treat
on equal footing positive and negative orientations for the
triangles, as BF theories do. The result is a complex amplitude,
as said, and not a delta function over flat connections as in BF
theory.\\

The above model seems to us to be a very promising starting point for
the construction of a GFT for quantum gravity in 4 dimensions,
both Riemannian and Lorentzian, based on the idea of gravity as a
constrained BF theory \cite{laurentkirillpuzio}. This type of
formulation has been central to the construction of all spin foam
models in 4d, and lots is known already about the constraints that
the bivectors have to satisfy in order to admit a true geometric
interpretation, i.e. to be interpretable as coming from a
discretized tetrad field, as it should be in a Palatini-like
formulation of gravity.\\

Most of the model building on GFTs and spin foam models from
constrained BF theory have used a well-motivated but rather
indirect procedure, we feel, based on the kinematical
identification of bivectors with Lie algebra generators, and thus
translating them in constraints on the Lorentz group
representations and on the intertwiners appearing in the spin foam
representation of the Feynman amplitudes \cite{SF}. This however,
as we discussed, resulted in amplitudes with a less than
straightforward relation with discrete gravity actions, and with a
symmetrization over opposite orientations that is not what we
should expect, we have argued, from a 3rd quantized gravity
perspective. Moreover, from the point of view of a path integral
quantization of BF theory, the identification of bivectors with
Lie algebra generators acting on representation spaces remains a
bit obscure, given that this holds at a quantum level while in a
path integral one integrates over classical variables only
(although not solutions to the classical equations of motion) and
does not refer to quantum states if not at the boundary. A similar
doubt concerns the more recent construction of spin foam models
\cite{EPR,eterasimone1,eterasimone2,laurentkirillnew} and related
GFTs \cite{iolaurentkirill}, based on coherent states. Here the
geometric picture behind the chosen implementation of the
constraints is much clearer, but again is justified at the quantum
level in terms of the coherent states basis in each representation
space. On the one hand this suggests a semiclassical validity only
of the identification; on the other hand, it results in quantum
amplitudes with a less than straightforward relation with
classical gravity actions, and with the same symmetrization over
opposite orientations as in usual models. In the end, the
procedure adopted may well result to be correct and our doubts
unfounded or settled, but we feel that further work is needed to
clarify the situation, and we see our new models as a useful
framework in which to do so.\\

In fact, the new 4d BF-like model, thanks to the explicit presence
of bivector variables and to their role in the discrete path
integral clearly analogous to that played by the B field in
continuum formulations, suggests that a much more straightforward
way of implementing the simplicity constraints is possible. This
is simply to constrain directly the integration over the bivector
variables of our 4d model. The simplest way of doing so is to
insert appropriate delta functions imposing the simplicity
conditions on bivectors, and one has just to make sure that this
is done consistently and as geometrically expected at the level of
each Feynman diagram. Alternatively, and preferably, one should
implement the constraints directly at the level of the new GFT
action, and for doing so one has to identify clearly which
constraints are needed in each 4-simplex (interaction term) and
which refer to the gluing of 4-simplices (kinetic term), or
whether one should instead constrain directly the field in both
terms, as it is done for the other spin foam models
\cite{SF,iolaurentkirill}. Work on this is in progress
\cite{iosimplicity}. The expected result of this new way of
implementing the simplicity constraints, starting from our 4d
model, is to obtain a constrained GFT whose Feynman amplitude are
given by a path integral for an action that could be directly
interpreted as the discretization of the Plebanski action for
classical gravity, i.e. of the form: $S= \sum_f A_f(B_f) \Theta_f
(g_{e*}) + \sum_v \lambda_v f(B_{f\mid v})$, where $\lambda_v$ are
Lagrange multipliers imposing the constraints $f(B_{f\mid v})$ on
the bivector variables $B_{f \mid v}$ associated to each dual face
(triangle) $f$ incident to each dual vertex (4-simplex) $v$ (we
have neglected here the quantum corrections to the 1st order Regge
action coming from the measure).\\

Let us stress that a result of this type would be of interest, we
believe, also from a purely simplicial gravity perspective. In
fact, recent progress in spin foam models has motivated work on
so-called \lq\lq Area Regge calculus\rq\rq
\cite{williams,areaRegge}, i.e. a formulation of classical and
quantum simplicial gravity in which the fundamental geometric
variables were the areas of the triangles of the simplicial
complex as opposed to the edge lengths as in traditional Regge
calculus. In fact, spin foam models based on constrained BF theory
ended up associating, as basic geometric variables, representation
labels to triangles, with the interpretation of areas of the same.
It was noted \cite{areaRegge}, however, that, while the
correspondence areas-edges works (almost) fine for a single
4-simplex, constraints on the areas variables are needed in order
to capture correctly the simplicial geometry as soon as more than
one 4-simplex is considered. The identification of these area
constraints have proven to be very difficult. Our model would
suggest that a better formulation of classical and quantum
simplicial gravity, directly following the continuum picture of
gravity as a constrained BF theory, would use our bivector
variables to determine the areas of triangles, and that the needed
constraints needed in order to encode the geometry are constraints
on these bivector variables, and not directly on the areas. More
precisely, the simplicity constraints will restrict the components
of the bivectors other than their modulus (area), and their
re-phasing in terms of constraints on areas is, if possible,
certainly not straightforward. In any case, the description of
simplicial geometry implicit in our models in both 3 and 4
dimensions deserves to be studied in more detail. We leave this
for future work.\\

The above 4d model, as well as its 3d version, in both Lorentzian
and Riemannian versions, would also be the natural starting point
for understanding in a more clear way the role that coherent
states for the Lorentz group play in spin foam models, and, in 4d,
for understanding better the justification for the procedure used
in the recently proposed spin foam models in order to impose the
simplicity constraints. What we would expect is that the
parameters labelling coherent states for the Lorentz group in both
3d and 4d models will be directly related, if not identified, with
the new (bi-)vector variables we introduce {\it in the
approximation in which our free field classical equations of
motions are satisfied}, i.e. in the approximation in which one
substitutes, in the amplitudes for our 3d and 4d models, the
generators of the Lorentz Lie algebra for the (bi-)vector
variables, which indeed represents a dominant (semiclassical)
contribution to our amplitudes, as it is clear from the structure
of our propagators. However, further work is needed to confirm or
refute this expectation. This work is currently in progress
\cite{iotimcoherent}.

\section{Conclusions and outlook}
We have presented a new class of GFT models for the dynamics of
quantum geometry, in any spacetime dimension and signature. The
construction was based on the extension of the GFT formalism to
include additional variables with the interpretation of a discrete
counterpart of the continuum B field in BF-like formulations of
gravity. The Feynman amplitudes for the new GFTs, i.e. the
corresponding spin foam models, have exactly the form of true
simplicial gravity path integrals, with a clear-cut relation with
discrete gravity actions, as opposed to other known models in
which the connection arises only in some asymptotic limit. In 3d
the new models are seen to provide a quantization of discrete
quantum gravity in 1st order (Palatini) form, in a local and
discrete 3rd quantized framework in which topology is allowed to
fluctuate. In 4d and higher, the new models have the form of a 3rd
quantized framework for BF theory, but with an additional
dependence of the amplitudes on the orientation of the simplicial
complex, of the type on would expect in a path integral
quantization of 1st order gravity.\\

The Lorentzian models also present a very nice interplay between
the two sets of discrete variables (B field and connection) which
leads automatically to a suppression of all the configurations
which do not match the simultaneous geometric interpretation of
both of them.\\

The GFT provides also a precise prescription for the quantum
correction to the classical Regge-like action (in 1st order form)
that have to be included in the corresponding path integral, in
absence of further restrictions to the models. These additional
terms in the action become negligible in both the continuum limit
(large number of simplices of small size) and in the
semi-classical limit (arbitrary number of simplices but large size
of simplices, thus large associated action), leaving only the
Regge action to contribute to the path integral, as one would
expect. In the general case, and as soon as one goes beyond these
limiting regimes, the simplicial action provided by our GFT models
turns into a generic $f(R)$ extended action for gravity. We feel
that this has several interesting implications at the simplicial
gravity level as well as from a more phenomenological perspective,
that deserve to be studied in more detail.\\

The way the large-P limit affects the amplitudes of the new models
sheds new light, we feel, on the usual large-J limit\footnote{The
large P regime is related to the large j regime, as we have seen,
in a \lq semi-classical approximation of the amplitudes\rq when
these are expressed in the (j,P) variables.} that brings usual
spin foam models in closer relation with simplicial gravity path
integrals, by allowing an approximation of the vertex amplitudes
with the cosine of the Regge action. Indeed, our models suggest
that this limit is a large distance limit which is equivalent, at
the discrete level, to a large action and thus semi-classical
limit (because of the way the Regge action, and its higher order
corrections, depends on the hinge volumes). As such, it has two
effects: it kills any quantum interference between opposite
orientations for the hinges, in the usual models only, in which
such opposite orientations are treated on equal footing; it kills
any short-distance effect such as $R^n$ corrections to the action,
leaving only the Regge term as the leading contribution to the
quantum amplitudes, with next to leading order contributions being
represented by inverse curvature terms $1/R^n$ which indeed modify
the IR physics of the corresponding classical discrete gravity
theory.\\

Let us also mention that the explicit presence of a discrete
analogue of the B field in our amplitudes allows a rather
transparent definition of a Wick rotation in simplicial quantum
gravity

\medskip

Our results, as we have discussed, support the view of GFTs as
local and discrete 3rd quantizations of gravity, providing a nice
field theoretic description of the quantum dynamics of the
fundamental building blocks of quantum space
\cite{iogft,iogft2,laurentgft, dgftreview}. Also, the new models seem to
implement nicely the ideas, discussed at length as a motivation
for the work we have presented in this paper, as well as in
\cite{causal,feynman,generalised}, on the notion of causality and
orientation dependence in quantum gravity, and to provide a
definition of {\it causal transition amplitudes} for quantum
gravity states, with all the expected properties.\\

At a more practical level, the new 4d models represent, in our
opinion, a very convenient starting point for the construction of
a GFT (and spin foam, and simplicial quantum gravity) formulation
of quantum gravity as a constrained BF theory, based on a more
straightforward and geometrically clean procedure of implementing
the so-called simplicity constraints that reduce BF theory to
gravity, than in other spin foam formulations, as we have
discussed above. Also, they offer a new context in which to study
the low energy physics of GFTs and loop quantum gravity, e.g.
graviton propagator calculations \cite{graviton}.\\

Even more importantly, maybe, the new models, and possible
modifications of the same, seem to provide the long sought for
{\it explicit} unifying framework for spin foam/loop quantum
gravity and simplicial quantum gravity approaches (quantum Regge
calculus and dynamical triangulations). Looking at these different
approaches from the proposed common GFT framework can offer, we
hope, new possibilities for mutual enrichment and
cross-fertilization between the various lines of research that are
currently pursued as separate avenues toward a common goal, in
particular regarding the outstanding issue of the continuum and
semiclassical approximation of the discrete picture of quantum
geometry they all seem to be based on \cite{ioemergence}.

\section*{Appendix}

Here we collect a few basic facts about groups \cite{knapp} used
in the discussion in the main text. \\

A Cartan subalgebra is a maximal abelian subalgebra of the Lie
algebra $\mathfrak{g}$. The image of the Cartan subalgebra under
the exponential map is called a Cartan subgroup. It is possible to
show that all Cartan subalgebras (subgroups) have the same
dimension r, which is called the rank of the group.\\

In a compact group all its Cartan subgroups are conjugate to each
other and any element of the group is conjugate to an element of
some fixed Cartan subgroup. This means that we can parametrize the
conjugacy classes of the group by the elements of the Cartan
subalgebra.\\

In a noncompact group the situation can be more complicated as on
one hand not all Cartan subgroups are conjugate. and the other,
even if we do select a representative from each conjugacy class of
Cartan subgroups it is not true in general that each element of
the group is conjugate to an element of one of the selected
subgroups. Fortunately, however, these exceptional elements form a
set of measure zero in the Haar measure in the cases that we are
interested in. At the practical level, what follows from the above
is that on a noncompact group there is no global \lq spherical
coordinate system' \cite{marinov}, i.e. it is impossible to
parametrize all the elements of the group by r parameters
(elements of a fixed Cartan subalgebra) like in the compact case.
Rather, there will be several different domains in general with
all elements in each domain being conjugate to a Cartan subgroup
of a specific topology, with the elements which don't lie in any
domain forming a set of measure zero. The elements in each domain
are parametrizable by a set of r parameters, which are just the
elements of the corresponding Cartan subalgebras.\\

Intuitively, if one uses a Cartan decomposition of $\mathfrak{g}$,
i.e. if one writes $\mathfrak{g} = \mathfrak{k} + \mathfrak{p}$,
with the Killing form being negative definite on the elements of
$\mathfrak{k}$ (the \lq compact' part as the 1-parameter subgroup
generated by any element in $\mathfrak{k}$ is just a circle $S^1$)
and positive definite on the elements of $\mathfrak{p}$ (the \lq
noncompact' part, the 1-parameter subgroup generated by any
element is diffeomorphic to the real line); and if one obtains a
Cartan subalgebra by taking e.g. k generators from $\mathfrak{k}$
and p ones from $\mathfrak{p}$ (with k + p = r of course). Then
the Cartan subgroup corresponding to this subalgebra is not
conjugate to the Cartan subgroup corresponding to a subalgebra
formed from a different relative proportion of compact and
noncompact elements. If we call the elements generated by
$\mathfrak{k}$ \lq rotations' and the ones generated by
$\mathfrak{p}$ \lq boosts', then the different Cartan subgroups
correspond to different relative number of rotations to boosts.
The r parameters which parametrize the elements in each domain are
split into two classes, periodic and aperiodic. The number of
periodic parameters is equal to the number of rotations while the
number of the aperiodic ones is equal to the number of boosts.

\end{document}